\documentclass[11pt,twoside]{article}
\usepackage{psfig}
\usepackage{amsmath}
\usepackage{amssymb}
\usepackage{cite}
\usepackage{array}
     \newcommand{\ds}{\displaystyle}
     \newcommand{\scs}{\scriptstyle}
\begin{document}\hbadness=10000\begin{center}
\ 

\thispagestyle{empty}\vskip 1cm
     \centerline{\bf STRANGE PARTICLES}
\centerline{\bf FROM} 
				\centerline{\bf DENSE HADRONIC MATTER}
\vskip 0.5cm
\uppercase{Johann Rafelski}\\ {
     Department of Physics, University of Arizona\\ 
     Tucson, AZ 85721}\\ \ \\
\uppercase{Jean  Letessier \lowercase{and} Ahmed Tounsi}\\ {
     Laboratoire de Physique Th\'eorique et Hautes Energies\\
     Universit\'e Paris 7, Tour 24, 2 Pl. Jussieu, F-75251
     Cedex 05, France}\\ \vskip 0.7cm
\vskip 0.5cm
\centerline{\it (Received April 3, 1996)}

 \end{center}  
\vspace*{-10cm}
\centerline{Vol. 27 (1996)\hfill {\it ACTA 
PHYSICA  POLONICA B}\hfill No 5}
\vspace*{9.5cm}

\begin{abstract}
After a brief survey of the remarkable accomplishments of
the current heavy ion collision experiments up to 200A GeV,
we address in depth the role of strange particle production in
the search for new phases of matter in these collisions. In particular, 
we show that the observed enhancement pattern of otherwise
rarely produced multistrange antibaryons can be consistently explained
assuming color deconfinement in a localized, rapidly disintegrating 
hadronic source. We develop the theoretical description of this source,
and in particular study QCD based processes of 
strangeness production in the deconfined, thermal quark-gluon plasma phase,
allowing for approach to chemical equilibrium and dynamical evolution.
We also address thermal charm production.
Using a rapid hadronization model we obtain
final state particle yields, providing detailed theoretical 
predictions about strange particle spectra and yields 
as function of heavy ion energy. 
Our presentation is comprehensive and self-contained:
we introduce in considerable detail the procedures used in 
data interpretation, 
discuss the particular importance of selected  experimental results 
and show how they impact the theoretical developments.
\vskip 0.5cm

\end{abstract}

\vfill
\centerline{(1037)}
\newpage

\baselineskip=12pt
\tableofcontents
\baselineskip=12.9pt

\section{Introduction
} \label{RHIC}
\begin{figure}[tb]
\vspace*{2.5cm}
\centerline{\hspace*{2.7cm}
\psfig{width=10cm,figure=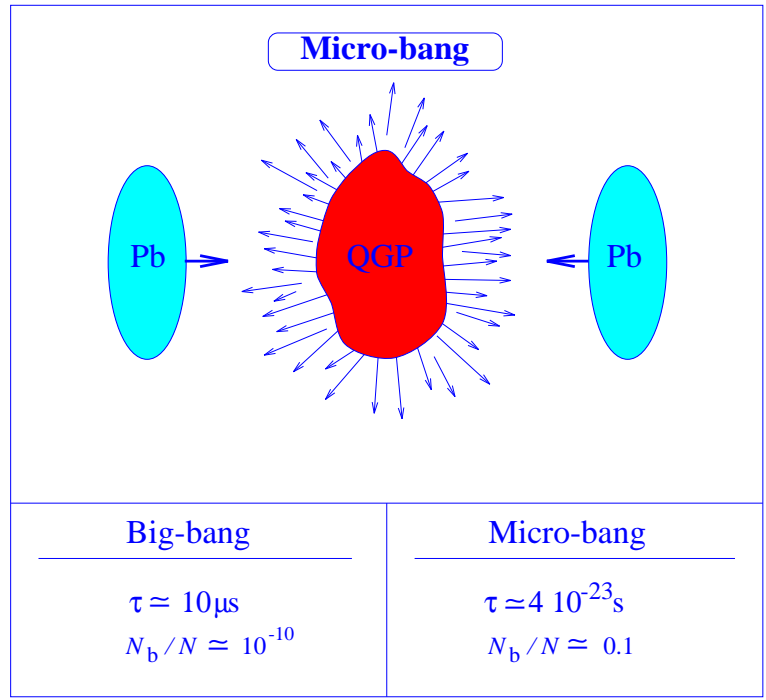}
}
\vspace*{-2.2cm}
\caption{ \small
Qualitative illustration of the relativistic nuclear 
collision and the differences between the Big-Bang 
and Micro-Bang.  \protect\label{figmicrobang} 
}
\end{figure}

Our interest is to study under laboratory conditions matter as it existed 
during the era of the early Universe at which temperatures
were in excess of 200 MeV, thought to be less than $10\,\mu$s 
after the big bang.  Beams of heaviest nuclei at relativistic energies
are the tools in this research program: in 
nuclear heavy ion collisions the
participating strongly interacting hadronic nuclear matter is
compressed and heated. Unlike the early Universe, the volume occupied by 
such laboratory  `micro bang' is small, see Fig.\,\ref{figmicrobang}. 
However, our hope and expectation is that  collisions
of largest nuclei which are now studied experimentally  
will allow us to explore 
conditions akin to infinite systems of hot hadronic matter. 
Another difference  with the early Universe condition is, 
as shown in Fig.\,\ref{figmicrobang}, that though 
one of the most characteristic features of 
these heavy ion collisions is the formation of many new particles,
their number per nucleon (baryon) remains considerably 
smaller than was present at the low baryon 
density of matter present in the early Universe. Much of the theoretical
effort in this field is thus devoted to the understanding 
and interpretation of the experimental data and in particular 
their extrapolation to conditions of long lived, statistically equilibrated 
matter with low baryon density.

The great  variety of hadronic particles  known (mesons, baryons) 
implies that the structure and properties
of their source will be very rich, and could comprise some new and 
unexpected phenomena. Our present discussion will, however, 
be limited to consideration of just two model phases of highly 
excited hadronic matter: 
\begin{enumerate}
\setlength{\itemsep}{-0.1 cm}

\item  the conventional, confined phase we shall call hadronic gas (HG),
consisting of hadronic particles of different type, (including short lived
resonances), such as $\pi$, $\rho$, $N$, $\Delta$, etc., with masses and
degeneracies in most cases well known. Along with Hagedorn  \cite{HAG},
we  presume that particle interactions in HG are accounted  for by giving
the resonances the status of independent fractions in the gas;
\item  the deconfined phase will be seen as a liquid of quarks and
gluons, interacting perturbatively, an approach which is properly
justified only in the limit of very high energy densities.  We shall call
this phase the quark-gluon plasma (QGP). Our specific objective is to 
discover and explore  this new form of matter \cite{HM96}.
\end{enumerate}

There is today considerable interest in the study of the transformation
of strongly interacting matter between these two phases. 
Considerable theoretical
effort is committed in the framework of finite temperature lattice gauge
theory\cite{lattice}  to perform simulations with the objective to obtain a
better understanding of the properties of quantum chromodynamics (QCD)
at finite temperature. In analogy to water-vapor transition we
expect and indeed see that two phases of hadronic matter are separated by 
a first order phase transition. However, the theoretical simulations also 
suggest that it could well be that there is no
transition, just a phase cross-over, similar to the situation
prevailing in the atomic gas transition to electron-ion plasma. 
A comparison between theory and experiment will be always subject
to the constraint that a true phase
transitions cannot develop in a finite system. However, 
the number of accessible degrees of freedom that are being excited
in collisions of heavy nuclei is very large, and this should allow us
to explore the properties of the true infinite matter
phase transition using finite nuclei as
projectiles and targets. It is for this reason that beams of largest
nuclei are the required experimental tools in this research program. 
 
Another way to look at the phase transformation between the two 
prototype phases, HG and QGP,
arises from the consideration of the nature and in particular the
transport properties of the vacuum state of strong 
interactions: in the QGP phase (the `perturbative vacuum state') it is
 allowing for the free propagation of quark 
and gluon {\em color} charges. The `true vacuum' in which we live
is a color charge insulator, only the color neutral mesons and baryons 
can propagate. Because at high temperature we cross to the conductive phase,
it is possible to consider the change in the properties of the vacuum akin
to the situation with normal matter. Moreover, from the theoretical
point of view,  the observation of the `vacuum
melting' and the study of the
properties of the perturbative and true vacuum is the
primary objective of the nucleus-nucleus
high energy collision experimental program --- high energy nuclear 
collisions are today the only known laboratory method allowing the study of
extended space-time regions containing a locally modified vacuum state.

Inside the domain of perturbative vacuum, at sufficiently high excitation 
energy we expect to encounter a quantum gas of quarks and
gluons subject to the QCD perturbative interactions characterized by
the (running) coupling constant $\alpha_{\rm s}$. Even though the
strength of the QCD interactions is considerably greater than the 
strength of electromagnetic interactions, $\alpha_{\rm QED}=1/137$,
the moderate magnitude
$\alpha_{\rm s}/\pi\le 0.3$ at the energy scales corresponding
to temperatures of $T\simeq 250$ MeV should permit us
to study the quark matter in a first estimate of its properties, as if it
consisted of a gas of quarks and gluons interacting
perturbatively. We will consider in this way the  
strangeness production, and also use 
perturbative expressions in $\alpha_{\rm s}$ 
to improve the free quantum gas equations of
state of quarks and gluons. 
We use the analytical expressions up to the region of the
phase cross-over to the confined hadronic gas world, hoping that
the qualitative features of the deconfined phase will be 
appropriately described in that way. Clearly, this is a domain
that will see in future more effort both in terms of 
improvements of the perturbative expressions, and also due to
further exploration of numerical lattice gauge theory results.

\begin{figure}[htb]
\vspace*{1.5cm}
\centerline{\hspace*{-0.5cm}
\psfig{width=8.5cm,figure=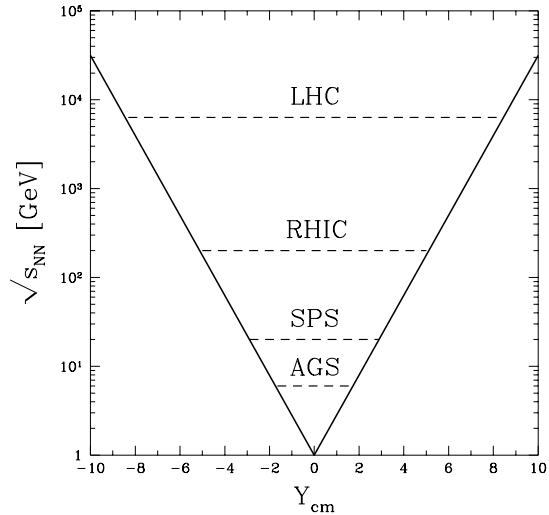}
}
\vspace*{-0.5cm}
\caption{ \small
For the nucleon-nucleon center of momentum frame (CM) 
$\protect\sqrt{s_{\rm NN}}$ energy the horizontal dashed 
lines show the maximum
rapidity gap between the projectile and target. 
 \protect\label{figrap} 
}
\end{figure}

Relativistic heavy ion experimental programs at the AGS accelerator at
the Brook\-haven National Laboratory (BNL) and at the SPS accelerator at 
the European Center of Nuclear Research (CERN) in Geneva, begun in 1986--87.
From the onset of the 
program it was assumed that the higher the collision energy
and the heavier the colliding nuclei, the greater
the energy density that could be created and hence more extreme
conditions of matter and, e.g., earlier the time since the beginning of our
Universe one expects to be able to study. Moreover, 
it is expected that as the
energy and thus the rapidity gap, see Fig.\,\ref{figrap}, 
is growing, the chances increase to 
form a truly baryon-free region on space time, more similar 
to the conditions prevailing in the early Universe, see 
Fig.\,\ref{figmicrobang}. For these reason new facilities were build and
before the end of the century, the next generation 
of experiments will exploit the 100A+100A GeV Relativistic
Heavy Ion Collider (RHIC) at BNL, sporting 10 times the CM-energy 
available today at SPS. The race is on, and at CERN collisions of
heavy ions up to about 3.5A+3.5A TeV will become feasible upon 
completion of the Large Hadron Collider (LHC). 
Here yet considerably more extreme 
conditions should be reached, as the collision energies are 400 times 
higher than accessible today  at SPS . 
The expected onset of the LHC program follows RHIC schedule by 
about 7 years. 
 
In this article we will address mainly strange particle production 
in collisions at 200A GeV. This center of interest arises from 
the prediction \cite{Raf82}, 
that the (enhanced) production of (multi)strange antibaryons is 
specifically related to color deconfinement: in the deconfined QGP phase we 
find enhanced production of strangeness flavor by thermal glue based 
processes, leading to high $\bar s$ densities, which in turn leads to highly
enhanced  production of strange  antibaryons. 
Moreover, the strange antibaryon particle production
mechanisms being very different from the usual ones, the behavior of the 
yields (cross sections) with energy will be shown here to differ considerably
from usual expectations.  The strange antibaryon particle signature of QGP
requires that the transition from the deconfined state to the confined
final hadronic gas phase consisting of individual hadrons occurs
sufficiently rapidly in order to assure that the 
memory of the high density of
strangeness in the early phase is not erased. 

As noted above the enhanced QGP strangeness yield depends to some
extend on {\it thermal} equilibrium gluon collision frequency. 
Many experimental results, which we survey
in the following suggest that the
particles  produced in heavy ion collisions are indeed thermal, i.e., that
either they have been produced by a thermal source, e.g., in a 
recombination of thermal constituents, or that they have had time to 
scatter and thermalize after formation. This means that 
the thermalization of the energy content in heavy ion reactions is rapid on
the time scale of the collision. The required mechanisms of such
a rapid thermalization and associated entropy production are
today unknown in detail, but plausible given the large number of accessible
degrees of freedom in high energy nuclear collisions. 
 
This survey comprises three logical parts

\vspace{-3mm}
\begin{itemize}
\setlength{\itemsep}{-0.2 cm}
\item[---] Experimental results and their analysis,
\item[---] Strange quark production and fireball dynamics,
\item[---] Final state particle production,
\end{itemize}
\vspace{-3mm}

\noindent and we outline briefly in the following three paragraphs their 
respective contents.

It is clearly not possible to present here a 
comprehensive discussion of the ten years of experimental effort. 
Rather, in following 
section  \ref{tools}, we will briefly address the highlights of the
experimental results. We then turn to the main topic of the paper 
in section \ref{secSTRoverv}. We introduce strange particle 
properties, discuss diagnostic tools in more detail and describe 
the  key  strangeness (antibaryon) experimental results obtained 
at SPS in subsection  \ref{expres}.  
In section \ref{secanalysis} we use the framework of the thermal 
fireball model to analyze the experimental 
strange antibaryon ratios and to derive the properties of the 
source. 

In section \ref{thermalfsec} we develop the equations of
state of the QGP-fireball and use these to determine  
the initial conditions which we expect to be formed in 
different collisions. Applying conservation of energy and baryon
number we also obtain the properties of the fireball at
different important instances in its evolution. This is done 
for all systems studied currently at CERN and BNL experimental 
facilities and we show that there 
is a profound difference between the 15A GeV BNL data and 
the 200A GeV CERN data, which precludes interpretation 
of the low energy results in terms of a (suddenly disintegrating) 
QGP-fireball. We then turn our attention in section \ref{strprod} 
to a  comprehensive
study of the QCD--QGP based thermal strangeness production and
also discuss briefly the related topics  of the thermal charm production. 
The production of strangeness being strongly dependent on 
the magnitude of the strong interaction coupling constant, we develop
in section \ref{runalfasec} the renormalization group based 
description of the appropriate value. 
In section \ref{gammas} we explore the variation of the phase space 
occupancy of strange and charmed particles in the different
collision environments.

We now turn our attention to the particle production yields: 
we discuss the hadronization constraints
and parameters in the section \ref{hadromod}, and present 
the excitation functions of multistrange particles and their ratios
in section \ref{results}. In the final section \ref{endrem} we give
a brief evaluation of our work.
 
\vspace{-3mm}
\section{Diagnostic tools} \label{tools}
The reader should be aware from the outset
that the observation of a transient new phase of
matter, formed and existing just for a brief instant in time, 
perhaps for no more than $10^{-22}$s, is only possible if time 
reversibility is broken in a shorter time, which is implicitly presumed in this
field of research. In our work this is implicitly assumed when we
introduce the thermal fireball. However, how this quantum decoherence 
occurs is one of the great open problems that challenges us today. 

\vspace{-3mm}

\subsection{Principal methods}
Given the assumption of rapid decoherence we can 
seek accessible observables which can 
distinguish between micro-bangs comprising the two
prototype phases, the HG, or the QGP that subsequently hadronizes.
Several useful experimental signatures of dense hadronic
matter and specifically the formation and properties of QGP have been
now theoretically and experimentally explored.  
These can be categorized as follows:
\begin{enumerate}
\setlength{\itemsep}{-0.1 cm}

\item {\bf Electromagnetic probes:}\\
$\bullet$ {\it direct photons} and {\it dileptons}  \cite{Feinberg,Wong}.
Since quarks are electromagnetically
charged, their collisions produce these particles, and the yields are
highly sensitive to the initial conditions, for example in relatively
`cold' matter the direct photons can be hidden by the
$\pi^0\to\gamma\gamma$ process.

\item {\bf Hadronic probes:}\\
$\bullet$ {\it strangeness}  
\cite{Raf82,KMR86,SH95,AIP,GR96} (and also charm) is
the topic of primary interest here --- 
theoretical considerations  show that in the deconfined 
quark-gluon plasma  (QGP) phase high local strange and 
antistrange particle density is reached permitting abundant formation 
of strange antibaryons. Furthermore, enhanced production of strangeness is 
expected comparing QGP-based theoretical strangeness yield to reactions 
involving cascades of interacting, confined hadrons. Experimental 
comparison between A--A and N--N collisions reveals indeed such an 
enhancement at 200 GeV A \cite{NA35S}, 
not seen in N--A interactions at 200 GeV \cite{Bia92}.

$\bullet$ Global observables such as {\it particle
abundance} measure the entropy produced 
in the collision \cite{entropy,Divonne}.

$\bullet$ {\it Hanbury Brown-Twiss} (HBT) interferometry allows to 
determine the particle source size  \cite{BGJ90}.
 
\item {\bf Charmonium:}\\ 
$\bullet$ even though the charmonium state 
$c\bar c=\Psi$ is a
hadronic particle, the way it is proposed as an observable it different
from the other hadronic probes, and its small yield also reminds us more
of an electromagnetic probe. Once produced in the initial interactions,
$\Psi$ is used akin to X-rays in the Roentgen picture: from the shadows
of remaining abundance we seek to deduce a picture of the hadronic matter
traversed~\cite{Matsui,Wong}.

\end{enumerate}

We believe that our recent advances\cite{analyze,dynamic}
in the study of strange particle production have brought about
the long aspired substantiation of the formation of deconfined 
and nearly statistically equilibrated QGP phase  at energies 
available at the SPS accelerator,  
$\sqrt{s_{\rm NN}}\simeq 9 + 9$ GeV. We reach this conclusion
because the observed abundances of strange antibaryons are closely
following the expected pattern characteristic for a rapidly hadronizing
deconfined phase, at the same time as an excess of entropy
\cite{entropy} characteristic
for melted hidden (color) degrees of freedom is recorded. 

We can reach this conclusion because unlike the other particle observables,
the final state observable `strangeness' is more than just one average 
quantity which is enhanced  when  one
compares nucleon-nucleon  (N--N)  and nucleon-nucleus (N--A) reactions with 
nucleus-nucleus (A--A) interactions. The interesting
aspects of this observable is that there are many different particles,
and that certain strange particles (strange antibaryons) appear much
more enhanced, since their production is rather suppressed
in conventional interactions. 
 

\vspace{-3mm}

\subsection{Particle spectra} \label{spectrasec}
One of the best studied observables are spectra of hadronic particles ---
it is  convenient to represent these using instead of
the longitudinal momentum, the rapidity  $y$ and to use the transverse
mass $m_\bot$ instead of the transverse momentum of a particle:
\begin{equation}
y= {1\over 2}\ln\left( {{E+p_z}\over {E-p_z}}\right)\,,\quad
E=m_\bot \cosh y\,,\quad m_\bot=\sqrt{m^2+p_\bot^2}\,,
\end{equation}
where `$\bot$' is perpendicular to the collision axis `$z$'. While 
$m_\bot$ is invariant under Lorentz transformations along the collision
axis, the particle rapidity  $y$ is additive, that is, it changes by the
constant value of the transformation for all particles. This allows to
choose the suitable (CM --- center of momentum) reference frame
characterized by its rapidity $y_{\rm CM}$ for the study of the particle
spectra.
  
\begin{figure}[htb]
\vspace*{-1.cm}
\centerline{\hspace*{.7cm}
\psfig{width=9.5cm,figure=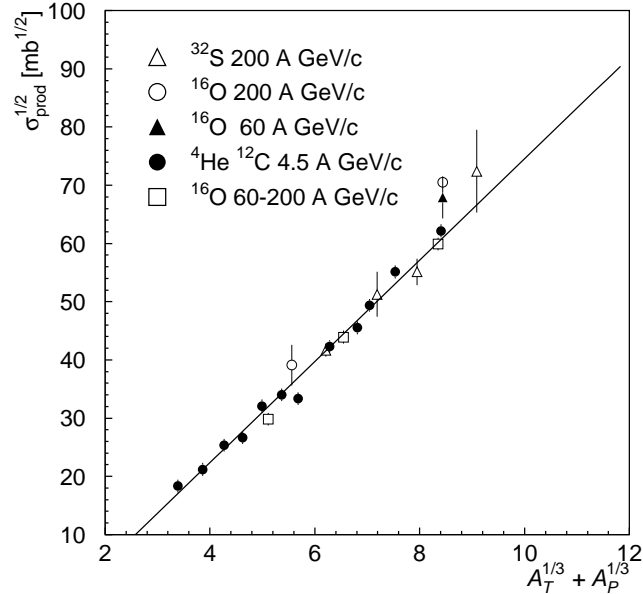}
}
\vspace*{-1.cm}
\caption{ \small
Root of the inelastic reaction cross section $\protect\sqrt{\sigma}$ 
as function of geometric size of interacting nuclei,
$A_{\rm P}^{1/3}+A_{\rm T}^{1/3}$, for different collision partners,
after Ref.\protect\cite{Ander89}.   \protect\label{figgeometry} 
}
\end{figure}

Simple kinematic considerations show that the center of momentum
frame in nuclear collisions is  for symmetric systems  just
is 1/2 of the projectile rapidity, and for asymmetric  collisions 
such  as  S--Au/W/Pb systems 
with the participating masses $A_{\rm P}$, $A_{\rm T}$  of 
the projectile and, respectively, target nuclei one finds \cite{Sch88} 
(neglecting small corrections):
\begin{equation}\label{eqycm}
y_{\rm CM}={y_{\rm P}\over 2}-{1\over 2}\ln {A_{\rm T}\over A_{\rm P}}\,.
\end{equation}
Assuming small impact parameter collisions with a suitable central
trigger,  all projectile nucleons participate while the target
participants $A_{\rm T}$ can be estimated from a geometric 
`interacting tube' model.  The geometric picture is well supported by 
the linear relation of the size of the reaction zone, defined to be 
$\sqrt{\sigma}$, the square root of the reaction cross section, and the 
geometric size of the interacting nuclei in central collision, 
$ A_{\rm P}^{1/3}+A_{\rm T}^{1/3}$. We show this in Fig.\,\ref{figgeometry}.
Once the central rapidity is confirmed experimentally, this allows the
determination of the CM-energy involved in the interaction. 
 
The geometric  approach reproduces well 
the value of central rapidity around which the
particle  spectra are centered. 
In the specific case of 200A GeV  S--Au/W/Pb
interactions one sees $y_{\rm CM}=2.6\pm0.1$ \cite{HM96}. 
However, the shape and the 
width of rapidity spectra is providing proof that much of the 
primary longitudinal momentum remains as collective longitudinal flow
which tends to expand the source in longitudinal direction.

\begin{figure}[htb]
\vspace*{0.2cm}
\centerline{\hspace*{-.2cm}
\psfig{width=8.6cm,figure=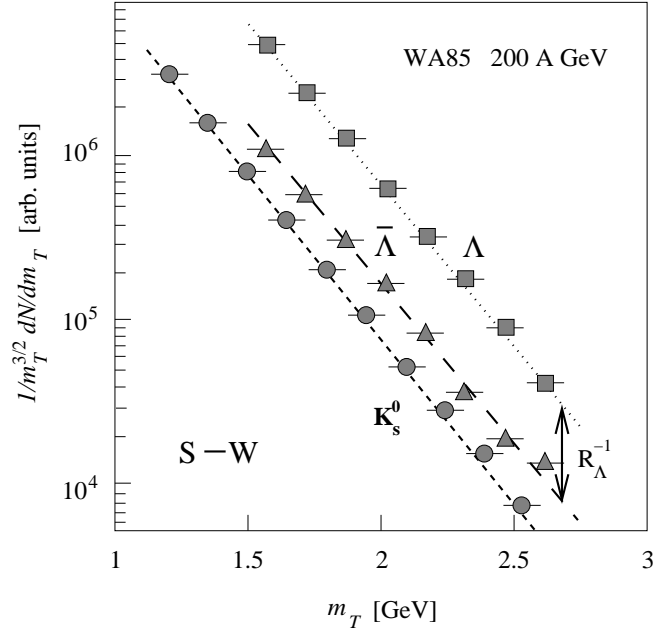}
}
\caption{ \small
Strange particle spectra for $\Lambda,\ \overline{\Lambda},\ 
{\rm K}_s$ . Line 
connecting the $\Lambda$ and $\overline{\Lambda}$ spectra, denoted 
$R_\Lambda^{-1}$, shows how the ratio $R_\Lambda$ of these particle 
abundances can be extracted. Experimental WA85 results from reference
\protect\cite{WA85T}.   \protect\label{specWA85} 
}
\end{figure}

The central rapidity WA85\cite{WA85T} transverse mass 
spectra $m_\bot^{-3/2}dN_i/dm_\bot$ of diverse strange particles
are shown in the  Fig.\,\ref{specWA85}. Similar results were also 
obtained by the NA35 collaboration  \cite{NA35S,NA35TPPI} 
and these temperatures are
consistent with the results considered here.
It is striking that within the observed interval $1.5<m_\bot<2.6$ GeV
the particle spectra are exponential, as required for a  thermal
source, irrespective of potential presence of longitudinal collision flow.
Very significantly, there is clearly a common inverse slope
temperature, with its inverse value around $T=232$ MeV. 
 
 These spectral $m_\bot$ shapes
lead to the suggestion that hadronic particles were produced by a thermal
source with temperature $T$ (particle spectra inverse slopes). 
We will discuss in more detail the many 
consequences of this simple remark in the following section \ref{TMod}.
Here, we emphasize that the observed temperatures vary for different systems
and energies. In collisions of S with Au nuclei at SPS, at 200A GeV,
record temperatures  of the magnitude of
$230$ MeV have been observed, which value is much greater than the
temperatures ${\mathcal O}$(160 MeV) that were noted in $p$--$p$
collisions  \cite{HAG}. The use of thermal models to describe the
$p$--$p$ collisions has been often critically scrutinized. However, the
experimental evidence in favor of thermal source in relativistic nuclear
collisions is overwhelming and the physics motivation considering the
large number of participating degree of freedom considerably stronger.

\begin{figure}[htb]
\vspace*{-1.5cm}
\centerline{\hspace*{0.7cm}
\psfig{width=10.6cm,figure=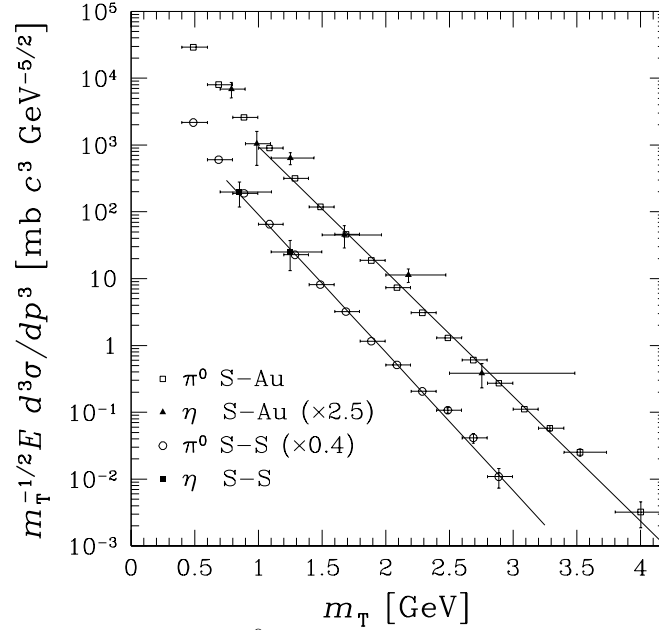}
}
\vspace*{-1.cm}
\caption{ \small
Neutral particle $\pi^0,\ \eta$ spectra (invariant cross
sections divided by $m_\bot^{1/2}$) in central rapidity interval
$2.1<y<2.9$. 
Upper solid line S--Au: thermal spectrum with temperatures $T=232$ MeV;
lower solid line S--S: $T=210$ MeV. Experimental data courtesy 
of the WA80 collaboration.   
\protect\cite{WA80spec}\label{specWA80} }
\end{figure}
It is remarkable that the same thermal behavior was seen
for this $m_\bot$ range by the WA80 
collaboration\cite{WA80spec} for the neutral hadrons $\pi^0,\,\eta$.
In Fig.\,\ref{specWA80} we have reploted the WA80 results 
multiplying the invariant cross
sections by the power $m_\bot^{-1/2}$, so that there is direct
correspondence between the data of experiments
WA85 and WA80, both experiments focus on the central region in rapidity 
$2.1<y<2.9$\,. The upper straight line 
in Fig.\,\ref{specWA80} corresponds to an eye-ball thermal fit 
(emphasized in the WA85 
$m_\bot$-interval $1.5<m_\bot<2.5$ GeV), with $T=232$ MeV for the
S--Ag system, the lower solid line is for S--S
collisions and was done with $T=210$ MeV. 
The choice of S--S temperature was based on the WA94 \cite{WA94}
results for spectra of strange antibaryons. Note that we separated
by factor 0.4 the $\pi^0$ S--S results from the S--Au results;
and that the relative $\eta$ to $\pi^0$ normalization
enhancement is 2.5, which factors makes the $\eta$ abundance fall
onto the $\pi^0$ yields. It is noteworthy that the WA80 particle
spectra span 7 decades. The rise in meson yield at low $m_\bot$
is due to the here unaccounted contribution of decaying
resonances produced very abundantly in hot hadronic matter. Similarly,
some of the concavity of the spectrum arises from non-trivial
and in the current approach unaccounted flow effects.

\vspace{-0.3cm}
\subsection{Electromagnetic probes} \label{secEMP}
\vspace{-0.1cm}

Photons and leptons are, on first sight, the most promising probes of dense
hadronic matter\cite{Feinberg}. 
Electromagnetic interactions are strong enough to lead
to an initial detectable signal, with secondary interactions being too
weak to alter substantially the shape and yield of the primary spectra.
Thus direct photons  and leptons  contain  information about the
properties of dense matter in the initial moments of the collision. Of
particular interest could be the exploration of the initial time period
leading to the formation of the thermal equilibrium. 
 
In all interactions in which we can form final state photons, also
dileptons can be produced in the decay of a off-mass shell photon:
$\gamma^*(M)\to l(p_l)\bar l(p_{\bar l})$. Here, the dilepton pair
produced at a given (central) rapidity $y$ is solely characterized by its
invariant mass $M^2=(p_l+p_{\bar l})^2$. Because the dilepton formation
requires one additional electromagnetic interaction, the dilepton yield
is considerably smaller, by a factor 300 or more, compared to the yield
of direct photons. However, the presence of numerous hadronic particles
that can decay into photons and/or dileptons implies that the
experimental sensitivity is also related to the strength of these
backgrounds, and in this respect experience has shown that dileptons hold
a small edge over direct photons. 
 
The photon backgrounds are substantial, essentially arising from neutral
meson decays. $\pi^0$ decay in flight produces also high energy photons,
and in SPS experiments even at several GeV (in the CM frame of a
fireball) the backgrounds are significant, covering up within today's
experimental precision in S--W/Pb all direct signals of dense matter.
On the other hand, the measured high energy $\gamma$ yield, given the
multi-segmentation of the WA80/93/98 $\gamma$-detector, allows to
reconstruct the spectrum of very high energy mesons  as shown above in
Fig.\,\ref{specWA80}, providing a rare comprehensive glimpse of the
hadronic particle spectrum over many decades of yield. 
 
It is not likely that the situation will change greatly in more
energetic, thus `hotter' interactions, unless a major change of the
reaction mechanism occurs: the radiance of direct photons and the entropy
content, which defines the final hadron multiplicity and thus secondary
yield of photons and leptons, are both rising in a similar way with
temperature. Consequently, the signal to background ratio is relatively
unaffected. What can have considerable impact is the life span of the hot
initial state: because we are far from equilibrium conditions for 
electromagnetic
probes, the direct photon yield is proportional to the life span of the
hot matter fireball. Since the size of the initial system is proportional
to the life span of a freely exploding fireball, considerable advantage
will result when the present experiments with the relatively small system
(S--Pb) are extended to the largest available Pb--Pb collisions. We
expect an enhancement by factor 4 of the direct photon signal, compared
to the hadron decay backgrounds. Consequently, if indeed the current
experimental situation, as has been repeatedly suggested, is just at the
sensitivity limit, a very strong direct photon signal of new physics
should be seen in the ongoing Pb--Pb CERN experiments.
 
The subtle advantage of dileptons over photons arises from the
possibility to consider the yield of dileptons as function of the
lepton pair invariant mass $M$: hadronic particle decays occur within well
defined regions of $M$ and hence one can expect windows of opportunity in
which the backgrounds are small. Specifically, in the rather wide
interval $m_\phi(s\bar s)<M<m_\Psi(c\bar c)$ there are no hadron
resonances contributing to the dilepton background. Thus in the middle of
this dilepton yield dip, around $M\simeq 2\pm 0.6 $GeV, 
any additional radiance is more easy to note  \cite{Kat92}.
 
\begin{figure}[tb]
 \vspace*{1.5cm}
\centerline{\hspace*{6cm}\psfig{width=10cm,figure=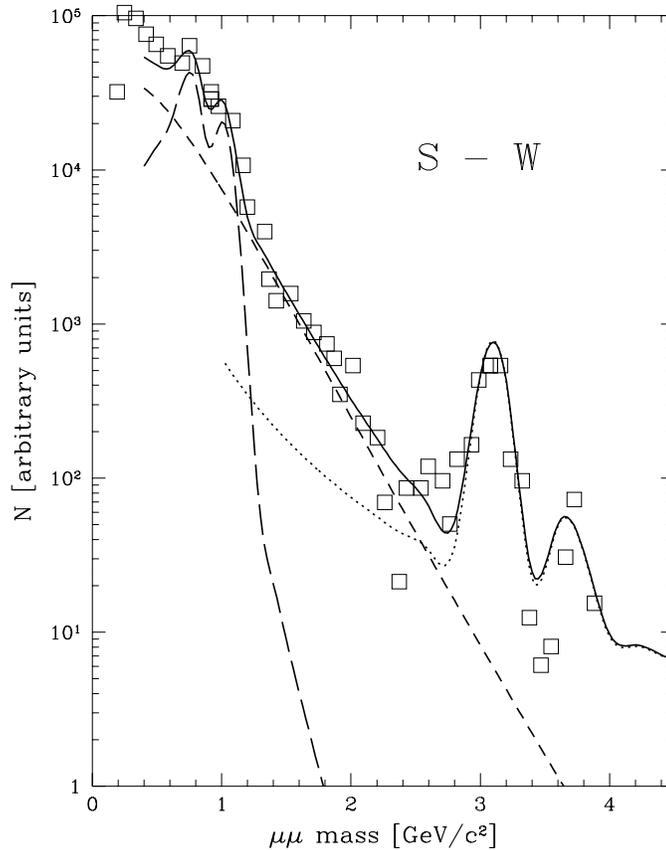}}
\vspace*{-4.5cm}
\caption{\small
Spectrum of dimuons as a function of the dimuon invariant mass
(arbitrary normalization), after Ref.\protect\cite{Kat92,dynamic}. 
The solid line is
the sum of the thermal QGP dimuons (short-dashed contribution),  the
hadron contribution (long-dashed component) and the Drell-Yan (with
$K=2$) together with normalized $J/\!\psi$ contributions (dotted line,
chosen to fit the $J\!/\!\psi$ peak). Experimental results (open squares)
are from NA34/3 experiment.
\protect \label{Fini3}}
\end{figure}

Several CERN experiments (NA34/3, NA38) have seen in S--Pb/U interactions
considerable dilepton yield above background in this kinematic domain
 \cite{Mas95,Tse95}. A typical calculation of the dimuon yields as
function of the invariant mass, taken from our earlier work
 \cite{Kat92,dynamic} is shown in Fig.\,\ref{Fini3}: the solid line is the
sum of the thermal QGP dimuons (short-dashed contribution),  the hadron
contribution (long-dashed component) and the Drell--Yan (with $K=2$)
together with renormalized $J\!/\!\psi$ contributions (dotted line,
chosen to fit the $J\!/\!\psi$ peak). 

Recently, NA34/3 and NA45 have furthermore reported considerable
enhancement of the dilepton yield in the low $M\simeq 0.5$ GeV region
 \cite{Mas95,Tse95,Wur95}. Though more spectacular, the physical meaning
of this result hinges on a comprehensive understanding of many possible
hadron resonance process that could produce low $M$-dileptons, and thus
is potentially primarily a probe of the confined hadronic rather than
deconfined QGP phase. 

All these dilepton enhancements reported do not contradict the lack of
associated direct photon signatures, which turn out to be just not
visible given the background levels. On the other hand, we can justly
expect that if the dilepton phenomena described here are indeed related
to primordial new physics effects, we should see them clearly also in photon
radiance, when the new data for the much larger Pb--Pb collisions is
analyzed. On the other hand, should the dilepton results not 
manifest themselves in the
direct photon enhancement, it is likely that these originate from some
not fully understood normal hadronic effect, and thus are of considerable
lesser physical relevance.

\vspace{-3mm}
\subsection{Charmonium suppression} \label{secCHARMO}
\vspace{-1mm}
 
Experimental results show the predicted \cite{Matsui}
suppression of the Charmonium yield after its
interaction with dense matter \cite{Ram95}. Despite early hope that the
interaction of charmonium $c\bar c=\Psi$ state with dense hadronic matter
will be able to distinguish the difference in structure of the confined
and deconfined matter, more recent and detailed theoretical studies
accounting for the composite structure of charmonium  \cite{KS95} have
revealed that the absorption/dissolution of charmonium in strongly
interacting matter is similar for the different structures here considered. 
Consequently,
the suppression of charmonium production in nuclear interactions
involving passage of $\Psi$ through the dense matter, can be now
accounted for both by interaction with confined and unconfined dense
matter. While we therefore cannot on the basis of this effect obtain
evidence for or against formation of QGP phase, the observed suppression
phenomenon provides in itself a very clear confirmation of the formation
of a rather small and localized, dense hadronic matter region. 
 
\subsection{HBT-interferometry} \label{secHBT}
Pion and kaon correlation functions are measured to study the space-time
evolution of the hadronic source. They are simply obtained from the ratio
of the two particle cross section to the product of the two single
particle cross sections, taken preferably from two different events to
assure exclusion of correlation effects. The resulting correlation $C_2$
is fitted to the convenient form valid for azimuthally symmetric sources:
\begin{equation}\label{fitHBT}
C_2=D\left[1+\lambda e^{-(q_o^2R_o^2+q_s^2R_s^2+q_l^2R_l^2+
2 q_l q_o R^2_{lo})}\right],
\end{equation}
where `$l$' (longitudinal) denotes the two particle momentum projection
onto the axis parallel to the beam, and `$o$' and `$s$' are the
directions
perpendicular to beam axis: `$o$' (out) is parallel and `$s$' (side) is
orthogonal to the two particle transverse momentum sum axis. The last
term in Eq.\,(\ref{fitHBT}), which mixes the components $q_o$ and $q_l$ 
has been often neglected \cite{HeinzHBT}.
 
Compared to HBT interferometry of stars, the situation in heavy-ion 
collisions is complicated by the finite lifetime and the strong dynamical
evolution of the particle emitting source. Thus the interpretation of the
observed correlations between the produced particles is in general
model-dependent, and a considerable amount of theoretical effort has been
spent on the question to what extent this intrinsic model dependence can
be reduced by a refined analysis~\cite{HeinzHBT}.
 
The HBT type interpretation of experimental results leads to the
following hypothesis regarding the particle source:
\begin{enumerate}
\setlength{\itemsep}{-0.1 cm}

\item   emission of particles is chaotic ($\lambda\to 1$),
\item   correlated particles do not arise primarily from
resonance decays,
\item   they do not interact subsequent to strong interaction
freeze-out --- corrections for Coulomb effects are often applied,
\item   kinematic correlations, e.g., energy-momentum
conservation, are of no relevance.
\end{enumerate}
 
Considerable wealth of available experimental results
 \cite{NA35HBT,NA44HBT} leads us to a few conclusions of relevance to the
understanding of the reaction mechanisms operating in relativistic
nuclear collisions.
\begin{itemize}
\setlength{\itemsep}{-0.1 cm}
\item The nuclear collision geometry determines the source size of pions
and kaons. No evidence is found for a major expansion of the hadronic
fireball, required, e.g., for a (long lived) mixed (HG/QGP) intermediate
phase. 
\item The size of the particle source is similar though a bit smaller for
strange (kaons) than non-strange (pions) particles.
\item There is proportionality of the central hadron multiplicity yield
to the geometric volume of the source.
\item Evidence is emerging for presence of transverse flow of the
particle source.
\end{itemize}
These results suggest to us that after its formation the (deconfined)
fireball expands and then rather suddenly disintegrates and
hadronizes, freezing out final state particles at the very early stage of
the evolution of strongly interacting matter.

\section{Strangeness} \label{secSTRoverv}

\subsection{Properties of strange particles}\label{stpartsec}
We now briefly survey the key properties 
of strange particles and  mention some prototype methods for their
detection.  Among strange baryons (and antibaryons) we record:\\

\begin{figure}[tb]
\vspace*{0.1cm}
\centerline{
\psfig{width=8cm,figure=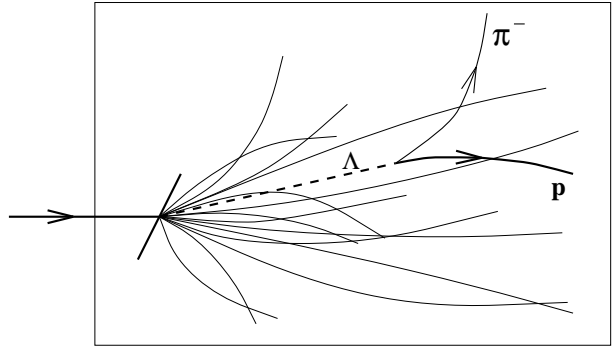}
}
\caption{ \small
Schematic representation of the $\Lambda$-decay topological 
structure showing as dashed line the invisible $\Lambda$ and the decay
`V' of the final state charged particles. Other directly produced 
charged particle tracks propagating in a magnetic field normal to 
the figure plane are also shown.\label{v-dec}}
\end{figure}

\noindent {\bf HYPERONS} $Y(qqs)$ and $\overline{Y}(\bar q\bar
q\bar s)$ comprising two types of particles\footnote{Here and 
below the valence quark content is 
indicated in parenthesis}, the isosinglet $\Lambda$ and
the isotriplet $\Sigma$. Among the hyperons we distinguish:

\noindent $\bullet$ The isospin singlet {lambda} $\Lambda(uds)$, 
a neutral particle of mass
1.116 GeV that decays weakly with proper path length
$c\tau$=7.9cm. The dominant and commonly observed decay is 
$$\Lambda\to p+\pi^- \quad 64\%\,,$$
the other important weak decay 
$$\Lambda\to n+\pi^0 \quad 36\%\,,$$ 
has only the hard to identify neutral particles in the final state. 
The decay of a neutral particle into a pair of charged particles forms 
a characteristic `V' structure shown in Fig.\,\ref{v-dec}.

Aside of the ground state (positive parity, spin 1/2) we encounter a 
spin $1/2^-$ resonance $\Lambda\ (1.405)$ and also $3/2^-$ 
state $\Lambda\ (1.520)$. These and higher excited resonance states (13 are
presently known with mass below 2.350 GeV) decay hadronicaly with the two
principle channels:
\begin{eqnarray}
\Lambda^*&\to& {\rm Y}+\mbox{meson(s)}\,,\nonumber\\ 
\Lambda^*&\to& {\rm N}+\overline{\rm K}\,.\nonumber
\end{eqnarray}
Since the hadronic decays have free space proper decay paths of 1--10 fm
(widths $\Gamma=16$--250 MeV), all these resonances contribute to the
abundance of the observed `stable' strange particles $\Lambda\,,{\rm K}$.
The practical approach to the observation of $\Lambda$ is to
observe the
(dominant) decay channel with two final state charged particles pointing to
a formation vertex remote from the collision vertex of
projectile and target. This approach
includes in certain kinematic region the events which originate from the
K$_{\rm S}$ decay (see below). The well established method of data
analysis has been reviewed elsewhere \cite{Que92}.

\noindent $\bullet$ The isospin triplet $\Sigma^0,\ \Sigma^\pm$ of 
mass 1.189  GeV. The decay of neutral 
$$\Sigma^0\to \Lambda+\gamma+76.9\ {\rm MeV}$$ 
occurs within $c\tau=$ 2.22 $10^{-9}$ cm, thus well away from the
reaction region, but for the observer in the laboratory this remains
indistinguishable from the interaction vertex. Consequently all
measurements of $\Lambda$ combine the abundances of $\Lambda$ and
$\Sigma^0$, and all the higher resonances that decay hadronicaly into
$\Sigma^0$. $\Sigma^0$ is taken to be produced with a thermally reduced
rate compared to the abundance of $\Lambda$:
\begin{equation}
N_{\Sigma^0}=\left({m_\Sigma\over m_\Lambda}\right)^a 
{\rm e}^{-(m_\Sigma-
m_\Lambda)/ T} N_\Lambda\,.
\end{equation}
Here the power $a$ depends on what precisely is measured. For example when 
$N$ stands for $Ed^3N/d^3p$ we have $a=1$; when this spectral
distribution is integrated over a wide region of rapidity, $N$ stands
for $dN/dM_\bot$ and we find $a=3/2$ since we have $m/T>>1$ ($a=0$ follows
when $m/T<<1$).
 
As with $\Lambda$ there are several (nine)
 heavier $\Sigma$ resonances known at $m\le
2.250$ GeV. When produced, they all decay hadronicaly producing
$\overline{\rm K},\,\Lambda\,,\Sigma\,.$
 
Turning briefly to the charged $\Sigma^\pm$ we note that there is only
one dominant decay channel for the $\Sigma^-$ decay: 
$$\Sigma^-\to n+\pi^-\qquad c\tau=4.43\ {\rm cm}\,.$$ 
Because there are two isospin
allowed decay channels of similar strength for the $\Sigma^+$:
\begin{eqnarray}
\Sigma^+&\to& p+\pi^0\qquad \,51.6\%\,,\nonumber\\ 
&\to& n+\pi^+\qquad 48.3\%\,,\nonumber
\end{eqnarray} 
the decay path here is nearly half as long, $c\tau=2.4$ cm. $\Sigma^\pm$
have not yet been studied in the context of QGP studies, as they are
relatively more difficult to observe compared to $\Lambda$ --- akin to
the $\Xi$ decay  (see below) there is always an unobserved  neutral particle 
in the final state, but unlike $\Xi$ the kink that is generated by the
conversion of one charged particle into another, accompanied by the
emission of a neutral particle, is not associated with subsequent 
decay of the invisible neutral particle accompanied by a `V' charged 
particle pair. 
 
It is generally subsumed that abundances of all three $\Sigma$ are equal.\\

\begin{figure}[tb]
\vspace*{-0.1cm}
\centerline{
\psfig{width=8cm,figure=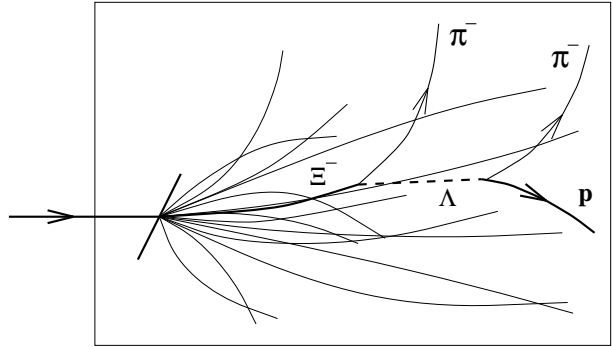}
}
\vspace*{0.5cm}
\caption{ \small
Schematic representation of the $\Xi^-$-decay topological 
structure showing as dashed line the invisible $\Lambda$ emerging 
from the decay kink and the decay `V' of the final state charged 
particles. Other directly produced 
charged particle tracks propagating in a magnetic field normal to 
the figure plane are also shown.\label{kink-dec}}
\end{figure}

\vspace{-3mm}
\noindent {\bf CASCADES} $\Xi(qss)$ and 
$\overline{\Xi}(\bar q\bar s\bar s)$\\
The double  strange cascade baryons and antibaryons $\Xi^0(ssu)$ and
$\Xi^-(ssd)$ are below the mass threshold for hadronic decays into
hyperons and kaons, also just below the weak decay threshold for
$\pi+\Sigma$ final state. Consequently we have one primary decay in each
case:
\begin{eqnarray}
\Xi^-(1321)&\to& \Lambda+\pi^-\qquad 
               c\tau=4.9\,{\rm cm}\,,\nonumber\\ 
  \Xi^0(1315)&\to& \Lambda+\pi^0\qquad \
               c\tau=8.7\,{\rm cm}\,.\nonumber\end{eqnarray}
The first of these reactions can be found in charged particle tracks
since it involves conversion of the charged $\Xi^-$ into the charged
$\pi^-$, with the invisible $\Lambda$ carrying the `kink' momentum. For
$\Xi^-$ to be positively identified it is required that the kink 
combines properly with an observed `V' of two charged particles which
identify a $\Lambda$ decay\,. This decay topology situation is 
illustrated in  Fig.\,\ref{kink-dec}. 

There are also several $\Xi^*$ resonances known, which (with one exception)
feed down into the hyperon and kaon abundances by weak decays. The
exception is the hadronic decay of the spin-3/2 recurrence of the 
spin-1/2 ground state:
$$\Xi(1530)\ \to\ \Xi+\pi\qquad \Gamma =9.5\,{\rm MeV}\,.$$
Since the 3/2 state is populated twice as often as is the spin 1/2 ground
state, the penalty due to the greater mass is almost compensated by the
statistical factor, in particular should the source of these
particles be at high (that is $T>180$ MeV) temperatures.\\ 
 
\noindent {\bf OMEGAS} $\Omega(sss)^-$ and $\overline{\Omega}(\bar s\bar
s\bar s)$\\
There are several primary weak interaction decay channels leading to the
relatively short proper decay path $c\tau=2.46$ cm:
\begin{eqnarray}
\Omega(1672)^-&\to\ \Lambda+{\rm K}^-\qquad \quad
                           &68\%\,,\nonumber\\
\phantom{\Omega(1672)^-}&\!\!\!\!\!\to\ \Xi^0+\pi^-\qquad &24\%\,,\nonumber\\
\phantom{\Omega(1672)^-}&\!\!\!\!\!\to\ \Xi^-+\pi^0\qquad &9\%\,.\nonumber
\end{eqnarray}
The first of these decay channels is akin to the decay of the $\Xi^-$,
except that the pion is now a kaon. In the other two options, after
cascading has finished, there is a neutral pion in the final state, which
makes the detection of these channels not practical. There is only one
known, rather heavy, $\Omega^*(2250)$ resonance.
 
It should be remembered that the abundance of $\Omega$ benefits from the
spin-3/2 statistical factor.\\
 
\noindent {\bf KAONS} K$(q\bar s)$, $\overline{\rm K}(\bar q s)$

$\bullet$ {\bf neutral kaons} K$_{\rm S}$, K$_{\rm L}$\\
This is not the place to describe in detail the interesting physics of
the short and long lived neutral kaons, except to note that both are
orthogonal combinations of the two neutral states $(d\bar s)$, $(\bar d s)$. 
The short lived combination has a $c\tau=2.676$ cm and can be observed in
its charged decay channel:
\begin{eqnarray}{\rm K}_{\rm S} &\to&\pi^+ + \pi^- 
               \qquad 69\%\,,\nonumber\\
    		& \to & \pi^0+\pi^0 \ \qquad 31\%\,.\nonumber\end{eqnarray} 
Care must be exercised to separate the K$_{\rm S}$ decay
from  $\Lambda$ decay, since in both cases there are two a
priori not identified charged particles in the final state, making a `V'
originating in an invisible neutral particle.
 
The long lived kaon K$_{\rm L}$ with $c\tau=1549$ cm has not been studied 
in relativistic  heavy ion collision experiments.
 
\noindent $\bullet$ {\bf charged kaons} K$^+(u\bar s)$, ${\rm K}^-(\bar u s)
=\overline{\rm K}^+$\\
Charged kaons can be observed directly since their mass differs
sufficiently from the lighter $\pi^\pm$ and the heavier
proton/antiproton. However, at the SPS energies the CM-frame has
rapidity 2.5--3 and thus the distinction between the different
charged particles is not easy, though not impossible, such that
directly measured spectra should become available in the near
future. K$^\pm(494)$ decay with $c\tau=371$ cm, with 
three dominant channels, of
which the one with only charged particles in final state
(smallest branching ratio) has been used in our field:
\begin{eqnarray}
{\rm K}^+&\to\ \mu^++\nu_\mu\qquad 
			\qquad &63.5\%\,,\nonumber\\   
\phantom{{\rm K}^+}&\to\ \pi^++\pi^0\qquad \qquad &21.2\%\,,\nonumber\\
\phantom{{\rm K}^+}&\!\to\ \pi^++\pi^++\pi^- 
\quad &5.6\%\,.\nonumber\end{eqnarray} In
general it is subsumed that the mean abundance of the charged kaons is
similar to the abundance of the neutral K$_{\rm S}$\,.\\

\noindent {\bf\mathversion{bold}$\phi$\mathversion{normal}-MESON} 
$\phi (s\bar s)$\\
The vector meson $\phi$ with mass 1019.4 MeV has a relatively
narrow full width $\Gamma_\phi=4.43$ MeV, since it is barely above
the threshold for the decay into two kaons. Consequently, the
total width and thus particle yield could be  
easily influenced by hadronic medium effects: these 
could facilitate induced decays. 
On the other hand the slow decays into two leptons
\begin{eqnarray} \phi&\to& e^++e^-\qquad\, 0.031\% \,,\nonumber\\   
&\to&\mu^++\mu^- \qquad 0.025\%\,,\nonumber\end{eqnarray}
which have partial widths 1.37 keV and 1.1 keV allow the 
determination of the number of
$\phi$-mesons that emerge from the interaction region. While
absolute particle yields may be difficult to determine, one can
compare the yield of $\phi$ to the yield of $\rho$(770)-meson, the
non-strange partner of the $\phi$.

\vspace{-3mm}

\subsection{Diagnostic signatures}\label{stdiag}
\vspace{-1mm}
Several effects combine to make strangeness a very interesting
diagnostic tool of dense hadronic matter. All strange matter has to be
made in inelastic reactions, while light $u,\ d$  quarks are  also
brought into the reaction by the colliding nuclei. The strange quarks $s$
are found abundantly in relativistic nuclear collisions\cite{enhALL}, 
at 200A GeV much more so than it could be expected based on 
simple scaling of $p$--$p$ reactions \cite{NA35S,NA35enh}, 
while this enhancement is not reported in $p$--A collisions \cite{pAres}. 
Because there are many different 
strange particles, we have a very rich field of observables with 
which it is possible to explore diverse properties of the source. 
This is a trivial but indeed the most important reason 
why `Strangeness' is such a very informative
observable of dense hadronic matter. Strange antibaryons were from the 
beginning recognized as being very important in the study of the 
dense hadronic matter. The high strange particle density in the
 QGP led to the prediction \cite{Raf82} 
of highly amplified abundance of multistrange antibaryons.
Considering that these particles are rarely produced in conventional
collisions, while they can be easily formed in a primordial dense soup
containing many $\bar s$ quarks, one is easily led to suggest that their
abundance is a significant signature of deconfinement. 
Because of the experimental complexity related to detection of strange
particles, only recently, results  
experiments became available, allowing a thorough test of the theoretical
ideas. These results obtained at CERN with 200A GeV projectiles 
\cite{WA85T,NA35S,Omega,NA35pbar} support the contention that
strange antibaryons are found in greatly 
anomalous abundance. It remains to be
seen if the systematic behavior as function of, e.g., collision energy
will confirm QGP-fireballs as the source of strange antibaryons. But it
can be safely concluded today that strangeness has fulfilled the 
high expectations about being a useful signature
of the nature of the dense hadronic phase.
 
Generally, not only (relative) abundances but also 
the spectra of several strange
particles K$^\pm$,  K$_{\rm S}$, $\phi$, $\bar p$, $\Lambda$,
$\overline{\Lambda}$,  $\Xi^-$, $\overline{\Xi^-}$, $ \Omega$, $
\overline{\Omega}$ are studied as function of rapidity and transverse
mass. We have included in above list the closely related antiprotons 
$\bar p$, which are also fully made in the collision. The 
classic observables based on these particles are their
abundance ratios: leaving out an overall normalization 
factor associated with the reaction 
volume,  and recalling that there are relations between the abundances
such as of kaons (K$^++$K$^-\simeq2$K$_S$) we have 9 independent 
normalization parameters describing the yields of K$^\pm$,  K$_{\rm S}$,
$\phi$, $\bar p$, $\Lambda$, $\overline{\Lambda}$,  $\Xi^-$,
$\overline{\Xi^-}$, $ \Omega$, $ \overline{\Omega}$\,. These can be
redundantly measured with the help of the $36=9\cdot 8/2$ independent
particle yield ratios. Aside of the yield normalization parameters, there
are in principle 11 different spectral shapes  which we presume to be
closely related to each other and to be governed by the same inverse
slope parameter ({temperature}) parameter. The experimental fact that
once effects related to particle decays and matter expansion (transverse
flow) are accounted for, the $m_\bot$ spectra of all these particles are
characterized by a common temperature, cannot be taken lightly and
suggest strongly some deeper connection between all these particles
that arise form quite different individual formation processes in the 
confined phase. Our point of view is that the source of all strange 
particles is a thermalized fireball permitting a  common mechanism
to govern the production of the very different strange particles, 
as well as $\bar p$. We will develop in full below, in our theoretical 
approach this picture of strange particle production. We will
presume that the strong interactions allow to
achieve local thermally  equilibrated fireball, a fact which is very much
in experimental evidence, but which is far from being understood, as we
stressed above. Our analysis based on the results obtained at 200A GeV
favors a picture of the reaction in which the hadronization occurs
rapidly such that the observed strange particles can have properties
representative of the  expected properties of the primordial phase.
 
Several global properties of the final state strange particle abundance 
carry such information. Consider that, when finite baryon density is
present, which breaks the particle/antiparticle symmetry, the exact
balance between $s$ and $\bar s$ quarks requires non-trivial relations
between the parameters characterizing the final state hadron abundances.
These strangeness conservation constraints imply different particle
distributions for different structures of the source. It turns out that
in the statistical approach the key parameter is the strange quark
chemical potential~$\mu_{\rm s}$:
\vspace{-3mm}

\begin{enumerate}
\setlength{\itemsep}{-0.2 cm}
\item In a deconfined state in which quark bonds are broken, the
strangeness neutrality implies $\mu_{\rm s}=0$\,, independent of
prevailing temperature and baryon density.
\item In any state consisting of locally confined hadronic clusters,
$\mu_{\rm s}$, for finite baryon density, is generally different from
zero, in order to compensate the
asymmetry introduced by the finite baryon content.
\end{enumerate}
\vspace{-3mm}

\noindent The vanishing strange quark potential $\mu_{\rm s}\simeq 0$ is a striking
result of different analysis of the today available data of the CERN
experiments WA85  and  NA35  \cite{Raf91,analyze,Heinzy,SH95}. 
This important conclusion arises from study of particle abundance 
ratios, which act as remote thermo- and chemico-meters of the 
particle source.  Aside of the strange quark chemical potential,
one also is able to derive the light ($u,\,d$) quark chemical potentials
from strange baryon abundances.
 
Other generic observables that determine abundance 
of the final state strange particles and thus can be derived 
from the particle abundances are:
\vspace{-5mm}

\begin{itemize}
\setlength{\itemsep}{-0.2 cm}
\item Specific (with respect to baryon  number $B$) strangeness yield 
$\langle \bar s\rangle/B$\\
Once produced strangeness escapes, bound in diverse hadrons, from 
the evolving fireball and hence  the total abundance observed 
is characteristic for the initial extreme conditions reached in the collision.
Theoretical calculations suggest that glue--glue
collisions in the QGP phase provide a sufficiently fast mechanism and
thus an explanation for strangeness enhancement comprised in this observable. 
 \item Phase space occupancy $\gamma_{\rm s}(t_{\rm f})$.\\
Strangeness freeze-out conditions at particle hadronization 
time $t=t_{\rm f}$, given  the initially produced abundance,
determine the final state observable phase space occupancy of strangeness 
$\gamma_{\rm s}(t=t_{\rm f})$. 
\end{itemize}
\vspace{-5mm}

\subsection{Highlights of strangeness experimental results}
\label{expres}
\vspace{-1mm}
We now briefly describe the key experimental 
results on which our here presented theoretical developments 
are based either in detail or/and conceptual design:\\

\begin{figure}[tb]
\vspace*{-.2cm}
\centerline{\hspace*{0.1cm}\psfig{width=6.cm,figure=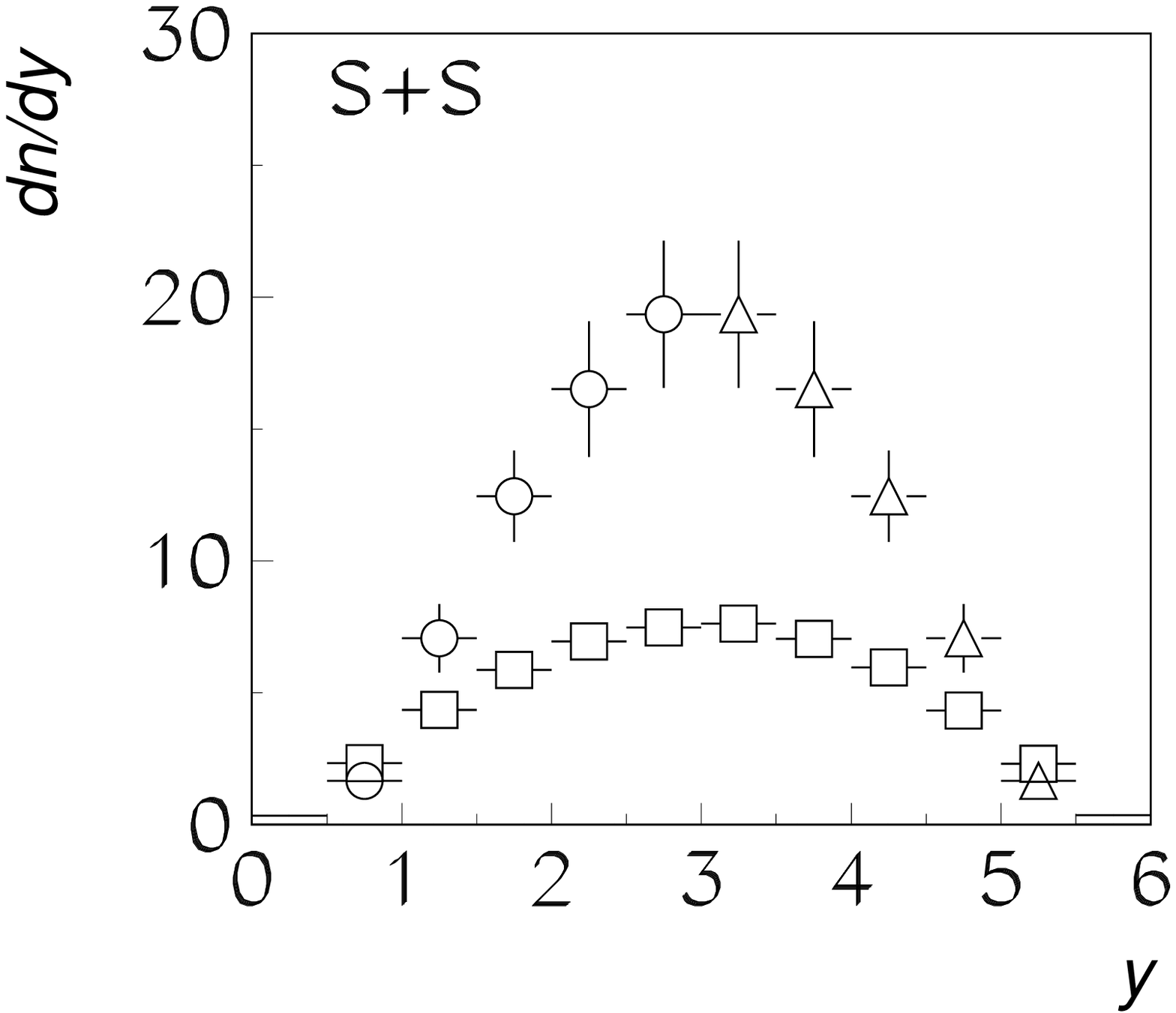}
\hspace*{0.3cm}\psfig{width=6.cm,figure=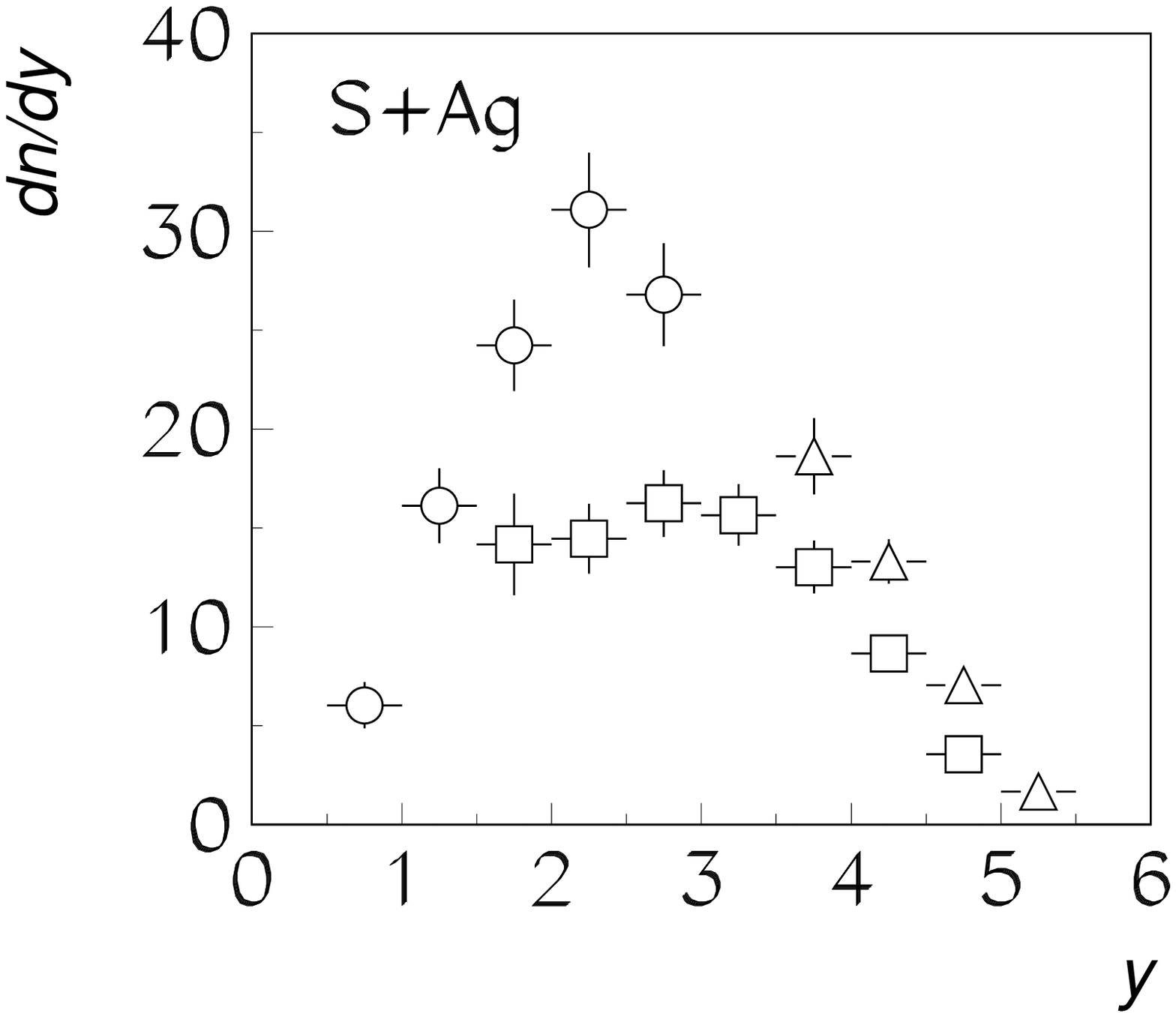}}
\vspace*{-0.4cm}
\caption{ \small
Abundance of $1.6 \Lambda+4{\rm K}_{\rm S}+1.6 \overline{\Lambda}$ 
as function of rapidity. 
On the left S--S, on the right S--Ag (open circles are the
directly measured data). The triangles are reflected data points
for S--S and reflected-interpolated data employing S--S and S--Ag.
The squares in S--S case are the results for N--N collisions
scaled up by the pion multiplicity ratio, for S--Ag these are the
scaled up p--S results.
Courtesy  of NA35 collaboration \protect\cite{NA35cent}.
\label{cenralstr}}
\end{figure}

\vspace{-3mm}
\noindent $\bullet$ {\bf Centrality of strangeness production}\\
We consider a measure of the abundance of $\langle s+\bar s\rangle$ in 
Fig.\,\ref{cenralstr}. We show here the integrated transverse mass 
$m_\bot=\sqrt{m^2+p_\bot^2}$ distribution for
$1.6 \Lambda+4{\rm K}_{\rm S}+ 1.6 \overline{\Lambda}$ as determined 
by the experiment NA35\cite{NA35cent}, as function of rapidity. For the
case of S--S the open circles are the measured data points, the
open triangles are the symmetrically reflected data points, and
squares are the results of N--N (isospin symmetric
nucleon-nucleon) collisions scaled up by pion
multiplicity; the difference, most pronounced at central rapidity
$y\simeq 3$ shows a new source of strangeness in the collision,
and the important lesson to be drawn from this result is that 
strangeness enhancement originates in the central rapidity
region.  We also show in
Fig.\,\ref{cenralstr} similar results for S--Ag collisions: here the
open circles are the measured points, open triangles are
estimates based on S--S and the `reflected' S--Ag results, and the open
squares are pion multiplicity scaled p--S results. \\

\vspace{-3mm}
\noindent $\bullet$  {\bf Anomalies of strange antibaryon abundances}\\
The WA85 collaboration has extensively studied in the central rapidity
region the relative abundance of the different strange baryons and 
antibaryons. The  particle spectra ratios have been
obtained at  $p_\bot\ge 1$ GeV. The results for relative abundances can
be presented both for the sum of abundance with $p_\bot\ge 1$ GeV or using
as cut a fixed value $m_\bot\ge 1.7$ GeV. In the thermal model this latter
set of values is of primary interest. However, given prior studies of
relative particle abundances one often identifies the anomalies using 
the fixed $p_\bot$ approach. Moreover such ratios correspond
more closely to the total particle abundance ratio, as we shall
see in section \ref{results}. The experiment WA85 \cite{WA85} 
has reported the  following ratios between same baryons and antibaryons:  
 \begin{equation} \begin{array}{lr}
   R_\Lambda& =\ 0.20\pm 0.01 \\ &\\  
      R_\Xi &=\ 0.41\pm0.05   
\end{array}
   \  \quad \mbox{ for } y \in (2.3,2.8) \mbox{ and } m_\bot>1.9\   
\mbox{GeV}.
 \label{cascade}
 \end{equation}
have been analyzed carefully in our recent work \cite{Raf91,analyze},
and the chemical properties of the source were derived.

\begin{figure}[hb]
\vspace*{-0.2cm}
\centerline{
\psfig{width=9cm,figure=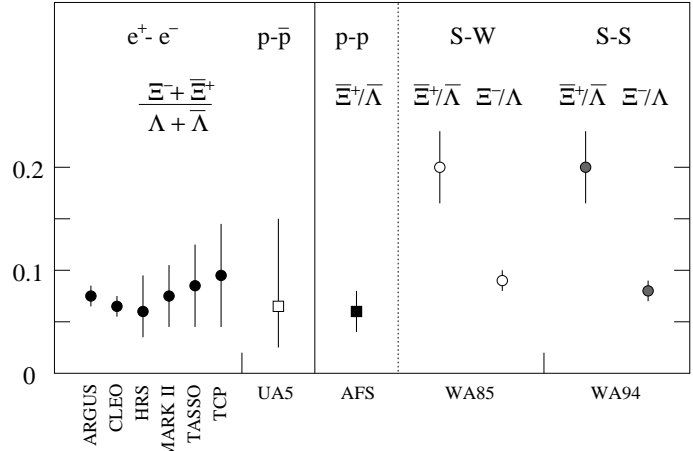}
}
\vspace*{-0.5cm}
\caption{ \small
Ratio (at fixed $p_\bot$) of (multi)strange baryon-antibaryon 
particle abundance, 
measured in the central rapidity region at 200A GeV S--S/W collisions, 
compared to ratios obtained in lepton and nucleon induced reactions. 
Data assembled by the WA85/94 collaboration\protect\cite{WA85}.
 \protect\label{ratiosWA85}}
\end{figure}

Strangeness abundance (phase space occupancy) at moment of particle
emission  is probed when ratio of
particles is considered that contains a different number of
strange quarks. In Fig.\,\ref{ratiosWA85} such a World sample of 
strange baryon and antibaryon data is presented. We note  the strong
enhancement of the ratios seen in heavy ion reactions (S--S/W at 200A GeV).
In the kinematic domain of Eqs.\,(\ref{cascade}) 
the experimental results reported by the WA85
collaboration are:   
\begin{equation}
   \frac{\overline{\Xi^-}}{\overline{\Lambda}+\overline{\Sigma^0}} =   
0.4\pm0.04 
     \, , \quad
   \frac{\Xi^-}{\Lambda+\Sigma^0} =0.19\pm0.01
               \, .
 \label{newa}
 \end{equation}
If the mass difference between $\Lambda$ and $\Sigma^0$ is neglected,
this implies  that an equal
number of $\Lambda$'s and $\Sigma^0$'s are produced, such that
 \begin{equation}
   \frac{\overline{\Xi^-}}{\overline \Lambda}
   = 0.8\pm 0.08 
     \, , \hspace{1.15cm}
   \frac{\Xi^-}{\Lambda} = 0.38\pm0.02 
               \, .
 \label{new1}
 \end{equation}
\begin{figure}[h]
\vspace*{-0.8cm}
\centerline{\hspace*{-1.cm}
\psfig{width=9cm,clip=,figure=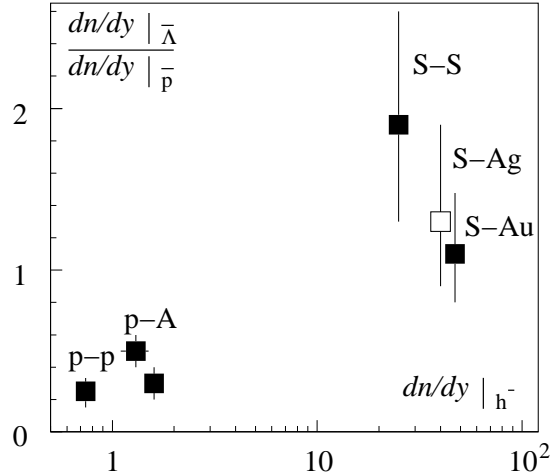}
}
\vspace*{-2.5cm}
\caption{ \small
 Ratio of the rapidity density $dn/dy$ for $\overline{\Lambda}/
\bar p$, measured at central $y$, as function of the
negative hadron central rapidity density $dn/dy|_{{\rm h}^-}$.
Courtesy of NA35 collaboration\protect\cite{NA35pbar}. 
\label{lbpbNA35}}
\end{figure}
\begin{figure}[b]
\vspace*{1.cm}
\centerline{\hspace*{0.cm}
\psfig{width=7.cm,figure=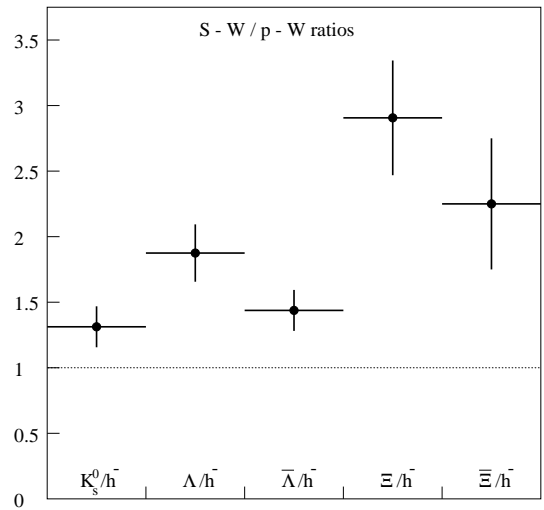}
}
\vspace*{-1.cm}
\caption{ \small
Ratio of $h^-$ normalized particle abundances: S--W results 
divided by $p$--W results at 200A GeV in the same rapidity window 
near to $2.5<y<3$. Dotted line: expected yields. Courtesy of 
WA85 collaboration\protect\cite{WA85}. \protect\label{colWA85} }
\end{figure}

The fact that the more massive and stranger anticascade practically
equals at fixed $m_\bot$ the abundance of the antilambda is most
striking. These results are inexplicable in terms of hadron-cascade
models for the heavy-ion collision \cite{Csernai}. The relative yield
of $\overline{\Xi^-}$ is 3.5 times greater than seen in the $p$--$p$
ISR experiment \cite{ISR} and all other values reported in the
literature, which amounts to a 4 s.d. effect \cite{WA85}.

Another most remarkable result related to these findings is due to the 
NA35 collaboration\cite{NA35pbar}: in
Fig.\,\ref{lbpbNA35} we show the ratio of the rapidity density $dn/dy$
at central $y$ of $\overline{\Lambda}/\bar p$, as function of the
negative hadron central rapidity density $dn/dy|_{{\rm h}^-}$. The
$p$--$p$ and $p$--A reactions are at small values of $dn/dy|_{{\rm h}^-}$,
while the S--S, S--Ag, S--Au reactions are accompanied by a relatively high
$dn/dy|_{{\rm h}^-}$. We observe that there is an increase in this ration
by nearly factor 5, and even more significantly, the abundance of the
{\it heavier} and {\it strange} $\overline{\Lambda}$ is similar if not
greater than the abundance of $\bar p$.\\
 
\vspace{-3mm}
\noindent $\bullet$ {\bf Collectivity of strange particle production}\\
The WA85 collaboration\cite{WA85} has shown that there is a
trend in these anomalous strange baryon abundances
in that the yields in nuclear collision S--W (normalized by h$^-$
abundance) when compared to the p--W collisions are increasing with the
strangeness content, as illustrated in Fig.\,\ref{colWA85}. This indicates
that strange particles are formed in some collective mechanism, which
favors the assembly of multiply strange hadrons. Comparable result
is reported by the NA38 collaboration \cite{NA38str} which has
shown that the ratio 
$${\phi\over{\rho+\omega}}\propto {{s\bar s}\over {q\bar q}}\,,$$ 
rises by nearly a factor three in S--U compared to p-W
reactions, in collisions with greatest particle density.\\
 
\vspace{-3mm}
\noindent $\bullet$ {\bf Thermal nature of (strange) particle spectra}\\
We have discussed these remarkable results
above and refer here in particular to Fig.\, \ref{specWA85}
and the accompanying discussion for further details.

\vspace{-3mm}

\section{Thermal fireball} \label{TMod}
\vspace{-3mm}
\subsection{Comparison to kinetic theory approach}
\vspace{-1mm}
The special virtue of the thermal fireball 
framework is that the spectra and
particle abundances can be described in terms of a few parameters which
have very intuitive meaning. In this the thermal model
analysis of the experimental results differs fundamentally from other
efforts made with individual particle cascade type models. These contain
as inputs detailed data and their extrapolations, and often also
assumptions about unknown reaction cross sections. The
attainment of thermal equilibrium is in these calculations
result of many individual particle-particle collisions. However,  
for the nucleon-nucleon (N--N) collisions we already know that
the appearance of the thermal particle distributions in the
final state is inexplicable in terms of dynamical microscopic 
models \cite{Divonne}. Consequently, there is no reason to
expect that some microscopic dynamical approach invoking multiple
series of  N--N type interactions lead to any better 
understanding of the thermalization process. Moreover, if the 
underlying and yet not understood thermalization processes are,
as is likely in view of the N--N situation, much faster than those
operating in the numerical cascade codes, these results would 
not be adequate.   

Such an uncertainty about the microscopic mechanisms does not 
beset the thermal approach, where we do not implement microscopic 
approach to thermalization, but rather analyze
the data assuming that, though not understood, thermalization is
the fastest, nearly instantaneous, hadronic process. 
The prize one pays in this approach is that under certain conditions one
looses the ability to describe some details of the collision evolution.
For example, we have not been able to identify within a thermal model a
method to determine the stopping fractions (i.e., energy
or baryon number deposition rate) governing the different collisions and we
extract this parameter in qualitative form from the data. In the microscopic
kinetic theory 
models one can in principle claim to `derive', e.g., the energy-momentum
stopping. This current deficiency of the thermal model disappears under
conditions which could lead to full stopping. In the near future we will see
up to which energy this may occur for the Pb--Pb reactions. The initially
studied maximum energy is 158A~GeV and we hope that in a very near
future the energy range between 40A and 158A~GeV can be explored.
 
It should be noted here that in a rough survey of the particle yields one
aught to observe the considerable impact of the surface of the colliding
nuclei, always present in symmetric systems. Consequently, it is no
surprise that many  observed particle rapidity yields are wider than
expected even in presence of full stopping --- the degree
of stopping reached can be more effectively 
explored considering the rapidity shapes of particles which cannot be
easily made in single hadron interactions (e.g., $\overline{\Lambda}$).

\vspace{-3mm}
\subsection{Thermal parameters}\label{paramthsec}
\vspace{-1mm}
\baselineskip=12.5pt

We now  discuss in qualitative terms the global parameters of the thermal 
fireball model. 
We suppose that the primordial source is a space-time localized region  
of thermal hadronic matter which is the source of all
particle Boltzmann type spectra. At relatively high $m_\bot$ the 
exponential spectral shape is relatively little deformed by resonance
decay and the fireball dynamics, here in particular 
flow phenomena. Thus this portion of the spectrum should be similar for
different particles, which would allow a reduction of all data
to just one basic spectral shape form:
\begin{equation}\label{spectra}
{dN\over d^3p}=N_i e^{-E^{(i)}/ T}=
     N_i e^{-\cosh (y-y_{\rm CM})\,{m^i_\bot/ T}}\,.
\end{equation}
The parameters of each particle distribution include the
inverse slope $T$ (`temperature') of the $m_\bot$ distribution, centered
around the $y_{\rm CM}$. 

The fireball is created in central symmetric collisions at  the 
CM-rapidity of the N--N system, which is for relativistic systems  just
is 1/2 of the projectile rapidity. For asymmetric collisions such  as 
S--Au/W/Pb the CM rapidity depends on the ratio of the participating masses
$A_{\rm P}$, $A_{\rm T}$  of 
the projectile and, respectively, target nuclei, see Eq.\,(\ref{eqycm}).

The relative abundance of particles emerging from the thermal
fireball is controlled the  chemical (particle abundance) 
parameters, the particle fugacities\cite{analyze}, which allow to
conserve flavor quantum numbers. Three fugacities are 
introduced since the flavors $u,\,d,\,s$ and as appropriate $c$ are
separately conserved on the time scale of hadronic collisions and can
only be produced or annihilated in particle-antiparticle pair production
processes\footnote{We will in general not introduce and/or discuss the
fugacities for quarks heavier than $s$. While we explore in
qualitative terms the charm production, it remains a rather small 
effect even at LHC energies.}.
 The fugacity of each hadronic 
particle species is the product of the valence quark fugacities, 
thus, for example, the hyperons have the fugacity 
$\lambda_{\rm Y}=\lambda_{\rm u}\lambda_{\rm d}\lambda_{\rm s}$.
Fugacities are related to the chemical potentials $\mu_i$ by:
\begin{equation}
\lambda_{i} =e^{\mu_{i}/T}\,,\quad 
\lambda_{\bar{\imath}}=\lambda_i^{-1}\qquad 
i={u,\,d,\,s}\, .   \label{lam}
 \end{equation}
Therefore,  the chemical potentials for particles and 
antiparticles are opposite  to each other, provided that there
is complete  chemical equilibrium, and if not, that the deviation from
the  full phase space occupancy is accounted for by  introducing a  
non-equilibrium chemical parameter $\gamma$ (see below).
 
In many applications it is sufficient to combine the
light quarks into one fugacity 
\begin{equation}
\lambda_{\rm q}^2\equiv\lambda_{\rm d}\lambda_{\rm u}\,,\quad 
\mu_{\rm q}=({\mu_{\rm u}+\mu_{\rm d}})/2\,. \end{equation}
The slight isospin asymmetry in the number of $u$ and $d$ quarks is described by
the small quantity 
\begin{equation}\label{delmu}
\delta\mu=\mu_{\rm d}-\mu_{\rm u}\,,\end{equation}
which may be estimated by theoretical considerations: 
we introduce the light flavor imbalance in the fireball:
 \begin{eqnarray}
   \delta q={\langle d-\bar d\rangle - \langle u-\bar u\rangle
        \over \langle d-\bar d\rangle + \langle u-\bar u\rangle}\, .
 \label{eq4}
 \end{eqnarray}
In a central S--W collisions, considering a tube with the transverse area of
the S projectile  swept out from the W target, and   in Pb--Pb collisions
one has 
$$\delta q^{\rm S-W}\simeq0.08\qquad\delta q^{\rm Pb-Pb}=0.15\,.$$
The value of $\delta\mu$ is
at each fixed $T$ determined by the value of $\delta q$, but depends
on the assumed structure of the source such as the HG and the QGP.
For the QGP~\cite{Raf91}, the ratio $\delta\mu / \mu_{\rm q}$ 
is independent of
$\lambda_{\rm s}$, due to the decoupling of the strange and
non-strange chemical potentials in the partition function. For
$\mu_{\rm q}<\pi T$ we find the simple relation:
 \begin{eqnarray}
   \delta q^{\rm QGP} \simeq 2 {\mu_{\rm d}-\mu_{\rm u}\over
				\mu_{\rm d}+\mu_{\rm u}}
                       ={\delta\mu\over \mu_{\rm q}}\, .
\end{eqnarray}
The relation between $\delta q^{\rm HG}$ and $\delta\mu$ 
was obtained numerically computing the partition function with 
all mesons and baryons up to 2 GeV mass\cite{analyze}. In a 
large region of interest to us here ($T\sim 150$--$200$ MeV) 
it was found that $\delta q^{\rm HG}\simeq \delta q^{\rm QGP}$. 
Thus irrespective of the state of
the source:
 \begin{eqnarray}\label{delq}
   {\delta\mu\over \mu_{\rm q}} =\delta q\quad
\simeq0.08\ \mbox{for\ S--W}\quad 
\simeq 0.15\ \mbox{for\ Pb--PB}\, .
\end{eqnarray}

Since a wealth of experimental data can be described with just a few 
model parameters, this leaves within the thermal model a considerable 
predictive power and a strong check of the internal consistency of the
thermal approach we develop. Specifically, in the directly hadronizing 
off-equilibrium QGP-fireball considered here there are 5
particle multiplicity parameters (aside
of $T$ and $y_{\rm CM}$) characterizing all particle 
spectra: the fireball size $V$, two  fugacities 
$\lambda_{\rm q},\,\lambda_{\rm s}$, of which the letter one is
not really a parameter in our approach, as we will set $\lambda_{\rm
s}=1$ because of strangeness conservation in the QGP phase, and two
particle abundance non-equilibrium parameters we will discuss at
length below in section \ref{hadromod}: the strangeness occupancy 
factor we call $\gamma_{\rm s}$ and the ratio $R^{\rm s}_{\rm
C}$, see Eq.\,(\ref{RsC}), of meson
to baryon abundances normalized to hadronic gas equilibrium. 
Only the last of these parameters is related to the mechanism 
governing the final state hadronization process, the others will be 
determined using a dynamical picture  of the collision, in which the
input is derived from more general qualitative conditions of the
colliding system, such as the energy content or stopping power.  Thus the
validity of thermal and (approach to) chemical equilibrium can
be conclusively  tested,
comparing the observed particle  spectra and yields with the theoretical
predictions. We can do this without the need and in particular, 
without the capability to modify and adapt the
theoretical description to each new experimental result. 
Therefore, the thermal hypothesis can
be relatively easily falsified, but so far this has not been the case.

\vspace{-4mm}

\subsection{Stages of fireball evolution}
\vspace{-1mm}

We now look at the different stages of the temporal
evolution\cite{dynamic} and the related parameters of the fireball. 
The scenario we adopt is in view of the current understanding of
hadronic physics the most natural one in qualitative terms, in
accord with the general properties of the strong
interactions and hadronic structure widely known and
accepted today and it is in quantitative agreement with experimental
results obtained in relativistic nuclear collisions,
see section \ref{tools}. 

When studying collisions up to maximum available SPS energies we
suppose that the relevant time development stages of the relativistic
nuclear collision comprise:
\vspace{-3mm}

\begin{itemize}\setlength{\itemsep}{-0.2 cm}
\item[\it 1.] The pre-thermal stage lasting perhaps 0.2--0.4 fm/$c$, during
which the thermalization of the initial quark-gluon distributions occur.
During this time most of the entropy obtained in the collision must be
created by mechanisms that are not yet understood --- this is also
alluded to as the period of de-coherence of the quantum collision system.
Our lack of understanding of this stage will not impact our results, as
the reason that we lack in understanding is that the
hadronic interactions erase the memory of this primordial stage,
except for the entropy content.
\item[\it 2.] The subsequent inter-penetration of the projectile and the
target lasting about $\sim 1.5$ fm/$c$, probably also corresponding to the
time required to reach chemical equilibrium of gluons $g$ and
light non-strange quarks $q=u,\,d$\,. 
\item[\it 3.]  A third time period ($\simeq 5$ fm/$c$) during which the
production and chemical equilibration of strange quarks takes place.
During
this stage many of the physical observables studied here will be
initiated.
\item[\it 4.] Hadronization of the deconfined state ensues: it is
believed that the fireball expands at constant specific entropy per
baryon, and that during this evolution or at its end it decomposes into
the final state hadrons, under certain conditions in an (explosive)
process that does not allow for re-equilibration of the final state
particles.  
\end{itemize}
In the sudden hadronization picture of the QGP fireball suggested
by certain features seen in the analysis  of the strange antibaryon 
abundances for the 200A GeV  nuclear collision 
data  \cite{analyze,Raf91}, the hadronic observables 
we study are not overly sensitive to the details of
stage {\it 4}. Akin to the processes of direct emission, 
in which strange particles are made in recombination--fragmentation 
processes \cite{RD87,BZ83}, the chemical conditions prevailing in the
deconfined phase  are  determining  many relative final particle yields. 
Recent theoretical models show that such a sudden hadronization
may occur \cite{sudden}. Furthermore
if the hadronization occurs as suggested by recent lattice
results \cite{lattice} at a relatively low temperature (e.g., 150~MeV), 
the total meson abundance which is determined by the
entropy contents of the fireball at
freeze-out of the particles, is found about 100\% above the hadronic  gas
equilibrium expectations \cite{entropy}. This is consistent with  the
source of these particles being the QGP  \cite{analyze,entropy}.  The
freeze-out entropy originates at early time in collision since aside of
strangeness production which is responsible for about 10\% additional
entropy there is no significant entropy production after the initial
state has occurred \cite{entropy}. 

The above remarks apply directly to the 200A GeV data. The
general features of particle multiplicities obtained at 15A GeV
are consistent with the thermal equilibrium hadronic gas state expectations 
\cite{BNL-AGSthermal,LetAGS}. However, the source of these particles
could also be a QGP fireball, provided that a slow
re-equilibration transition occurs under these conditions,
leading to the equilibrium state among many final hadron gas particles.
 
The temperature of the fireball evolves in time and within our 
schematic model we introduce here a few characteristic values 
which have both intuitive meaning and are useful in future 
considerations. We characterize the above described stages by 
the following temperatures:
\vspace{-0.1cm}
\begin{center}
\begin{tabular}{cl}
$T_{\rm th}$ & temperature associated with the 
initial thermal equilibrium,\\

$\downarrow$ & {\it production of} $q,\ {\bar q},\ G$;
\\

$T_{\rm ch}$ & chemical equilibrium for non-strange 
quarks and gluons, \\

$\downarrow$ & {\it production of} $s,\ {\bar s}$ {\it quarks and
fireball expansion;} \\

$T_{\rm 0}\ $ &  condition of maximal chemical equilibrium: 
`visible' temperature,\\

$\downarrow$ &  {\it fireball expansion/particle radiation}; \\

$T_{\rm f, s}$ & temperature at freeze-out for non-strange or 
strange particles.
\end{tabular}
\end{center}
\vspace{-0.1cm}

\noindent We encounter a considerable drop in temperature and 
obviously $T_{\rm th} > T_{\rm ch} > T_{\rm f}$. However,  
the entropy content which determines the final particle multiplicities 
evolves more steadily, indeed it remains nearly constant: aside of the
initial state entropy formation, in our model additional entropy
increase is due to the formation of the strangeness flavor.
Thus strangeness formation processes are acting like a viscosity 
slowing down the transverse flow of hadronic matter. 

Initially, temperature decreases rapidly from $T_{\rm th}$ to
$T_{\rm ch}$ since there is rapid quark and gluon production
which establishes the chemical equilibrium, as we have 
shown \cite{cool} these processes 
generate  little entropy. We will explicitly compute the values of
$T_{\rm ch}$ for different systems balancing the
energy per baryon and the collision pressure.

If the final state particles emerge 
directly, without re-equilibration, from the fireball\cite{Raf91,RD87}, 
this  observed temperature $T_\bot$ in the particle spectra would be closely 
related to the full chemical equilibration temperature $T_{\rm
0}\ $: 
  
In the transverse mass spectra of strange (anti)baryons an
inverse temperature slope $T_\bot$ ($=232\pm5$ MeV in S--A
collisions at 200A GeV) is found, and the important matter 
is to relate this observed value to
the initial $T_{\rm ch}$ condition of the fireball. It is to
this end that we have introduced above the quantity $T_0$ which
arises from $T_{\rm ch}$ when we relax
strangeness to  (nearly) full chemical equilibrium, keeping the 
entropy content of gluons and light flavor unchanged. $T_0$ 
is always somewhat smaller than $T_{\rm ch}$ since energy has been 
spend to produce strangeness \cite{dynamic}. Even more energy is
spend into the transverse expansion and thus the
temperature at freeze out is nearly certainly considerably 
lower than $T_0$.  When  the final state 
particles emerge from the flowing surface, they are blue-shifted by  the
flow velocity. This Doppler shift effect restores the high apparent 
$T_\bot$  in high $m_\bot$ particle spectra \cite{ULI93}:
\begin{eqnarray} 
T_{\bot}\simeq\sqrt{1+v_{\rm f}\over 1-v_{\rm f}} T_{\rm
f}\,,\label{doppler} \end{eqnarray}
and $T_\bot$ is found in model calculations to be close if not
exactly equal to the value $T_0$ that would be present in the 
chemically equilibrated fireball, provided that no reheating 
has occurred in a strong
phase transition of first order. Despite our still considerable 
ignorance of the dynamics of fireball and particle
freeze-out mechanisms and conditions, we believe that the uncertainty  in
the value of the  temperature $T_0$ as derived from the value of
$T_{\bot}$  is not large. Namely, if QGP phase is directly
dissociating by particle emission, this is trivially so, since
we see what happened in a direct observation. If, as is
generally assumed, there were to
be substantial flow, one can assume some temperature $T_{\rm 0}$, and 
given equations of state (EoS), obtain the hydrodynamic radial
expansion \cite{Kat92}; especially at the high $m_\bot\simeq 2$
GeV the resulting inverse slope temperature $T_\bot$ of the
particle is found smaller but almost equal to $T_{\rm 0}$. 

\vspace{-3mm}
\subsection{Analysis of properties of the strange particle source} 
\label{secanalysis}
\vspace{-1mm}

In the thermal fireball model with sudden non-equilibrium
hadron formation, the observed particle 
yields can be relatively easily related  
to the physical properties of the fireball.

The abundance of particles emerging is, according to
Eq.\,(\ref{spectra}), determined by the normalization constant:
\begin{equation}\label{norm}
N_j=C_{j={\rm M,B}}\,V\prod_i n_i\,,\qquad n_i=g_i\lambda_i \gamma_i \,,
\end{equation}
where it is assumed that the final state particle of type $j$ contains 
the quark valence components of type $i$ and these are counted using 
their statistical degeneracy $g_i$, fugacity $\lambda_i=\exp(\mu_i/T)$ 
and the chemical equilibration factor $\gamma_i$. 
$ V$ is the emission source volume. Particle fragmentation 
has been found to change the recombination results in a minor way 
\cite{RD87}, because the fragmentation enhances the number of all quarks, 
and thus contributes in a similar way to all flavors, and further, 
since in the ratio of particle abundances a partial cancelation of 
fragmentation effect occurs. Moreover, fragmentation, by its intrinsic 
nature primarily increase the yield of particles at small $m_\bot$.
 
Once chemical non-equilibrium features are accounted for by three
significant chemical non-equilibrium abundance factors 
$\gamma_{\rm s}(t_{\rm f})$, the strangeness phase space occupancy, 
and $C_{\rm M,B}$, meson and baryon particle yield compared to chemical 
equilibrium yield in hadronic gas, see section \ref{hadromod}, 
the chemical potentials for
particles and antiparticles are opposite to each other and the particle
and antiparticle  abundances are related, see Eq.\,(\ref{lam}).
As indicated in Eq.\,(\ref{norm}),  the fugacity of each final state 
hadronic species is the product of the valence quark fugacities.

Thus the ratios of strange antibaryons to strange baryons {\it
of same particle type\/}: 
$$R_\Lambda=\overline{\Lambda}/\Lambda\,,\quad 
R_\Xi=\overline{\Xi}/\Xi\quad \mbox{and}\quad
R_\Omega=\overline{\Omega}/\Omega\,,$$ 
are in our approach simple functions of the quark 
fugacities. For the available two ratios in experiment WA85 one has
specifically
 \begin{eqnarray}
  R_\Xi =  {{\overline{\Xi^-}}\over {\Xi^-}} =
   {{\lambda_{\rm d}^{-1} \lambda_{\rm s}^{-2}} \over
    {\lambda_{\rm d} \lambda_{\rm s}^2}} \, ,
\qquad 
  R_\Lambda = {\overline{\Lambda}\over \Lambda} =
{{\lambda_{\rm d}^{-1} \lambda_{\rm u}^{-1}
                          \lambda_{\rm s}^{-1}} \over
    {\lambda_{\rm d} \lambda_{\rm u} \lambda_{\rm s}}} \, .
 \label{ratio}
 \end{eqnarray}
These ratios can easily be related to each other, in a way which shows
explicitly the respective isospin asymmetry factors and strangeness
fugacity dependence. Eq.\,(\ref{ratio}) implies:
 \begin{eqnarray}
   R_\Lambda R_\Xi^{-2} =  e^{6\mu_{\rm s}/T}\cdot
e^{2\delta\mu/T}\, ,   
\qquad 
   R_\Xi R_\Lambda^{-2} = e^{6\mu_{\rm q}/T} \cdot
e^{-\delta\mu/T}\, .   \label{R2}
 \end{eqnarray}
Eq.\,(\ref{R2}) is generally valid, irrespective of the
state of the system (HG or QGP), as long as the momentum spectra of
the radiated particles are ``thermal'' with a common temperature
(inverse slope). We see that once the left hand side is known
experimentally, it determines rather accurately the values of
$\mu_{\rm q},\mu_{\rm s}$ which enter on the right hand side with a
dominating factor 6, while the (small) flavor asymmetry $\delta\mu$, 
Eq.\,(\ref{delmu}), plays only a minor, but significant role, 
given the precision of the experimental results \cite{analyze}. 
This explains how, by applying these
identities to the early WA85 data \cite{WA85}, it has been possible
\cite{Raf91} to determine the chemical potentials with considerable
precision in spite of the still relatively large experimental errors
on the measured values of $R_\Lambda$, $R_\Xi$.
 
We obtain the following values of the chemical
potentials for S--W central collisions at 200A GeV:
\begin{eqnarray}
  {\mu_{\rm q}\over T} &&\hspace{-0.6cm}=  {\ln R_\Xi/R_\Lambda^2      
            \over 5.94}=0.39\pm0.04 \, , 
		\qquad \lambda_{\rm q} = 1.48\pm0.06
\label{muq}\\
{\delta\mu\over T}&&\hspace{-0.6cm}={\mu_{\rm q}\over T}\delta q= 
    0.031\pm 0.003 \, ,      
\label{dmu}\\
{\mu_{\rm s}\over T} &&\hspace{-0.6cm}=  {\ln R_\Lambda/R_\Xi^2\     
-0.062\over 6} = 0.02\pm 0.05\, . 
\label{mus}
\end{eqnarray}
Where $\delta q$, see Eq.\,(\ref{delq}) is valence quark flavor asymmetry.
In our dynamical description of the collision \cite{dynamic}, 
see section  \ref{thermalfsec}, we have been able to determine the value
$\lambda_{\rm q}$ reached in the collision. Naturally, as long as 
a QGP fireball is rapidly hadronizing, we have
$\mu_{\rm s}\simeq 0$ that is  $\lambda_{\rm s}\simeq 1$.
We find in section \ref{results} below substantial variation 
of $\lambda_{\rm q}$ with fireball energy content. 
Therefore the agreement of our here presented analysis with these
theoretical results can not be seen as being accidental.

We now show how in the thermal model the ratios between antibaryons with
different strange quark content are dependent on the degree 
of the strangeness saturation. Now it is important to remember that
our evaluation of the ratios is  at fixed $m_\bot$.
Up to cascading corrections 
a complete cancelation of the  fugacity and Boltzmann 
factors occurs when we form the product of the abundances 
of baryons and antibaryons,  comparing this product 
for two different particle kinds\cite{Raf91}, { e.g.}:
 \begin{equation}
   \left. {\Xi^-\over\Lambda}
   \cdot {\overline{\Xi^-}\over \overline{\Lambda}
   }\right\vert_{m_\perp>m_\perp^{\rm cut}} =\gamma_{\rm s}^2 \,,
  \label{gam1}
 \end{equation}
where we neglected resonance feed-down contribution in first 
approximation, which are of course considered in numerical 
studies \cite{analyze}..
Similarly we have  \begin{equation}
   \gamma_{\rm s}^2 = \left.
     {\Lambda\over p} \cdot {\overline{\Lambda}\over \overline p}      
  \right\vert_{m_\perp>m_\perp^{\rm cut}}
                    = \left.
     {\Omega^-\over 2\Xi^-} \cdot {\overline{\Omega^-} \over
                                            2\overline{\Xi^-}}
                      \right\vert_{m_\perp>m_\perp^{\rm cut}} \, ,
\label{gam3}
 \end{equation}
where in the last relation the factors 2 in the denominator correct
for the spin-3/2 nature of the $\Omega$.

Combining the experimental result Eq.\,(\ref{new1}) with
Eqs.\,(\ref{gam1}), we find the value $\gamma_{\rm s}=0.55 \pm 0.04$\,.
In a full analysis\cite{analyze} which accounts more precisely for resonance
decay and flow, this result becomes\begin{equation}
\gamma_{\rm s}=0.75\pm0.15\,. 
 \label{gams} \end{equation}

In part, the error stems from the dependence of the 
resonance cascading on the temperature $T_{\rm f}$ at which the final 
state hadrons are formed, assuming that the relative 
population of different hadrons is determined 
by the thermal populations. The calculation of the resonance 
decay effect is actually not simple, since resonances at
different momenta and rapidities contribute to a given daughter 
particle $m_\bot$. As the experimental measurements often sum 
the $m_\bot$ distributions with $m_\perp \geq m_\perp^{\rm cut}$
it is convenient to consider this integrated abundance for particle `i'
at a given (central) rapidity $y$:
 \begin{equation}
   \left. {dN_i \over dy} \right\vert_{m_\perp \geq
                                 m_\perp^{\rm cut}}
   =  \int_{m^{\rm cut}_\perp}^\infty dm_\perp^2
  \left\{ {dN_i^{0}(T) \over dy\, dm_\perp^2} +
 \sum_R b_{R\to i} {dN_i^R(T) \over dy\, dm_\perp^2} \right\}\, ,
\label{reso}
 \end{equation}
showing the direct `0' contribution  and the daughter
contribution from decays into the observed channel $i$) 
of resonances $R\to i$\,, with branching ratio $b_{R\to i}$\,, 
see Ref.\cite{analyze,sollfrank}. Extracting the degeneracy factors
and fugacities of the decaying resonances, we write shortly
 \begin{equation}
  N^R_i \equiv \gamma_R \lambda_R \tilde N^R_i\,,
 \label{ntilde}
 \end{equation}
and imply that particles of same quantum numbers are comprised in each 
$N^R_i$. Here $\gamma_R $ is the complete non-equilibrium factor of 
hadron (family) $R$. Between particles and anti-particles 
we have the relation
 \begin{equation}
N_{\bar i}^{\bar R} =  \gamma_R\, \lambda_R^{-1}\, {\tilde N}_i^R      
         = \lambda_R^{-2}\, N_i^R \, .
 \label{ntildea}
 \end{equation}
Thus the above considered particle ratios now become:
 \begin{eqnarray}
\left. {\overline{\Xi^-} \over \Xi^-}
\right\vert_{m_\perp\geq m_\perp^{\rm cut}}
         &&\hspace{-0.6cm}= {\gamma_{\rm s}^2 \lambda_{\rm q}^{-1} 
\lambda_{\rm s}^{-2} 
     {\tilde N}_\Xi^{\Xi^*} +
            \gamma_{\rm s}^3 \lambda_{\rm s}^{-3}
            {\tilde N}_\Xi^{\Omega^*} \over
            \gamma_{\rm s}^2 \lambda_{\rm q} \lambda_{\rm s}^2
            {\tilde N}_\Xi^{\Xi^*} +
            \gamma_{\rm s}^3 \lambda_{\rm s}^3
            {\tilde N}_\Xi^{\Omega^*} }\ ,
 \label{rxi}\\
\left. \hphantom{^-}
               {\overline{\Lambda} \over \Lambda}
             \right\vert_{m_\perp\geq m_\perp^{\rm cut}}
&&\hspace{-0.6cm}=
{             \lambda_{\rm q}^{-3}{\tilde N}_\Lambda^{N^*}+
     \gamma_{\rm s} \lambda_{\rm q}^{-2} \lambda_{\rm s}^{-1}
            {\tilde N}_\Lambda^{Y^*} +
    \gamma_{\rm s}^2 \lambda_{\rm q}^{-1} \lambda_{\rm s}^{-2}
           {\tilde N}_\Lambda^{\Xi^*}
\over
            \lambda_{\rm q}^3 {\tilde N}_\Lambda^{N^*} +
            \gamma_{\rm s} \lambda_{\rm q}^2 \lambda_{\rm s}
            {\tilde N}_\Lambda^{Y^*} +
           \gamma_{\rm s}^2 \lambda_{\rm q} \lambda_{\rm s}^2
            {\tilde N}_\Lambda^{\Xi^*}
}\, , \label{rlam}\\
\left. {\Xi^- \over \Lambda}
           \right\vert_{m_\perp\geq m_\perp^{\rm cut}}
&&\hspace{-0.6cm}=
{             \gamma_{\rm s}^2 \lambda_{\rm q} \lambda_{\rm s}^2       
             {\tilde N}_\Xi^{\Xi^*} +
            \gamma_{\rm s}^3 \lambda_{\rm s}^3
            {\tilde N}_\Xi^{\Omega^*}
\over
           \lambda_{\rm q}^3{\tilde N}_\Lambda^{N^*}+
            \gamma_{\rm s} \lambda_{\rm q}^2 \lambda_{\rm s}
            {\tilde N}_\Lambda^{Y^*} +
            \gamma_{\rm s}^2 \lambda_{\rm q} \lambda_{\rm s}^2
            {\tilde N}_\Lambda^{\Xi^*}
}\, .\label{rs}
 \end{eqnarray}
$\tilde N_\Lambda^{Y^*}$ contains also (in fact as its most important
contribution) the electromagnetic decay $\Sigma^0 \to \Lambda +
\gamma$.

Three different cases were considered \cite{analyze}
and results are presented in the table~\ref{sollresult1}:
 \begin{itemize}
 \item[{\bf A}:] a thermal model without flow, $\beta_{\rm f}=0$,
where the temperature $T_{\rm f}$ is assumed to correspond directly to
the apparent value $T_{\rm  app}=232$ MeV following from the slope of the
transverse mass spectra of high-$m_\bot$ strange (anti-)baryons;
 \item[{\bf B}:] a model with a freeze-out temperature of $T_{\rm
f}\simeq150$ MeV, {\it  i.e.,\/} a value consistent with the kinetic 
freeze-out criterion developed in \cite{ULI93} and with 
lattice QCD data \cite{lattice} on
the phase transition temperature, which entails a flow velocity at
freeze-out of  $\beta_{\rm f}=0.41$ in order to allow for the blue-shift
of the transverse particle spectra inverse slope to the value $T=232$ MeV;
 \item[{\bf C}:] in order to maintain zero net strangeness in the HG
fireball without additional off-equilibrium population factor
characterizing the relative chemical equilibrium between strange meson and
baryon abundances, the case $T_{\rm f}=190$ MeV with
$\beta_{\rm  f}=0.20$ was also explored.
 \end{itemize}

{\begin{table}[t]
\caption{\small
Thermal fireball parameters extracted from the
WA85 data \protect\cite{WA85} on strange baryon and anti-baryon production,
for three different interpretations of the measured $m_\bot$-slope.
Resonance decays were included. For details see text.
\protect\label{sollresult1}
}
\vspace{0.cm}
\begin{center}
\begin{tabular}{c|ccc} \hline\hline
        &
        {\bf A}                $\!\!\!\phantom{\Big\vert}$
        &        {\bf B}        &        {\bf C} 
\\
 
 T(MeV)       &       232         &        150        &        190
\\
 
 $\beta_{\rm f} $       &       0         &    0.41       & 0.20
\\

                  
                    $\lambda_{\rm s}$
                 &  1.03 $\pm$ 0.05
                 &  1.03 $\pm$ 0.05
                 &  1.03 $\pm$ 0.05
\\
                 $ \mu_{\rm s}/T$
               &  0.03 $\pm$ 0.05
               &  0.03 $\pm$ 0.05
               &  0.03 $\pm$ 0.05
\\
              $\mu_{\rm s}$ (MeV)
           &  7 $\pm$ 11
           &  4 $\pm$ 7
           &  6 $\pm$ 9
\\

                 
              $\lambda_{\rm q}$
            & 1.49 $\pm$ 0.05
            & 1.48 $\pm$ 0.05
            & 1.48 $\pm$ 0.05
\\
               $\mu_{\rm q}/T$
            &  0.40 $\pm$ 0.04
            &  0.39 $\pm$ 0.04
            &  0.39 $\pm$ 0.04
\\
               $\mu_{\rm B}$ (MeV)
            &  278 $\pm$ 23
            &  176 $\pm$ 15
            &  223 $\pm$ 19
\\

                 
              $\gamma_{\rm s}$
            &  0.69 $\pm$ 0.06
            &  0.79 $\pm$ 0.06
            &  0.68 $\pm$ 0.06
\\

                 
                              $\varepsilon$
                 &  $-0.22$
                 &   0.37
                 &    0
\\

                  
                              ${\cal S/B}$
                 & 18.5 $\pm$ 1.5
                 & 48 $\pm$ 5
                 & 26 $\pm$ 2.5
\\
                   $D_{\rm Q}$
                 & 0.135 $\pm$ 0.01
                 & 0.08 $\pm$ 0.01
                 & 0.12 $\pm$ 0.01

\end{tabular}
\end{center} 
 
\end{table}}
Several interesting results can be deduced by inspection of the 
table \ref{sollresult1}:
\vspace{-3mm}

 \begin{itemize}\setlength{\itemsep}{-0.2 cm}
\item[1.] The value of  $\lambda_{\rm q}$ are little
affected by resonance decays and by the origin of the slope of the
$m_\perp$-spectrum (thermal or flow). The absolute value of the
associated chemical potentials does by definition depend on the
freeze-out temperature $T_{\rm f}$.
\item[2.] The conclusion of \cite{Raf91} that the WA85 data on
strange baryon and anti-baryon production from 200 GeV\,A S--W
collisions establish  a vanishing strange quark chemical potential is
firmly confirmed, this result was found to remain  stable 
under large variations in the freeze-out 
temperature and  transverse flow velocity.
\item[3.] In the strangeness saturation factor
$\gamma_{\rm  s}$ the effects from resonance decays and flow can be
clearly seen and are of magnitude 15\%. However, no final state
condition could be found allowing the chemical equilibrium 
value $\gamma_{\rm s}^{\rm eq} = 1$.
 \end{itemize}
\vspace{-3mm}

Particle production at $T_{\rm f}=190$ leads to strangeness 
balanced emission, as indicated by the value of strangeness 
asymmetry in the produced particle population $\varepsilon$:
 \begin{equation}
   \varepsilon \equiv {\langle \bar s \rangle - \langle s \rangle
\over \langle s \rangle}\, ,
 \end{equation}
If the fireball disintegrates into equilibrium abundances of meson and 
baryons, which is not necessarily the case in a sudden reaction picture, 
then the high $T$ case (A) would lead to preferential emission of $s$-quarks, 
distilling a residue of $\bar s$-nuggets,
while the low $T$ case (B) would distill an $s$-nugget. This change in 
distillation properties of the evaporating fireball was noted long 
ago \cite{Raf87}. On the other hand, a suddenly disintegrating fireball 
should not be seen as leading to equilibrium meson to baryon abundances 
and this effect can restore the symmetric evaporation of strange and 
antistrange quarks, at the same time as the excess entropy content of 
the deconfined phase is resolved. 
We will discuss further the nonequilibrium hadronization features in
section \ref{hadromod}, while the entropy content of the fireball will 
preoccupy our attention in section \ref{entropysec}; the relevant variables 
are  shown at the bottom of the table~\ref{sollresult1}.

An avid reader of this analysis will of course observe that there are
three independent particle ratios we have considered: 
$$\Lambda/\overline{\Lambda},\,\quad
\Xi/\overline{\Xi},\, \quad \overline{\Xi}/\overline{\Lambda}$$ 
(the forth ratio is a product $ \Xi/\Lambda=\Xi/\overline{\Xi}\cdot 
\overline{\Xi}/\overline{\Lambda}\cdot \overline{\Lambda}/\Lambda$) 
but we also introduced here three parameters 
to be measured, $\lambda_{\rm q},\,\lambda_{\rm s},\, \gamma_{\rm s}$\,.
The  question which can be posed is: are the properties we derive {\it 
consistent} with the other particle abundances? Is there a dynamical 
model which will lead to the here `measured' values of the three 
parameters? 

Looking at the baryon-antibaryon sector, there were two recent results that 
we have already described, that should be consistent with this analysis, 
the NA35 $\overline{\Lambda}/\bar p$ ratio\cite{NA35pbar} and 
the first determination of the $\Omega$-sector\cite{Omega}. In both instances,
the observed yields are as predicted when our original analysis was 
made \cite{analyze} --- this provides strong support for the usage of strange 
baryons as chemico-meters of rapidly dissociating fireball. We will return
to discuss these results when we present the dynamical model which indeed 
leads to prediction of chemical properties with agreement with the results
of this analysis (`as has been measured'). 

\baselineskip=14pt
\vspace{-3mm}
\section{Thermal QGP fireball}\label{thermalfsec}

\vspace{-3mm}
\subsection{QGP equations of state}\label{EOSQGP}
\vspace{-1mm}

The QGP equations of state (E0S) are of considerable relevance
for the understanding of the magnitudes of different variables we consider
here. We use a rather standard, perturbative/nonperturbative  QCD
improved set of relations based on the Fermi/Bose liquid model with
thermal particle masses. The partition function of the interacting 
quark-gluon phase can be written as:
\begin{eqnarray}\label{ZQGP}
\hspace*{-0.cm}\ln Z^{\rm QGP}=\sum_{i\in \rm QGP}\! 
{{g_i(\alpha_s) V}\over{2\pi^2}} \int
\!\pm\ln\left(1\pm \gamma_i\lambda_i e^{-\sqrt{m_i^2(T)+p^2}/T} 
\right) p^2\,dp\,,
\end{eqnarray}
where $ i=g,\ q,\ \bar q,\ s,\ \bar s$, with $\lambda_{\bar\imath} =
\lambda_i^{-1}$ and $\gamma_{\bar\imath}=\gamma_i$. We take into account
the QCD interactions between quarks and gluons by allowing for thermal
masses 
\begin{equation}
m_i^2(T)=(m_i^0)^2+(c\,T)^2\,.
\end{equation} 
For the current quark masses we take: 
$$m_{\rm q}^0=5\  {\rm MeV},\quad m_{\rm s}^0=160\  
{\rm MeV},\quad  m_{\rm g}^0=0\,.$$
We have $c^2\propto\alpha_{\rm s}$, $\alpha_s$ being the QCD coupling 
constant. We fix $c=2$, arising for $\alpha_s\sim 1$ 
(the exact value was not of essence), while also
allowing for another effect of the QCD-interactions, 
the reduction of the number of effectively available 
degrees of freedom: we implement the following effective counting of 
gluon and quark degrees
of freedom, motivated by the perturbative QCD formul\ae:
\begin{eqnarray}
g_{\rm g}=16\quad&\to&\quad 
   g_{\rm g}(\alpha_s)=  16\left(1- {15\alpha_s\over 4 \pi}\right)\,,
\nonumber\\
g_{i-{\rm T}}=6\quad&\to&\quad  
 g_{i-{\rm T}}(\alpha_s)=    6\left(1-{50\alpha_s\over 21\pi}\right)\,,
\label{eq12}\\ 
g_{i-{\rm B}}=6\quad&\to&\quad  
  g_{i-{\rm B}}(\alpha_s)=   6\left(1-2{\alpha_s\over \pi}\right)\,,
\nonumber
\end{eqnarray}
where $i=u,\ d\,$ (we do not correct the strange quark degeneracy).
In Eq.\,(\ref{eq12}) two factors are needed for quarks:  the factor 
$g_{i-{\rm T}}$ 
controls the expression when all chemical potentials vanish 
(the $T^4$ term in the partition function for massless quarks) 
while $g_{i-{\rm B}}$ is taken as coefficient of the
additional terms which arise in presence of chemical potentials.
We took $\alpha_s=0.6$ which turned out to be the value best suited  for
the experimental data points.
We explore the physical properties of
the QGP fireball by considering the constraint between $T,\lambda_{\rm q}$
arising from a given initial specific energy content\footnote{Here 
$B$ is the baryon number. To avoid confusion, below the bag constant is 
denoted~${\cal B}$.} $E/B$. The collision energy gives us the values of the 
constraints to consider:
\begin{equation} \label{ECM}
{E\over B}= {\eta_{\rm E}{E_{\rm CM}}\over {\eta_{\rm B}A_{\rm part}}}
\simeq {E_{\rm CM}\over A_{\rm part}}\,,
\end{equation}
where $\eta_{\rm E}$ and $\eta_{\rm B}$ are respectively the stopping 
fraction\cite{stop} of energy and baryonic number and
$A_{\rm part}$ is the number of nucleons participating in the
reaction. The last equality follows when the stopping fractions are 
equal --- the experimental particle spectra we are addressing here,  and
in particular the visible presence of baryons in the central 
rapidity region, are implying
that this is a reasonable assumption for the current experimental
domain. In consequence, the energy per baryon in the fireball is to be
taken as being equal to the kinematic energy available in the collision.
In the current laboratory target experiments we have the
following kinematic energy content:
\begin{center}
\begin{tabular}{ll}
Au--Au  at 10.5A GeV  \quad & $\to$\quad $E/B = 2.3$ GeV ,\\ 
Si--Au at 14.6A GeV  & $\to$\quad  $E/B=2.6$ GeV, \\
A--A at 40A GeV  & $\to$\quad  $E/B=4.3$ GeV, \\ 
Pb--Pb at 158A GeV   &  $\to$\quad  $E/B =8.6$ GeV,\\
S--W/Pb at 200A GeV   &  $\to$\quad  $E/B =8.8$ GeV ,  \\
S--S at 200A GeV  & $\to$\quad  $E/B= 9.6$ GeV ,\\
\end{tabular}
\end{center} 
\begin{figure}[htb]
\vspace*{-0.7cm}
\centerline{\hspace*{-.7cm}
\psfig{width=14cm,figure=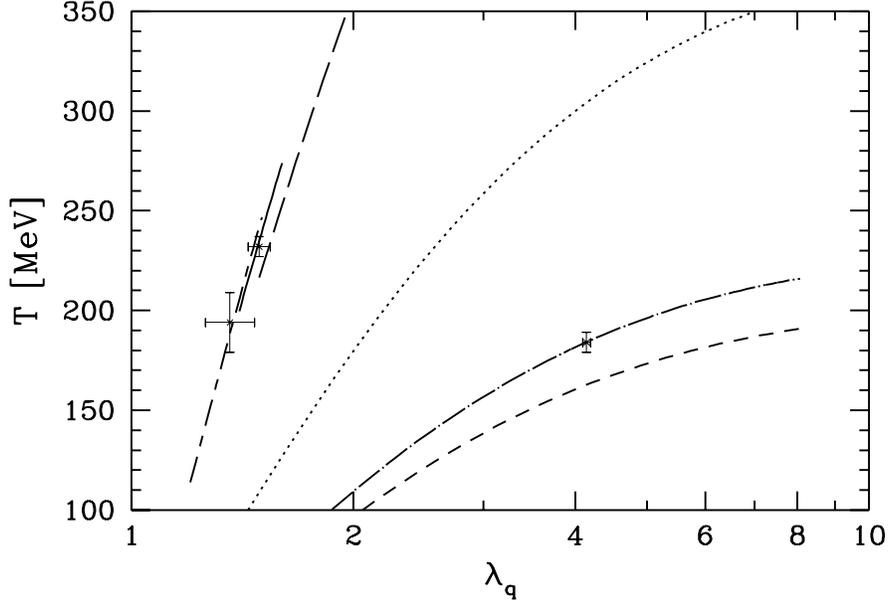}
}
\vspace*{-0.5cm}
\caption{ \small 
QGP-EoS constraint between temperature 
$T$ and light quark fugacity $\lambda_{\rm q}$ for a given fireball
energy content per baryon $E/B$ appropriate for the AGS and SPS
collision systems. Left to right: 2.3 (Au--Au), 2.6 (Si--Au),
4.3 (A--A), 8.6 (Pb--Pb), 8.8 (S--PB/W) and 
9.6 (S--S) GeV. See text for a discussion of experimental
point.\label{energy}
}
\end{figure}
Note that above we assumed collision with the geometric target 
tube of matter\cite{Sch88}, see section \ref{tools},
when the projectile is smaller than the target. In Fig.\,\ref{energy} we show
in the $T$--$\lambda_{\rm q}$ plane  the lines corresponding to this
constraint on the QGP-EoS. In the middle  the line corresponding to the 
lowest SPS accessible energy, 4.3 GeV, is depicted, which bridges the  
current SPS 
domain  shown to the left to the  BNL region on the lower right.
The experimental crosses show the values of  $\lambda_{\rm q}$ arising  
in our data analysis  \cite{analyze,BNL-AGSthermal,LetAGS}, combined 
with the inverse slope temperatures, 
extracted from transverse mass particle spectra. The fact that the 
experimental results  fall on the lines shown in  Fig.\,\ref{energy} 
is primarily due to  the choice $\alpha_s=0.6$ --- as this is the usual 
value in this  regime of energy it
implies for a QGP fireball EoS hypothesis that the assumption that 
stopping of energy and baryon number is similar deserves
further consideration.

\subsection{Initial conditions and fireball evolution}
There now remains  the issue what physical constraint or principle
determines which of the possible pair of $T,\ \lambda_{\rm q}$ values 
along the individual curves depicted in Fig.\,\ref{energy} (see 
experimental crosses shown) is actually initially reached in the 
reaction. We have explored the properties of the QGP
phase along these lines of constant energy per baryon and have  noticed
that with increasing $T$ the pressure in the QGP phase increases, and
that the experimental points coincide  with the dynamical pressure 
generated in the collision.  This gives birth to the intuitive idea that the 
initial conditions reached in the central fireball arise from the 
equilibrium between the fireball internal thermal pressure and the external 
compression pressure.

This condition takes the form \cite{dynamic}: 
\begin{equation}\label{Pbal}
P_{\rm th}(T,\lambda_i,\gamma_i)=P_{\rm dyn}+P_{\rm vac}\,.
\end{equation}
The thermal pressure follows in usual way from the partition function
\begin{equation}\label{Ptherm}
P_{\rm th}={T/V}\ln Z
(T,\lambda_{\rm q},\lambda_{\rm s};\gamma_{\rm g},
\gamma_{\rm q},\gamma_{\rm s})\,,
\end{equation} 
where aside of the temperature $T$, 
we encounter the different fugacities $\lambda_i$
and the chemical saturation factors $\gamma_i$ for each particle. For the
vacuum pressure we will use:
\begin{equation}\label{Pvac}
P_{\rm vac}\equiv{\cal B}\simeq0.1\ \mbox{GeV/fm}^3\,.
\end{equation}
 
\begin{figure}[b]
\vspace*{-0.5cm}
\centerline{\hspace*{-1.cm}
\psfig{width=15.5cm,figure=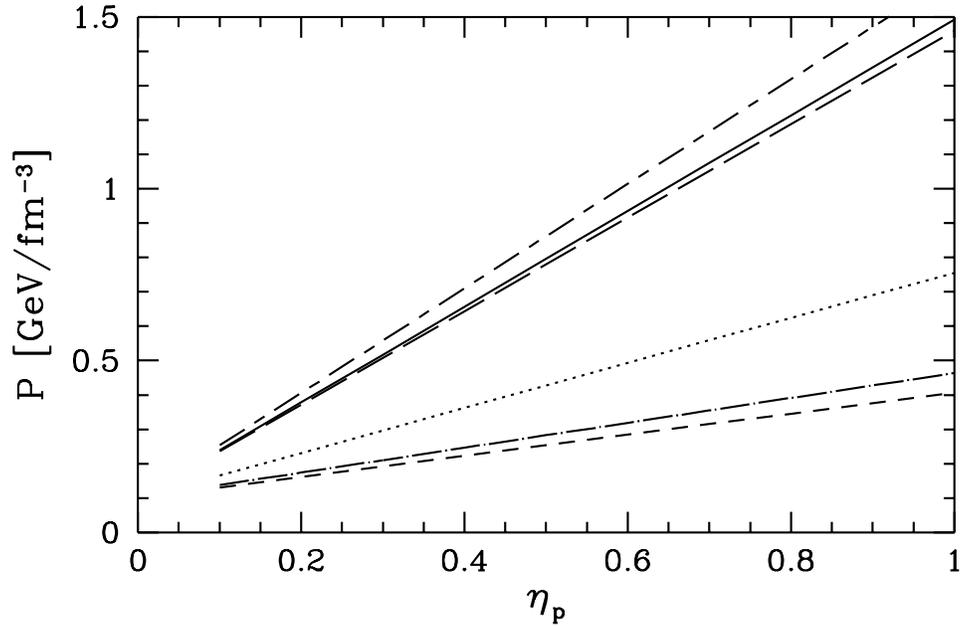}}
\vspace*{-0.5cm}
\caption{ \small 
The collision pressure $P$ as function of momentum stopping $\eta_{\rm
p}$ for different values of $E/B$ --- 2.3, 2.6, 4.3, 8.6, 8.8 and
9.6 GeV (from bottom to top, solid line is for 8.8 GeV), 
from Ref.\cite{dynamic}.
 \label{Pevolution}}
\end{figure}
The pressure due to kinetic motion follows from well-established
principles, and can be directly inferred from  the pressure
tensor \cite{deG80}
\begin{equation}\label{Tij}
T^{ij}(x)=\int\! p^iu^j f(x,p){\rm d}^3\!p\,,\quad i,j=1,2,3\,.
\end{equation}
We take for the phase-space distribution of 
colliding projectile and target nuclei
\begin{equation}\label{fxp}
f_{\rm P,T}(x,p)=\rho_{\rm P,T}(x)\delta^3(\vec p\pm\vec p_{\rm CM})\,,
\end{equation}
and hence in Eq.\,(\ref{Tij}) $u^j=\pm p^j_{\rm CM}/E_{\rm CM}$. 
We assume that the nuclear density is uniform within the nuclear size, 
$\rho_{0}=0.16$ /fm$^3$.
 
To obtain the pressure exerted by the flow of colliding matter, we
consider  the pressure component $T^{jj}$, with $j$ being  
the direction of $\vec v_{\rm CM}$. This gives
\begin{equation}\label{Pdyn}
 P_{\rm dyn}=\eta_{\rm p} \rho_{0}\frac{p_{\rm CM}^2}{E_{\rm CM}}\,.
\end{equation}
Here it is understood that the energy $E_{\rm CM}$  and the momentum
$p_{\rm CM}$ are given in the nucleon--nucleon CM frame and $\eta_{\rm
p}$ is the momentum stopping fraction --- only this fraction 
$0\le\eta_{\rm p}\le 1$ of the incident CM momentum can be used by a
particle incident on the central  fireball  (the balance remains in the
unstopped longitudinal motion) in order to exert dynamical pressure. For
a target transparent to the incoming flow, there would obviously be no
pressure exerted.  The simple expression Eq.\,(\ref{Pdyn}) is
illustrated in Fig.\,\ref{Pevolution} as function of the stopping fraction.
At current energies with stopping being above 50\% we explore
the conditions above 0.7 GeV/fm$^3$.

\begin{table}[tb]
\caption{Properties and evolution of different collision systems.
\protect\label{bigtable}}
\vskip 2mm
\begin{center}\small
\begin{tabular}{c|c|cccccc}
\hline\hline
Phase &  & \multicolumn{6}{c}{\phantom{$\displaystyle\frac{E}{B}$}$E/B$
[GeV]
\phantom{$\displaystyle\frac{E}{B}$}}\\
\raisebox{1mm}{space} & \raisebox{2mm}{$<\!s-\bar s\!>=0$}
                         & 2.35  &  2.6  &  4.3  &  8.8  &  8.6 & 8.6 \\
$\!$occupancy$\!$ & $\lambda_s\equiv 1$  & $\eta=$ 1
&$\eta=$1& $\eta=$ 1 & $\eta\!=\! 0.5$ & $\!\!\eta\!=\! 0.75\!\!$&$\eta=$ 1\\
& &$\!$Au--Au$\!$ &$\!$Si--Au$\!$&$\!$Pb--Pb$\!$ & S--Pb 
&$\!$Pb--Pb$\!$&$\!$Pb--Pb$\!$\\
\hline\hline
 & $T_{th}$ [GeV] & 0.238  & 0.260  & 0.361  & 0.410  & 0.444  & 0.471\\
$\gamma_q=$ 0.2 & $\lambda_q$
                         & 13.3  & 9.95  & 3.76  & 1.78 &  1.91  & 2.00\\
 & $n_g/B$               & 0.15  & 0.20  & 0.54  & 1.55  & 1.36  & 1.25\\
 & $n_q/B$               & 3.00  & 3.00  & 3.13  & 5.12  & 3.89  & 3.77\\
$\gamma_g=$ 0.2 & $n_{\bar q}/B$
                         & 0.00  & 0.00  & 0.13  & 2.12  & 0.89  & 0.77\\
 & $n_{\bar s}/B$        & 0.02  & 0.02  & 0.06  & 0.16  & 0.14  & 0.13\\
 & $\!P_{th}\!$ [GeV/fm$^3$]
                         & 0.42  & 0.46  & 0.76  & 0.79  & 1.12  & 1.46\\
$\gamma_s=$ 0.03 & $\rho_{\rm B}$
                         & 3.34  & 3.34  & 3.30  & 1.70  & 2.44  & 3.18\\
 & $S/B$                 & 10.7  & 11.8  & 18.8  & 40.0  & 35.8  & 33.4\\
\hline
& $T_{ch}$ [GeV]        & 0.200  & 0.212 & 0.263 & 0.280 & 0.304& 0.324\\
$\gamma_q=$ 1 & $\lambda_q$
                         & 4.92  & 4.14  & 2.36  & 1.49  & 1.56  & 1.61\\
 & $n_g/B$               & 0.47  & 0.56  & 1.08  & 2.50  & 2.24  & 2.08\\
 & $n_q/B$               & 3.06  & 3.11  & 3.51  & 5.16  & 4.81  & 4.62\\
$\gamma_g=$ 1 & $n_{\bar q}/B$
                         & 0.06  & 0.11  & 0.51  & 2.16  & 1.81  & 1.62\\
 & $n_{\bar s}/B$        & 0.04  & 0.05  & 0.11  & 0.25  & 0.22  & 0.21\\
 & $\!P_{ch}\!$ [GeV/fm$^3$]
                         & 0.42  & 0.46  & 0.76  & 0.79  & 1.12  & 1.46\\
$\gamma_s=$ 0.15 & $\rho_{\rm B}$
                         & 3.34  & 3.35  & 3.31  & 1.80  & 2.45  & 3.19\\
 & $S/B$                 & 11.0  & 12.3  & 19.7  & 41.8  & 37.4  & 34.9\\
\hline
 & $\gamma_s$ & 1  & 1 & 1 & 0.8 & 1 & 1\\
$\gamma_q=$ 1 & $T_0$ [GeV]
                         & 0.176  & 0.184  & 0.215  & 0.233  & 0.239  &
0.255\\
 & $\lambda_q$           & 4.92  & 4.14  & 2.36  & 1.49  & 1.56  & 1.61\\
$\gamma_g=$ 1 & $n_g/B$  & 0.47  & 0.56  & 1.08  & 2.50  & 2.25  & 2.09\\
 & $n_q/B$ & 3.11          & 3.06  & 3.51  & 5.12  & 4.81  & 4.60\\
 & $n_{\bar q}/B$        & 0.06  & 0.11  & 0.51  & 2.12  & 1.81  & 1.62\\
$\gamma_s=$ 0.8 & $n_{\bar s}/B$
                         & 0.29  & 0.34  & 0.68  & 1.27  & 1.43  & 1.33\\
or & $\!P_0\!$ [GeV/fm$^3$]
                         &0.28  & 0.30  & 0.41  & 0.47  & 0.54  & 0.71\\
$\gamma_s=$ 1 & $\rho_{\rm B}$
                         & 2.29  & 2.17  & 1.80  & 1.05  & 1.19  & 1.56\\
 & $S/B$                 & 12.9  & 14.5  & 24.0  & 49.5  & 46.5  & 43.4

\end{tabular}
\end{center}
\end{table}


We now can determine the initial conditions reached in heavy ion
collisions, since the two constraints, energy per baryon and
pressure allow to fix the values of $\lambda_{\rm q}$ and $T$,
provided that  we make a hypothesis about the degree of chemical 
equilibration of the state considered. 
In order to have some understanding of the conditions prevailing in the
early stages of the collision process, when the thermal equilibrium is
reached, but the chemical equilibrium for all components is still far
away, we take  20\% occupancy for gluons and light quarks, and 3\% for
strange quarks and solve the EoS for $T_{\rm th}$, and the associated
$\lambda_{\rm q}$\, which are shown along with other interesting
properties of the fireball (number of gluons per baryon, number of light
quarks and antiquarks per baryon, number of anti-strange quarks per
baryon, the pressure in the fireball, baryon density and the entropy per
baryon) in the top section of the table~\ref{bigtable}. Because the QGP 
phase is strangeness neutral we have always $\lambda_{\rm s}=1$. 
The columns of
table correspond to the  cases of specific experimental
interest,  in turn:  Au--Au and Si--Au collisions at AGS, 
possible future Pb--Pb collisions at SPS with 40A
GeV, S--Pb at 200A GeV, and for the Pb--Pb collisions at 158A GeV we
considered two possible  values of stopping, see Eq.\,(\ref{Pdyn}):
$\eta=0.75$ and  $\eta=1$\,. 
 
\begin{figure}[p]
\vspace*{-0.3cm}
\centerline{\hspace*{-3.5cm}
\psfig{width=13.8cm,figure=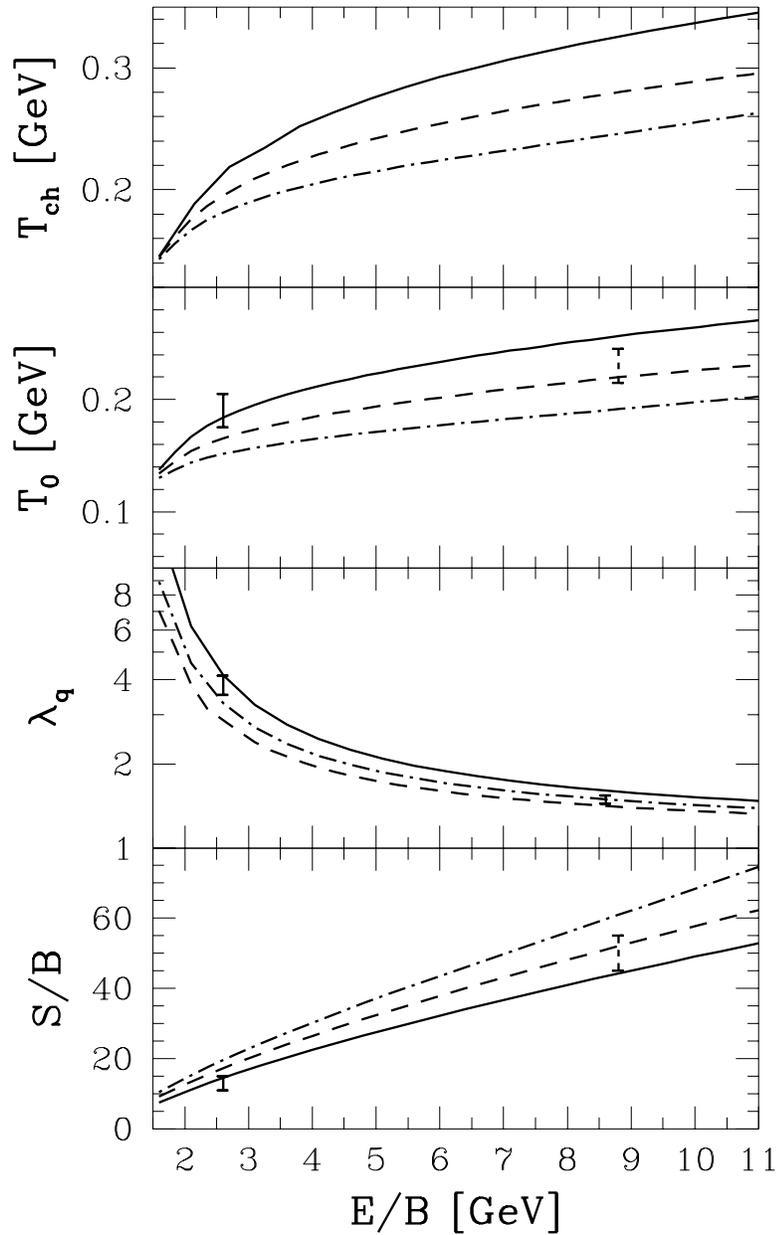}}
\vspace*{-1.2cm}
\caption{ 
Initial fireball temperature $T_{\rm ch}, T_0$, light quark
fugacity $\lambda_{\rm q}$
and entropy per baryon $S/B$ at the time of maximum
chemical equilibration, as function of the QGP-fireball energy content 
$E/B$; stopping $\eta=1$ (solid line), 1/2 (dot-dashed line) and 1/4
(dashed line). See text for comparison with analysis results.
\label{fig1S95}
}
\end{figure}
Next in our consideration of the system is the configuration when the $u$,
$d$ quarks and gluons have reached their chemical equilibrium abundances,
$\gamma_{\rm q}\to 1$, $\gamma_{\rm g}\to 1$. 
$T_{\rm ch}$ and  $\lambda_{\rm q}$ are shown in the middle section of
the table~\ref{bigtable}\,. It is worth observing that the baryon density
in the fireball introduces from the onset a rather large quark density,
which thus needs not to be produced, and thus the approach to chemical
equilibrium of the light quarks is here  faster than in the baryon-free
central region environments expected at much higher RHIC/LHC energies. It
can be argued that this partial (excluding strangeness) chemical
equilibrium occurs at the end of the nuclear penetration, about 1.5 fm/c
after the beginning of the collision. Though of major physical interest
this observation has no relevance to the results we present below. There
is no change in the pressure between top and middle sections of table
\ref{bigtable}, as the dynamical compression with the given $P$ is
present at this stage of the fireball evolution. But we see here that
$T_{\rm th}>T_{\rm ch}$, since the number of quarks and gluons present is
considerably lower in the early stages of the collision. 
 
At $T_{\rm ch}$ the strange flavor is still far from equilibrium and we
considered $\gamma_{\rm s}(t=1.5\,{\rm fm})\simeq 0.15$, 
appropriate for strange quark relaxation time 7 times larger than
the light quark one \cite{RM82}. The exact initial value is of
little consequence for the final yields, since we find near saturation of
strangeness abundance.  

After the collision has ended, for times $1.5\le t\le 5$--$10$ fm/c, 
 the strange quarks relax to their equilibrium
abundance and the temperature drops from $T_{\rm ch}$ to the value $T_0$,
shown along with other properties in the bottom section of table
\ref{bigtable}. We make exception to this full chemical equilibrium for
the S--W case, for which we assume that strange quarks have reached 80\%
of phase space occupancy  as suggested by the experimental results
\cite{analyze,eb}. During the formation of the strangeness flavor there
is already evolution of the fireball outside of the collision region and
we allow for this by keeping $\lambda_{\rm q}=$ Const. This effectively
freezes the entropy content of gluons and light quarks,  allowing for
significant drop in pressure and some cooling due to conversion of energy 
into strangeness. Aside of this chemical cooling \cite{cool}, there is 
cooling due to (adiabatic) expansion of the fireball, 
in which $\lambda_{\rm q}=$ Const., such that $T$ decreases from $T_{\rm ch}$
to the full chemical equilibrium value $T_0$\,.  We consider also
in the simple model calculations devoted to the study of the
strangeness production in section \ref{gammas}, see Fig.\,\ref{figvarRT}. 
 
For the S--Pb/W collisions the temperature values shown in the 
bottom portion of the table are similar to the inverse slopes 
observed in particle spectra and shown in Fig.\,\ref{energy}. 
Remarkably, the values of temperature $T_0$
found for the case of $E/B=8.6$ GeV at $\eta=0.5$ is 
just $233$ MeV, which corresponds nearly exactly to the reported 
inverse slopes of the WA85 results  \cite{WA85},
and  $\lambda_{\rm q}=1.49$ also agrees exactly with the results of our 
analysis  \cite{analyze}, also shown in Fig.\,\ref{energy}. 
Even though there are a number of tacit and explicit parameters 
(in particular $\eta=0.5, \alpha_{\rm s}=0.6$) we believe that this
result supports strongly the validity of our model involving the
QGP fireball.

It is of interest for many applications to determine the initial
fireball conditions systematically as function of the specific
energy. In Fig.\,\ref{fig1S95} we show as function of the
specific energy content $E/B$, in top portion the behavior of 
temperature $T_{\rm ch}$ at which light quarks and gluons have
reached chemical equilibrium. Below it, we show values of $T_0$,
determined by requiring that also strange quarks are
in chemical equilibrium. In the next segment of the figure the
fireball light quark fugacity $\lambda_{\rm q}$ and in the
bottom section the entropy per baryon $S/B$ at maximum chemical
equilibration in the QGP fireball. 
The experimental bars show for high (8.8 GeV) energy the result
of the data analysis \cite{analyze} discussed above, and those for low
energy (2.6 GeV) are taken 
from the analysis of the AGS data \cite{BNL-AGSthermal,LetAGS}. 
The range of the possible values as function of stopping $\eta$ 
is indicated by showing the results for $\eta=1$ (solid
line), 1/2 (dot-dashed line) and 1/4 (dashed line). 
These results are in many respects fulfilling our expectations.
We note the drop in temperature with decreasing energy and
stopping; for a given specific energy the value of $\lambda_{\rm q}$
is relatively insensitive to the stopping power; there is a (rapid) rise
of specific entropy with $E/B$.

\subsection{Difference between AGS and SPS energy range}
\label{AGSsec}

In the analysis of the collisions of 
S--ions at 200A GeV  with different
nuclear targets carried out at SPS we have shown in the framework of the
thermal model that the strange-quark fugacity is
$\lambda_{\rm s}\simeq 1$, i.e., $\mu_{\rm s}=0$, see section \ref{expres}. 
Even a cursory look \cite{BNL-AGSthermal,LetAGS} at the AGS results 
\cite{newdata1,newdata2,newdata3} shows that
 $\mu_{\rm s}\ne0$\,, actually $\lambda_{\rm s}\simeq 1.7$ and 
$\lambda_{\rm q}\simeq 3.6$\,. This implies that in Si--Au 15A GeV
collisions the final state particles are not displaying the required 
symmetry properties expected for a deconfined source. This implies that:
\begin{itemize}

\vspace{-0.3cm}
\item {the deconfined phase was not formed at all in these `low' energy
collisions, or,}
 
\vspace{-0.3cm}
\item {that complete re-equilibration 
occurs when the primordial deconfined high baryon density 
matter hadronizes.}
 
\vspace{-0.3cm}
\end{itemize}
\begin{figure}[htb]
\vspace*{-1.6cm}
\centerline{\hspace{0.2cm}\psfig{width=8cm,figure=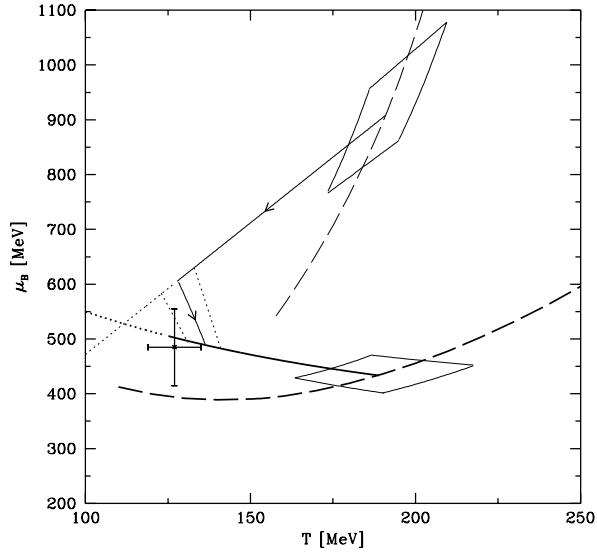}}
\vspace*{-2.3cm}
\caption{ \small
$\mu_{\rm B}$--$T$ plane with thick lines: HG model; thin lines: QGP model. 
Dashed lines: fixed energy per baryon $E/B=2.55$ GeV, lower for HG, 
upper for QGP. Solid lines: fixed specific entropy $S/B=13$, upper for 
QGP, lower for HG model. Trapezoidal regions enclose the initial 
condition, given the experimental uncertainty (see \protect\cite{LetAGS}). 
Solid line  connecting QGP and HG with dotted lines left/right: 
possible phase transition at 
$P=0.04\pm0.01$ GeV/fm$^3$. Note that continuations of $S/B=13$ lines 
beyond transition/hadronization are shown dotted, after
Ref.\protect\cite{LetAGS}.
\protect\label{F3AGS}}
\end{figure}

In Fig.\,\ref{F3AGS} we show in the $\mu_{\rm B}$-$T$ plane for the HG case
(thick lines) and QGP case (thin lines) the two different 
hypothetical histories of the collision. The dashed lines show the 
constraint arising from consideration of the fixed energy per baryon 
2.55 GeV A in the CM frame for HG (upper-) and QGP (lower line), 
while the solid lines are for fixed specific 
entropy per baryon $S/B=13$, deduced from the particle 
abundance observed in the final state \cite{LetAGS}. 
Where the solid and dashed lines meet, within a 
trapezoidal region determined by one unit error in entropy and an
error of 0.15 GeV in CM energy, we have a consistency conditions satisfied 
between the initial  specific energy and the final state entropy, 
thus presumably these are the initial values
of thermal parameters for the two phases. The initial HG
state (for $S/B=13,\, E/B=2.55$ GeV) has an energy density 
$\varepsilon_0^{\rm HG}=2.3$ GeV fm$^{-3}$, and baryon density
$\rho_0^{\rm HG}=6\rho_N$.
The initial pressure is $P_0^{\rm HG}=0.3$ GeV fm$^{-3}$.
In the QGP phase we find 
$\varepsilon_0^{\rm QGP}=1.2$ GeV fm$^{-3}$, 
$\rho_0^{\rm QGP}=2.9\rho_N$, $P_0^{\rm QGP}$=0.39 GeV fm$^{-3}$.
Somewhat surprisingly, the QGP is the more dilute phase at these condition.
Kinetic HG simulations such as ARC \cite{ARC} also reach such
rather high baryon and energy densities in these collisions. 

We also obtain very large  difference in $\mu_{\rm B}$ which takes an 
initial value  $\mu_0^{\rm HG}=440\pm40$~MeV 
in the HG scenario and $\mu_0^{\rm QGP}=910\pm150$~MeV 
in the QGP case --- not shown in the
Fig.\,\ref{F3AGS} is that the strangeness conservation requirement 
leads in the HG to an {\it initial} value 
$\mu_{\rm s,0}^{\rm HG}=0$, just as is in the case of the QGP. 
We see in Fig.\,\ref{F3AGS} that the initial temperatures $T_0$ for QGP
and HG scenarios are practically equal. For $E/B=2.55\pm0.15$ GeV 
and $S/B=13\pm1$ we have $T_0=190\pm30$ MeV.
The QGP fireball at $S/B=13$ (thin solid line) evolves  practically at fixed 
$\lambda_{\rm q}=4.8$ and $\lambda_{\rm s}=1$. However, for the 
HG fireball at fixed  $S/B=13$ (thick solid line) there is a strong 
variation in both these  fugacities but  the ratio 
$R_\lambda^{\rm HG}=\lambda_{\rm q}/\lambda_{\rm s}=2.17$ remains 
practically constant, 
assuring that the specific entropy is constant \cite{LetAGS}. 
The `experimental' cross is set at $\mu_{\rm B,f}=485\pm 70$~MeV 
and $T_{\rm f}=127\pm8$~MeV corresponding to the freeze-out 
conditions (with $\mu_{\rm s,f}=68$ MeV) deduced from the final state
particle spectra and abundances \cite{LetAGS}. However, 
in a three dimensional display including $\lambda_{\rm s}$ we would see
that only the HG is consistent with the freeze-out point. 

In Fig.\,\ref{F3AGS} the connecting nearly 
vertical lines between the two evolution paths (QGP/HG)
denote a possible phase transformation from 
QGP to HG. This was obtained   assuming  a first order phase transition 
and using the equations of state of both phases with bag pressure ${\cal B}=
0.1\ {\rm GeV\,fm}^{-3}$ (corresponding to ${\cal B}^{1/4}=170$ MeV)
 --- we note
that there is minor re-heating occurring while the baryochemical potential 
drops by 15\% --- however, the major re-equilibration is in the jump from 
$\lambda_{\rm s}=1$ in the plasma to $\lambda_{\rm s}\sim1.7$ in the HG phase
Thus the `short' connection between the
QGP to HG paths would be considerably stretched in full three dimensional 
display, reflecting on the need to well re-equilibrate the matter 
in transition, due to substantial differences in the properties of the 
QGP and HG phases reached at AGS energies. 

Given considerable differences in the statistical 
parameters in the two evolution scenarios, and the different initial
baryon and energy densities that would be reached, we 
believe \cite{LetAGS} that it is possible to distinguish between 
HG and QGP reaction alternatives at AGS energies, though a critical 
test has not been proposed yet.

\section{Thermal flavor production}
\label{strprod}
\subsection{Population evolution}
\label{spopsec}
The production of heavy flavor is a considerably
slower process compared to the multitude of different reactions 
possible in a quark-gluon gas, which are leading to
redistribution of energy between the available particles and
lead to thermal equilibrium. Thus even if we assume without 
microscopic understanding that thermal equilibration is rapid, 
we should not expect the chemical (i.e., particle
abundance) equilibrium to be present, especially so for heavy
flavor. A well studied example of this situation is strangeness production  
which constitutes a bottleneck in (chemical)
equilibration of strongly interacting confined matter.

We will evaluate in the following the dominant particle fusion
contributions to the relaxation constant $\tau_{\rm s}$ of strangeness. 
The first order strangeness production 
processes at fixed values of $\alpha_{\rm s}=0.6$ and $m_{\rm
s}=160$--180 MeV, have been studied  14 years ago \cite{BZ83,RM82}.
Thermal non-perturbative effects were more recently explored 
in terms of thermal temperature dependent particle
masses \cite{BLM90}. After the new production rates, including the now
possible thermal gluon decay,  were added up, the total strangeness
production rate was found little changed compared to the free space rate.
This finding was challenged \cite{AS93}, but a more recent reevaluation
of this work  \cite{BCDH95} confirmed that the rates obtained with
perturbative glue-fusion processes are describing precisely the
strangeness production rates in QGP,  for the here 
relevant $T>250$ MeV temperature range. A fuller discussion of 
this matter is given in a recent review  \cite{SH95}. Thus we 
can safely assume today that the first order  strangeness production
processes are  dominating the strangeness production rates in QGP, with
$\tau_{\rm s}\simeq 2$ fm/$c$ for the here relevant $T>250$ MeV
temperature range, see Fig.\,\ref{figtauss} below. In the next 
section \ref{runalfasec}
we will address the higher order production processes. 
 
\begin{figure}[tb]
\vspace*{0.cm}
\centerline{\hspace*{0cm}
\psfig{width=9.5cm,angle=-90,figure=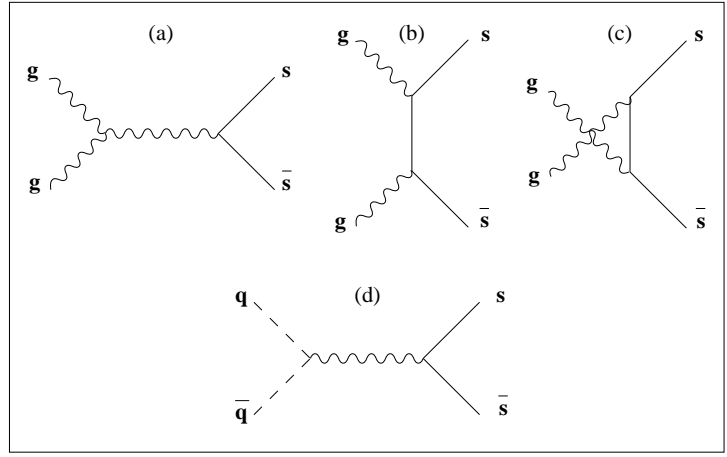}
}
\caption{ \small
Lowest-order
Feynman diagrams for production of $s\bar s$ (and similarly
$c\bar c$) by gluon fusion and 
quark pair fusion.\label{ssprod}}
\end{figure}

\begin{figure}[tb]
\vspace*{-0.7cm}
\centerline{\hspace*{-0.4cm}
\psfig{width=11cm,figure=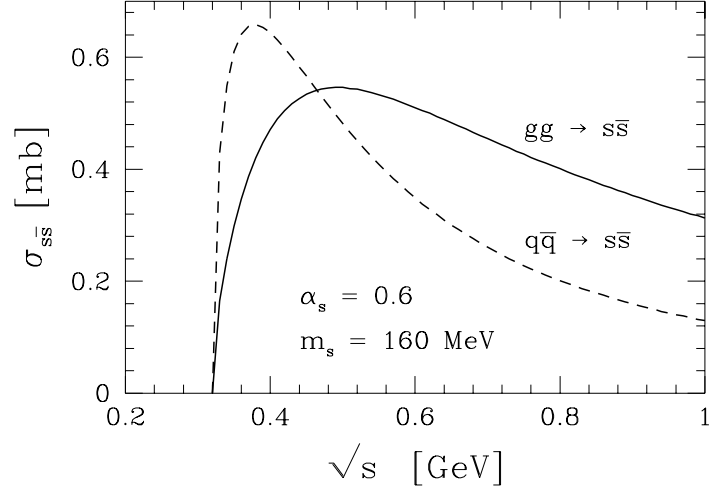}
}
\vspace*{-0.cm}
\caption{ \small
Strangeness
production cross sections for $\alpha_{\rm s} = 0.6$,  $m_{\rm s}
= 160$ MeV.}\label{sigss}
\end{figure}

We first consider the angle averaged flavor production cross sections. 
The evaluation of the lowest order 
diagrams shown in Fig.\,\ref{ssprod} yields\cite{Com74}:
\begin{eqnarray}
\bar\sigma_{gg\to s\bar s}(s) &=& {2\pi\alpha_{\rm s}^2\over 3s} \left[ 
\left( 1 + {4m_{\rm s}^2\over s} + {m_{\rm s}^4\over s^2} \right)
{\rm tanh}^{-1}W(s)\right.\nonumber\\
&&\hspace*{3.5cm}\left.-\left({7\over 8} + {31m_{\rm s}^2\over
8s}\right) W(s) \right]\,,\label{gl}\\
\bar\sigma_{q\bar q\to s\bar s}(s) &=& {8\pi\alpha_{\rm s}^2\over 27s}
\left(1+ {2m_{\rm s}^2\over s} \right) W(s)\,.\ 
\label{gk}
\end{eqnarray}
where $W(s) = \sqrt{1 - 4m_{\rm s}^2/s}$\,. We see in Fig.\,\ref{sigss}
that the magnitude of both cross sections is similar. 

With the production cross sections known, the 
net change in the strange quark abundance (and similarly
charm, though here the annihilation rate is negligible) is given
by the difference between the production and annihilation rates.
Thus the evolution of flavor abundance in the QGP can be quite
simply described by the population equation:
\begin{equation}\label{drho/dt}
{d\rho_{\rm s}(t)\over dt} = {dN(gg,q\bar q \to s\bar s)\over
d^3x\;dt} \;-\; {dN(s\bar s\to gg, q\bar q)\over d^3x\;dt}\, .
\end{equation}
This can be expressed in terms of the thermally averaged cross
sections $\langle\sigma v_{\rm rel}\rangle_T$ and particle densities
$\rho$:
\begin{eqnarray}
{d\rho_{\rm s}(t)\over dt} &=& 
\rho_g^2(t)\,\langle\sigma v\rangle_T^{gg\to s\bar s}\nonumber\\ & & 
 +\;
\rho_q(t)\rho_{\bar q}(t)
\langle\sigma v\rangle_T^{q\bar q\to s\bar s} -
\rho_{\rm s}(t)\,\rho_{\bar{\rm s}}(t)\,
\langle\sigma v\rangle_T^{s\bar s\to gg,q\bar q}\,.
\label{drho/dt1}
\end{eqnarray}
In chemical equilibrium, the strange quark density is a constant in time.
Setting the left hand side of Eq.\,(\ref{drho/dt1}) equal to 0, we
find the detailed balance relation for $t\to\infty$:
\begin{equation}
(\rho_g^\infty)^2\,\langle\sigma v\rangle_T^{gg\to s\bar s} +
\rho_q^\infty\rho_{\bar q}^\infty
\langle\sigma v\rangle_T^{q\bar q\to s\bar s}
= \rho_{\rm s}^\infty\,\rho_{\bar{\rm s}}^\infty\,
\langle\sigma v\rangle_T^{s\bar s\to gg,q\bar q}\, .\ 
\label{drho/dt2}\end{equation}
Eq.\,(\ref{drho/dt2}) relates the thermally averaged strangeness
annihilation rate to the production rate. We substitute it into
Eq.\,(\ref{drho/dt1}). Furthermore, since the kinetic and chemical
equilibration of light quarks and gluons occurs on a considerably
shorter time scale than the production of strangeness,  we can 
assume that the gluon and
light quark density is continually replenished through other
channels so that 
$$\rho_g(t) \to \rho_g^\infty;\quad \rho_q(t) \to
\rho_q^\infty;\quad \rho_{\bar q}(t) \to \rho_{\bar q}^\infty\,,$$ and we
obtain inserting this also into Eq.\,(\ref{drho/dt1}):
\begin{equation}\label{drho/dt3}
{d\rho_{\rm s}(t)\over dt} \equiv {dN_{\rm s}(t)\over {dVdt}} = 
(A_{gg}+A_{q\bar q})\left[1-\left(\rho_{\rm s}(t)\over \rho_{\rm
s}^{\infty}\right)^2\right]\, ;
\end{equation}
where we also have made use of the fact that $\rho_{\rm s}
\rho_{\bar{\rm s}}=\rho_{\rm s}^2$ in QGP, and $A$ is as defined by
\begin{equation}\label{prodgen}
A_{\rm AB}=\langle \sigma^s_{\rm AB}v_{\rm AB} \rangle_T 
               \rho_{\rm A}^\infty\rho_{\rm B}^\infty\, ;
\end{equation}

We can easily solve Eq.\,(\ref{drho/dt3}) analytically, when it is
possible to assume that the (invariant) production rate $A=A_{q\bar
q}+A_{gg}$ per unit volume and time is a constant in time:
\begin{eqnarray}\label{rhos(t)}
\gamma_{\rm s}(t)\equiv{\rho_{\rm s}(t)\over \rho_{\rm s}^\infty}
&=&{\rm tanh}(t/2\tau_{\rm s})\quad {\rm for}\ A={\rm Const.}\\
  &\simeq&(1-2e^{-t/\tau_{\rm s}})\quad{\rm for}\ t>\tau_{\rm s}\,.
\nonumber\end{eqnarray}
We see that the asymptotic limit is approached from below 
exponentially. $\tau_{\rm s}$ is referred to as the
relaxation time constant, here  for strangeness (and similarly charm)
production in QGP and is given by:
\begin{equation}\label{tauss}
\tau_{\rm s}\equiv
{1\over 2}{\rho_{\rm s}^\infty\over{(A_{gg}+A_{qq}+\ldots)}}\,,
\end{equation}
\begin{figure}[b]
\vspace*{-2.cm}
\centerline{\hspace*{-.7cm}
\psfig{width=10cm,figure=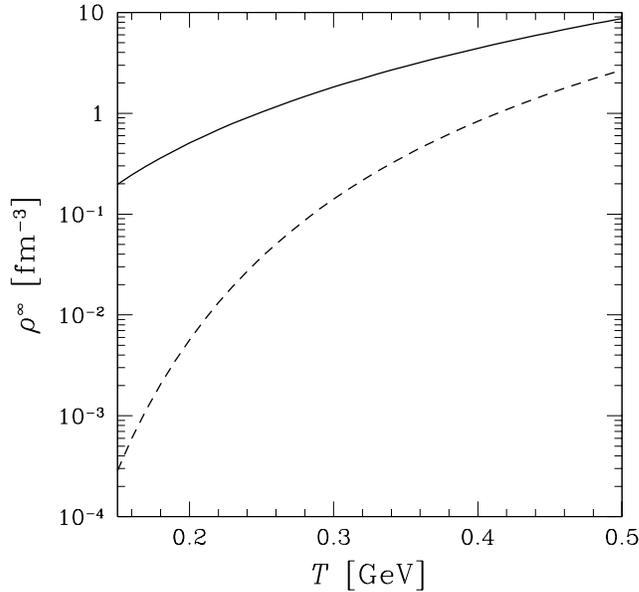}
}
\vspace*{-0.6cm}
\caption{ \small
 The statistical equilibrium density of strange or antistrange
quarks with $m_{\rm s}=160$ MeV (solid line) and charmed or
anticharmed quarks 
with $m_{\rm c}=1500$ MeV (dashed line) as
function of temperature $T$.}\label{eqsc}
\end{figure}
where the dots indicate that other mechanisms may contribute to the heavy
flavor production, further reducing the relaxation time. The equilibrium
abundance of heavy quarks and antiquarks in the QGP $\rho_{\rm s}^\infty$
is given by the convergent series expansion:
\begin{equation}\label{Nsinfty}
N_{\rm s}^\infty={3\over\pi^2}V T^3\,x^2{k}_2(x)\,,
\quad x={m_{\rm s}\over T}\,,
\end{equation}
with:
\begin{equation}\label{kl2}
{k}_2(x)\equiv \sum_{l=1}^\infty {(-)^{l+1}\over l}{\rm K}_2(lx)\,.
\end{equation}
The first term in the expansion Eq.\,(\ref{kl2}) leads to the 
Boltzmann approximation. The equilibrium density of strange ($m_{\rm
s}=160$ MeV) and charmed ($m_{\rm c}=1500$ MeV) quarks is shown in
Fig.\,\ref{eqsc}. We note that for $T\simeq250$ MeV strangeness equilibrium
abundance exceeds one $\bar s$-quark for each fm$^3$ of matter, which 
charm reaches for $T\ge 450$ MeV. However, as we shall see, charm production 
is too slow to reach the equilibrium within the life span of the dense matter
and hence this remark is presently only of academic interest.
 
\subsection{Thermal strangeness production}
\label{THavsec}
\baselineskip=14pt
We now determine  the thermal strangeness production rate:
\begin{equation}
A_{\rm s}\equiv A_{gg}+A_{u\bar u}+A_{d\bar d}=
\sum_{AB}\langle\sigma v_{\rm AB}\rangle_T
\rho_{\rm A}^\infty\rho_{\rm B}^\infty
={dN(gg,q\bar q\to s\bar s)\over d^3{x}\,dt}\,. \label{Ass}
\end{equation}
Thus the general expression for $A_{\rm s}$ is: 
\begin{eqnarray}
A_{\rm s}&=&\int_{4m_{\rm s}^2}^{\infty}ds
2s\delta (s-(p_{\rm A}+p_{\rm B})^2)
\int{d^3p_{\rm A}\over(2\pi)^32E_{\rm A}}\int{d^3p_{\rm
B}\over(2\pi)^32E_{\rm B}} \nonumber\\
\label{qgpA}
&&\times\left[{1\over 2} g_g^2f_g(p_{\rm A})f_g(p_{\rm B})
\overline{\sigma_{gg}}(s) + n_{\rm f}g_q^2 f_q(p_{\rm A}) 
f_{\bar q}(p_{\rm B})\overline{\sigma_{q\bar q}}(s)\right]\,.
\end{eqnarray}
where in principle the particle distributions $f_i$ could be 
different from the thermal Bose/Fermi functions we will use here.
The bar over the cross sections indicates that we use angle-averaged
expressions. In order to obtain the above form, we have introduced
a dummy integration over $s$ and have employed for the relative velocity
between two particles 
\begin{eqnarray}
v_{\rm AB}2E_{\rm A}2E_{\rm B}\equiv &2&\lambda^{1/2}(s)\nonumber\\
=&2&\sqrt{s - (m_{\rm A}+ m_{\rm B})^{2}}\sqrt{s - (m_{\rm A} -
m_{\rm B})^{2}}\to 2s\,,
\label{relvel}
\end{eqnarray}
where the last limit holds for (nearly) massless particles.

We are interested to understand at which values of
$\sqrt{s}$ the actual production processes occur, in order to
establish the value of $\alpha_{\rm s}$ we should employ. 
We rewrite the thermal
production rate Eq.\,(\ref{qgpA}) as an integral over the differential rate
$dA/ds$:
\begin{equation}
\label{dAds}
A_i \equiv \int_{4m_{\rm s}^2}^\infty\,ds\,{dA_i\over ds}
\equiv \int_{4m^2}^\infty\,ds\,\overline{\sigma_i}(s)\, P_i(s) 
\qquad i= g,q\,.
\end{equation}
Here $P_g(s)ds$ is the number of gluon collisions within the
interval of invariant mass ($s,s+\,ds$) per unit time per unit
volume, with a similar interpretation applying to $P_q(s)$. From
Eq.\,(\ref{qgpA}) we find:
\begin{equation}
\label{gma}
P_g(s) = {1\over 2}g_{\rm g}^2\int\!\!
{d^3p_{\rm A}f_g(p_{\rm A})\over (2\pi)^32E_{\rm A}}
{d^3p_{\rm B}f_g(p_{\rm B})\over (2\pi)^32E_{\rm B}}
2s\delta(s-(p_{\rm A}\!+p_{\rm B})^2)\, ,
\end{equation}
\begin{equation}
\label{qma}
P_q(s) = n_{\rm f}g_q^2\int\!\!
{d^3p_{\rm A} f_q(p_{\rm A})\over (2\pi)^32E_{\rm A}}
{d^3p_{\rm B}f_{\bar q}(p_{\rm B})\over (2\pi)^32E_{\rm B}}
2s\delta(s-(p_{\rm A}\!+p_{\rm B})^2)\, ;
\end{equation}
where $P_q$ includes both $u,d$ collisions in the factor $n_{\rm f}$ in 
an incoherent way, and hence $g_q=2\cdot 3$. 
For gluons we have $g_{\rm g}=2\cdot8$\,. Assuming that the particle
distributions depend only on the magnitude of the momentum, and using
\begin{equation}\label{grprob}
\delta(s-(p_A+p_B)^2) = {1\over 2p_Ap_B}\delta\left(\cos\theta - 
     1+{s \over 2p_Ap_B}\right)\,,
\end{equation}
we can carry out the two angular integrals to obtain:
\begin{eqnarray}\label{gsprob}
P_g \hspace*{-0.6cm}&& ={4\over\pi^4}s\, 
\int_0^{\infty}dp_A\,\int_0^{\infty}dp_B\, 
     \Theta (4p_Ap_B-s)f_g(p_A)f_g(p_B)\,, \\
\label{gtprob}
P_q \hspace*{-0.6cm}&& ={9\over 4\pi^4}s\,
\int_0^{\infty}dp_A\,\int_0^{\infty}dp_B\, 
     \Theta (4p_Ap_B-s)f_q(p_B)f_{\bar q}(p_B)\,.
\end{eqnarray}
The step function $\Theta$ arises because of the limits on the value of
$\cos\theta$ in Eq.\,(\ref{grprob}). To proceed, we assume thermal Bose and
Fermi distribution for the particle distributions in the fireball rest
frame. Possible $\vec x$-dependence is implicitly contained in $T$ and
$\mu_q$: 
\begin{eqnarray}
f_g(p) &=& {1\over e^{p/T} -1}\,,
\label{gbose}\\
f_q(p) &=& {1\over e^{(p-\mu_q)/T} + 1}\,,
\label{qfermi}\\
f_{\bar q}(p) &=& {1\over e^{(p+\mu_q)/T} + 1}\,.
\label{bqfermi}
\end{eqnarray}
The integrals in Eqs.\,(\ref{gsprob}, \ref{gtprob}) can be carried out
analytically, although only for $\mu_q = 0$ in the latter case. In this
limit we have:
\begin{eqnarray}
\int_0^{\infty}\!\!dp_Adp_B  
{\theta(4p_Ap_B-s) \over (e^{p_A/T}\mp 1)(e^{p_B/T}\mp 1)}
&&\nonumber\\
&&\hspace*{-3cm}= \sum_{n=1}^{\infty}(\pm)^n
    \int_0^{\infty}{dp_A \over (e^{p_b/T}\mp 1)}\;
    \int_{s/4p_B}^{\infty}\!\!dp_A\,e^{-np_A/T}\nonumber\\
&&\hspace*{-3cm}= \sum_{n,l=1}^{\infty} (\pm)^{n+l}\; {T\over n}
    \int_0^{\infty}dp_B\,
e^{-l\scs\frac{\scs p_B}{T}}e^{-n\frac{\scs s}{\scs 4p_BT}}\,.
\label{expand} \end{eqnarray}
This integral type is well known \cite{Gra80a}:
\begin{equation}\label{expprodint}
\int_0^{\infty}dx\,e^{-\beta /4x}e^{-\gamma x} = \sqrt{\beta /\gamma}
\; {\rm K}_1(\sqrt{\beta\gamma})\,.
\end{equation}
\baselineskip=15pt
We obtain  for the gluon case:
\begin{equation}
\label{Pg}
P_{g} = {4Ts^{3/2}\over\pi^4}\sum_{l,n = 1}^{\infty}
{1\over\sqrt{nl}}\;{\rm K}_1\left({\sqrt{nl\,s}\over T}\right)\,.
\end{equation}
Similar expression follows for quark processes when the chemical
potentials vanish: 
\begin{equation}\label{Pq0}
P_q|_{\mu_q=0} = {9Ts^{3/2}\over 4\pi^4}\sum_{l,n = 1}^{\infty} 
{(-)^{n+l} \over\sqrt{ln}}\; {\rm K}_1({\sqrt{nl\,s}\over T})\,.
\end{equation}
 
\begin{figure}[p]
\vspace*{1.2cm}
\centerline{\hspace*{-.7cm}
\psfig{width=9cm,figure=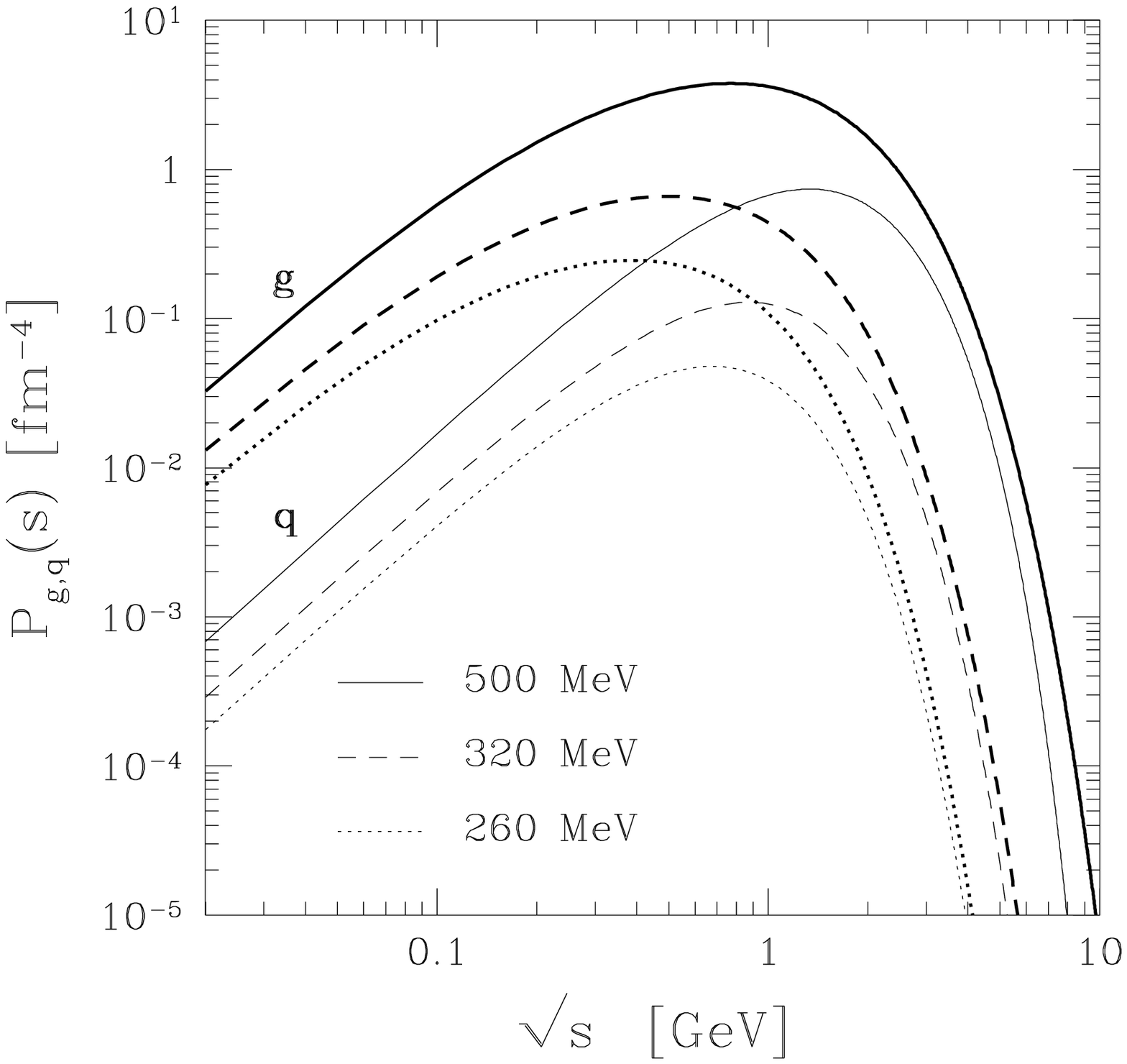}
}
\vspace*{-0.2cm}
\caption{ \small
The collision distribution functions for gluons (thick lines) 
and quarks (thin lines) as function $\protect\sqrt{s}$. Computed for 
temperature $T=260$ MeV (dotted lines) and $T=320$ MeV (dashed lines).
For quarks $\lambda_q=1.5$  was used in these two cases. The solid lines
show the RHIC domain,  $T=500$ MeV and $\lambda_q=1$ \,.\label{Psqg} }
\vspace*{2.2cm}
\centerline{\hspace*{-.3cm}
\psfig{width=9cm,figure=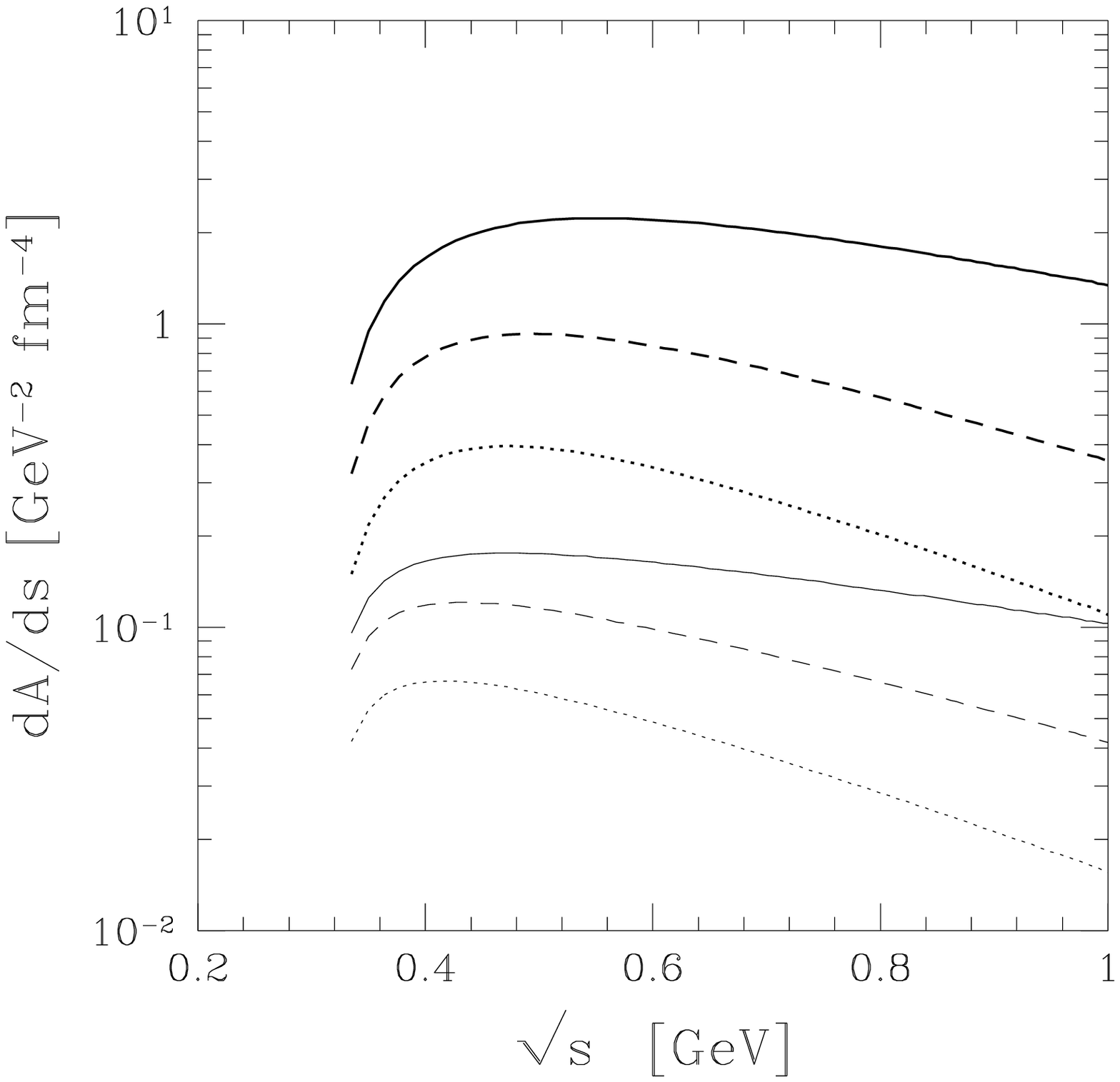}
}
\vspace*{-0.2cm}
\caption{ \small
Differential thermal strangeness
production rate $dA{_s}/ds = P(s)\sigma(s)$, with $T$ = 260 (dotted) and 320
MeV (dashed) and $m_{\rm s}$ = 160 MeV, for gluons (thick) and $q\bar q$
pairs (thin), with $\lambda_q=1.5$\,,  $\alpha_{\rm s} = 0.6$; and for 
$T=500$ MeV with $\lambda_q=1$ and $\alpha_{\rm s} = 0.4$ (solid line). 
\label{figdAds}}
\end{figure}

The case with $\mu_q> 0$ is of greater physical interest in the present
context of baryon-rich fireballs. In this case only the antiquark
distribution Eq.\,(\ref{bqfermi}) can be expanded in terms of a geometric
series for all values of the quark momentum. Keeping the quark Fermi
distribution, we obtain an expression containing one (numerical)
integration:
\begin{equation}
\label{Pq}
P_q={9T\over 4\pi^4}s\sum_{l=1}^{\infty}
{(-)^{l+1}\over l\lambda_{\rm q}^l}
\int_0^{\infty}dp_{\rm A}\,{e^{-l{\scs s\over\scs 4Tp_{\rm A}}} \over 
\lambda_{\rm q}^{-1}e^{p_{\rm A}/T}+1} \,,
\end{equation}
where $\lambda_{q} = e^{\mu_{q} /T}$ is the quark number
fugacity. The remaining integral over $dp_{\rm A}$ has to be solved
numerically. In Fig.\,\ref{Psqg} we show the 
collision distribution functions
Eqs.\,(\ref{Pg}, \ref{Pq}) describing the probability that 
a pair of gluons (thick lines) or a light quark $q$--$\bar
q$-pair (thin lines) collides 
at a given $\sqrt{s}$, for $T= 260$ (dotted) and 320 MeV (dashed),
which we expect to be appropriate limits on initial  fireball
temperatures for the S--W and Pb--Pb collisions. For
quarks we have taken $\lambda_q=1.5$, which properly accounts for the
baryon abundance in the fireball, see table~\ref{bigtable}. We also 
show  $T=500$ MeV (solid lines), with $\lambda_q=1$\,, 
which choice, as we hope, is exploring the future conditions at RHIC/LHC. 

The thermal differential production rates
$dA_i/ds$\,, Eq.\,(\ref{dAds}) for strangeness
are shown in Fig.\,\ref{figdAds}. 
We note that for the gluon fusion to strangeness 
processes the peak of the production occurs at 
relatively low energies $\sqrt{s}\simeq 0.5$ GeV, 
and it is slightly more peaked and higher in energy than seen for quark 
pair processes.  The dominance 
of the gluon channel in favor production 
arises primarily from the greater statistical 
probability to collide  two gluons in plasma at a given $\sqrt{s}$,  as
compared to the probability of $q+\bar q$ collisions, see 
Fig.\,\ref{Psqg} as well as from contributions at $\sqrt{s}$ away from 
production threshold. 

\baselineskip=12.9pt
\begin{figure}[tb]
\vspace*{1.6cm}
\centerline{\hspace*{-.3cm}
\psfig{width=10cm,figure=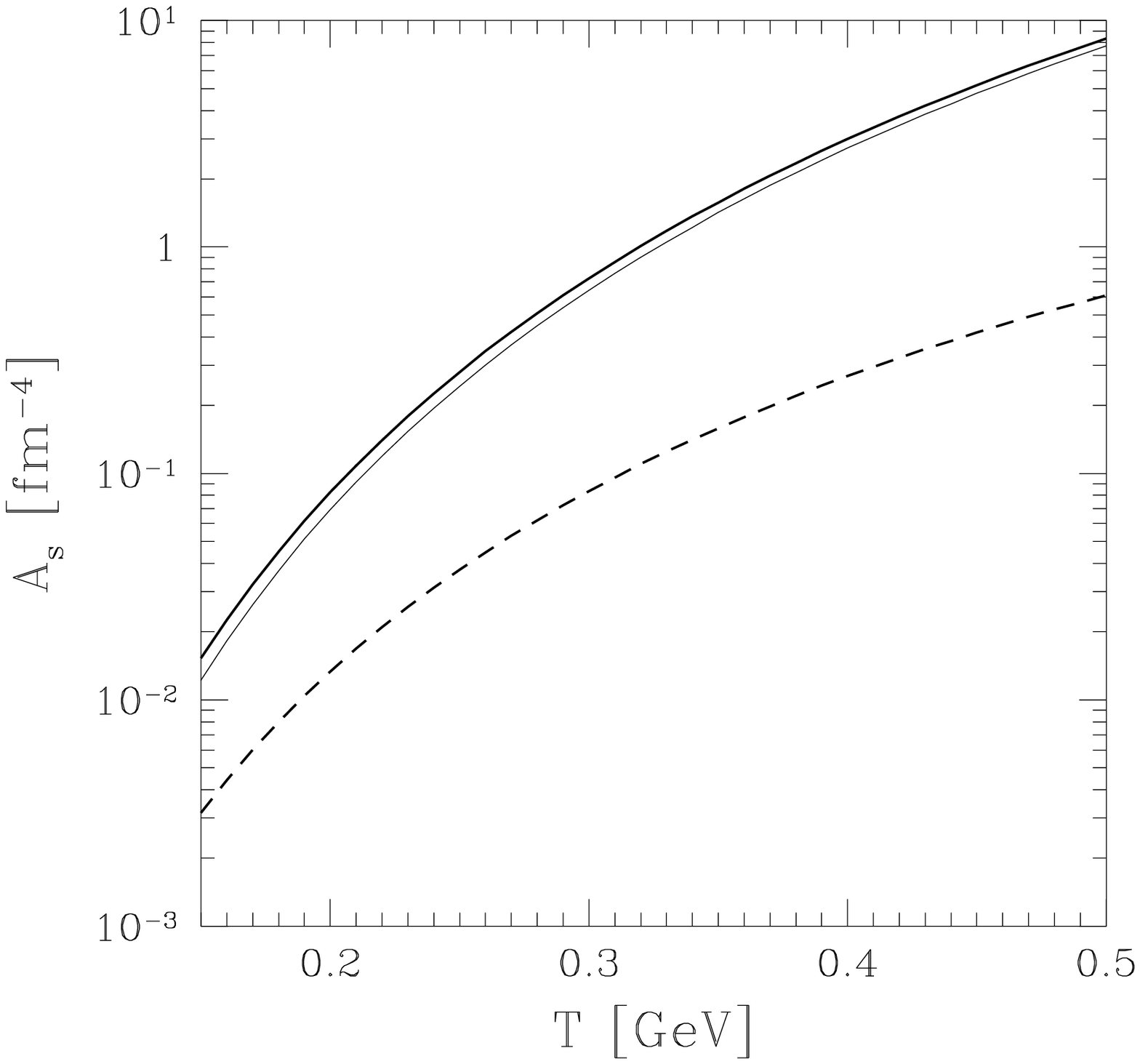}
}
\vspace*{-0.2cm}
\caption{ \small
Thermal strangeness
production rates $A_{\rm s}$ in QGP: total (thick solid line), gluons
only (thin solid line), and light quarks only (dashed line), calculated
for $\lambda_{\rm q}=1.5$\,, $m_{\rm s}=160$ MeV, $\alpha_s= 0.6$
as function of temperature. \label{figAss}} 
\end{figure}
The differential production rate can  be easily integrated, and we show
the result in Fig.\,\ref{figAss}. 
These results depend, of course, on the choice of the value of the 
strange quark mass, assumed here to be
$160$ MeV. 
The production rates in Figs.\,\ref{figAss} and \ref{figAcc}, 
when inserted into Eq.\,(\ref{tauss}), provide the relaxation time 
constants $\tau_{\rm s}$,
$\tau_{\rm c}$. In Fig.\,\ref{figtauss}
we show strangeness relaxation
constant using the same conventions and parameters as in 
Figs.\,\ref{figAss}.  The dominance of gluon fusion over 
quark fusion for strangeness production process can be now more
easily appreciated, and we note that as function of temperature
in the interesting interval the relaxation time drops by an
order of magnitude. This in particular explains the phenomenon, 
that when the
QGP fireball cools, the abundance of strangeness freezes out,
i.e., strangeness once produced is not reannihilated significantly. 

\begin{figure}[ptb]
\vspace*{1.3cm}
\centerline{\hspace*{-.3cm}
\psfig{width=9cm,figure=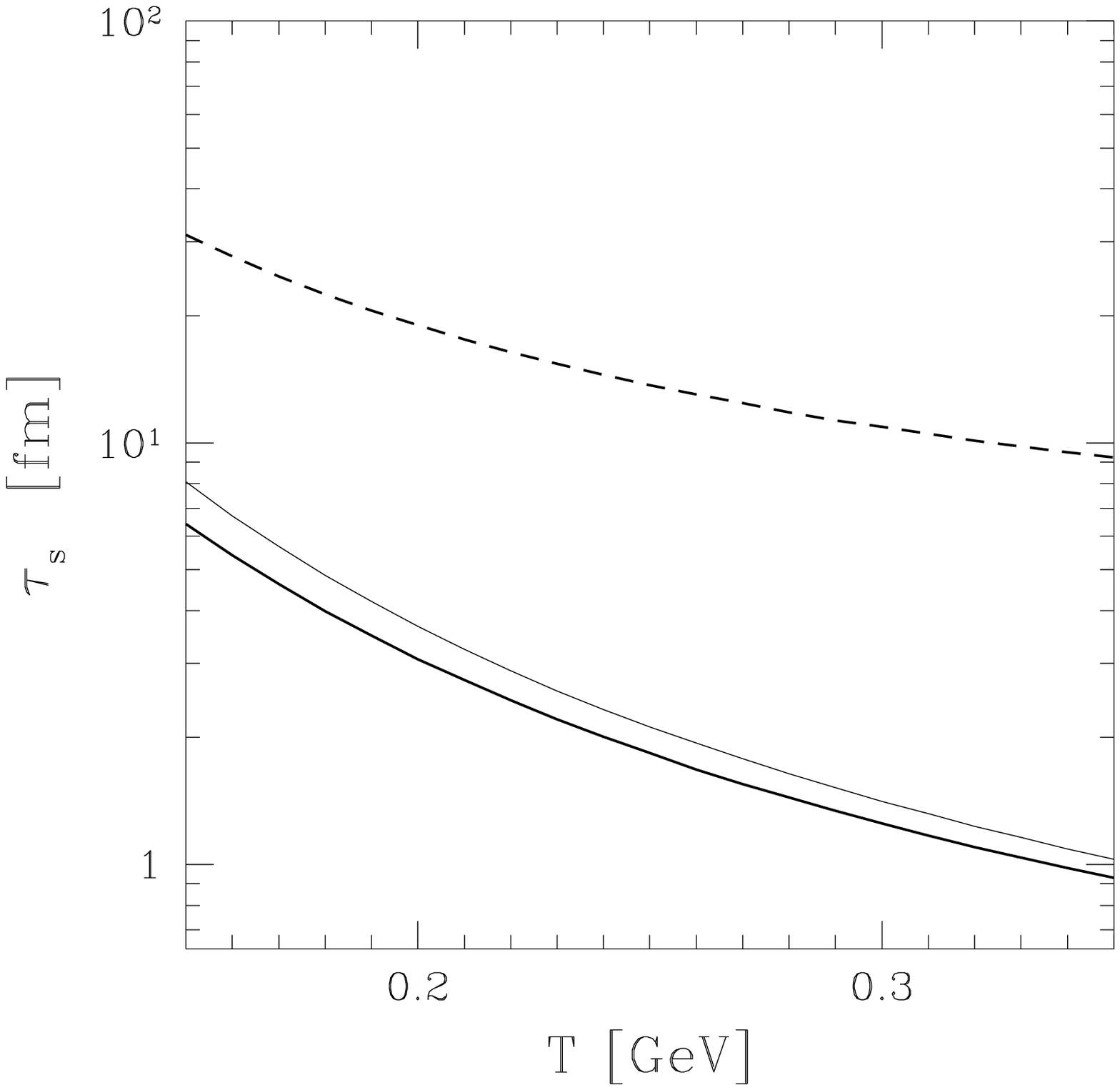}}
\vspace*{-0.5cm}
\caption{ \small
Thermal strangeness relaxation constants in QGP: same
conventions and parameters as in Fig.\,\protect\ref{figAss}.\label{figtauss}}
\vspace*{1cm}
\centerline{
\psfig{width=8cm,figure=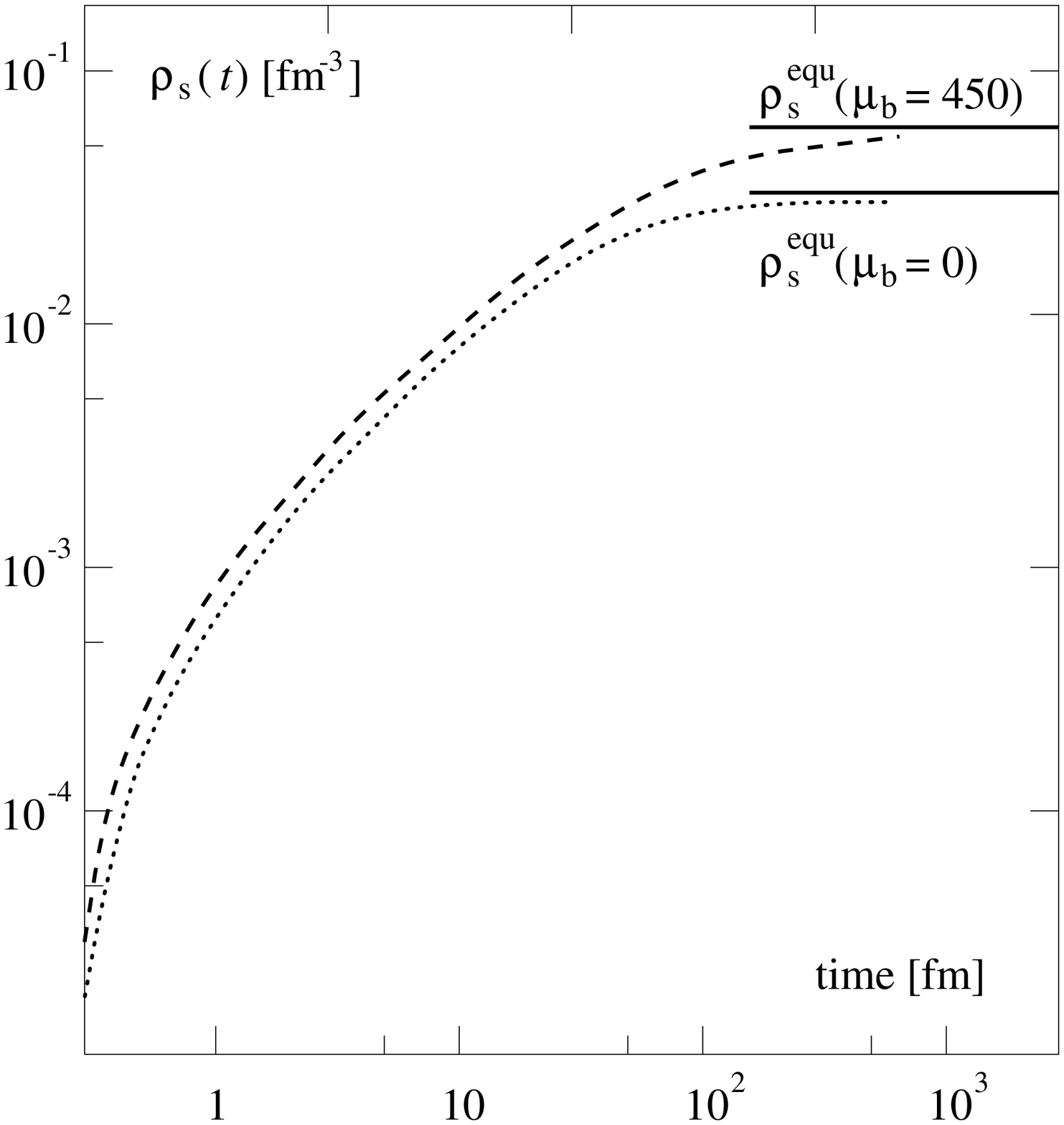}
}
\vspace*{-0.3cm}
\caption{ \small
Thermal strangeness production as function of time in a confined
hadron gas at $T=160$ MeV. Results for two values of baryochemical 
potential ($\mu_{\rm B}=0$ and 450 MeV are shown.
 After Koch et al. \protect\cite{KMR86}.\label{hadgasre}}
\end{figure}
We wish to record here that the strangeness phase space
saturation seen in SPS-relativistic heavy ion collision
experiments cannot be a simple
result of totally conventional physics. In the dense state of
highly excited confined HG fireball, there are many different strangeness
production channels and a full discussion is beyond the scope of this
presentation. The key results were described in detail elsewhere
\cite{KMR86,KR85}: in a gas consisting of particle states
with normal properties, strangeness saturation time scales
are very much longer, as is shown in Fig.\,\ref{hadgasre}, where the
approach to equilibrium abundance as function of time takes nearly 100
fm/$c$. Thus if this was the actual situation, then 
the strange particle abundance would be  largely
result of pre-thermal collisions, and thus could be easily
described by folding of a geometric microscopic collision model 
with the experimental N--N results. There is considerable ongoing effort to
simulate, using microscopic models, this initial phase of nuclear
collisions, and while these efforts can produce appropriate yields of some
particles, the overall reaction picture \cite{Csernai}, in particular
considering the multistrange baryons and antibaryons is so far not
satisfactory, supporting at least the claim that strangeness enhancement
requires some new physics phenomenon, if not QGP as we are
arguing here. Models that include microscopic deconfinement,
such as the dual parton model
\cite{Cap95}, but which do not assume thermalization, 
require the introduction of mechanisms
to fit the multistrange particle yields at
central rapidities. 

\vspace{-4mm}
\subsection{Running \mathversion{bold}$\alpha_{\rm s}$ 
\mathversion{normal}and flavor production
} \label{runalfasec}
\vspace{-2mm}

One of the key, and still arbitrary parameters we used above is the 
choice we have made $\alpha_{\rm s}=0.6$\,. While the value seems 
reasonable, others may prefer a smaller value, and just choosing 
$\alpha_{\rm s}=0.3$ would lengthen the relaxation time of strangeness
$\tau_{\rm s}\propto \alpha_{\rm s}^{-2}$ by factor 4, from 2 fm to 8 fm 
and  thus beyond the expected lifespan of the QGP fireball. 
We exploit here the  recent precise  determination 
of $\alpha_{\rm s}(M_Z)$ \cite{Lan95,Vol95,Ellis96} which
allow to eliminate assumptions about $\alpha_{\rm s}$ from our 
calculations.  

Running QCD methods can be used \cite{impact} to 
obtain the proper value of $\alpha_{\rm s}$ allowing to 
reevaluate strangeness production in a 
thermal QGP fireball and to justify the choice of the 
coupling constant we have made.  
Moreover, since the running QCD resummation is performed, 
we are able to account for a large class of contributing diagrams. 
However, the knowledge of the strange quark mass remains  limited 
and  will need further refinement, even allowing for the 
running-QCD effects considered here. Another remaining shortcoming 
is that  up to day there has not 
been a study of the importance of the final state (radiative gluon)
or initial state three body effects in the entrance channel. 
Such odd-$\alpha_{\rm s}$ 
infrared unstable processes have environment induced infrared cut-off 
(Landau-Pomeranchuck effect) which in dense matter
eliminates modes that are softer than the collision frequency.

The running coupling constant $\alpha_{\rm s}$ and quark mass
satisfy the QCD renormalization group 
equations\footnote{Caution should be exercised not to confuse
the chemical potentials with the variable $\mu$ used here
without and index, which 
denotes the energy scale of running QCD variables.}:

\begin{eqnarray}\label{dmuda}
\mu \frac{\partial\alpha_{\rm s}}{\partial\mu}
&=&\beta(\alpha_{\rm s}(\mu))\,,\\
\label{dmdmu}
\mu {\frac{\partial m}{\partial\mu}} &=&-m\,
\gamma_{\rm m}(\alpha_{\rm s}(\mu))\,.
\end{eqnarray}
These functions $\beta$ and $\gamma_{\rm m}$ are today
known for the SU(3)-gauge theory with $n_{\rm f}$ fermions, but only
in a perturbative power expansion in $\alpha_{\rm s}$; three
leading terms are known for $\beta$ and two for the $\gamma_{\rm
m}$\cite{QCD95}:
\begin{eqnarray}\label{betaf}
\beta^{\rm pert}&=&-\alpha_{\rm s}^2\left[\ b_0
   +b_1\alpha_{\rm s}       +b_2\alpha_{\rm s}^2 +\ldots\ \right] \,,\\
\label{gamrun}
\gamma_{\rm m}^{\rm pert}&=&\phantom{-}\alpha_{\rm s}\left[\ c_0
+c_1\alpha_{\rm s} + \ldots\ \right]\,,
\end{eqnarray}
with
\begin{eqnarray}
b_0&&\hspace*{-0.6cm}=  {1\over 2\pi}\!\left(\!11-{2\over 3}n_{\rm f}\!\right)\,,
\quad
b_1= {1 \over 4\pi^2}\!\left(\!51-{19\over 3}
        n_{\rm f}\!\right)\,,\\
c_0&&\hspace*{-0.6cm}={2\over \pi}\,,
\quad\quad\quad\quad\quad\quad
c_1={1\over 12\pi^2}
        \left(\!101-{10\over 3}n_{\rm f}\!\right)\,.
\end{eqnarray}

The $b_2$-coefficient in Eq.\,(\ref{betaf}) is renormalization
scheme dependent, and is known, e.g., in the modified minimum subtraction
dimensional renormalization ($\overline{MS}$) scheme:
\begin{equation}
b_2^{\rm MS}=
 {1 \over 64\pi^3} \left(2857-{5033\over 9}n_{\rm f}+{325\over 27}
        n_{\rm f}^2 \right) \label{beta2}\,.
\end{equation}
Since there is no renormalization scheme dependence in the full
non perturbative QCD process, cancelation between the
different renormalization group terms  must occur.  
The number  $n_{\rm f}$ of fermions that can be excited, depends
on the energy scale $\mu$. The form appropriate for the terms linear
in $n_{\rm f}$ is:
\begin{equation}\label{nfs}
n_{\rm f}(\mu)=2+\sum_{i=s,c,b,t}\sqrt{1-\frac{4m_i^2}{\mu^2}}
  \left(1+\frac{2m_i^2}{\mu}\right)\Theta(\mu-2m_i)\,,
\end{equation}
with $m_{\rm s}=0.16\,{\rm GeV},\,m_{\rm c}=1.5\,{\rm GeV},\,
m_b=4.8\,{\rm GeV}$\,. 
There is very minimal impact of the running of the masses
in Eq.\,(\ref{nfs}) on the final result. The bottom  mass
uncertainty has the greatest 
impact, since small perturbation of the $\beta$-function at 
$\mu\simeq 10$ GeV is enhanced strongly when the error 
propagates to $\mu=1$ GeV or $\mu=100$ GeV, which are the 
values being connected to each other.

Even if the perturbative expansion is leading to an adequate theoretical
description at small values of $\alpha_{\rm s}$, the extrapolated value of 
$\alpha_{\rm s}$, as the scale $\mu$ decreases, can approach and
exceed unity, where use of perturbative expansion is not easily
justified, and thus effort has to be made to incorporate all known terms
in the perturbative expansion of the $\beta$-function.
Recent work suggests that Pade approximants in QCD expression 
could improve the precision and enlarge the 
circle of convergence of the perturbative expansion 
\cite{Ellis95}. A suitable Pade-approximant 
for the $\beta$-function is:
\begin{eqnarray}\label{betpade}
\hspace*{-0.cm}\beta\to\beta^{(0,2)}\equiv \alpha_{\rm s}^2 b_0 
        \frac{1}{1-u\alpha_{\rm s}+v\alpha_{\rm s}^2}=\alpha_{\rm s}^2 b_0
  \frac{b_0}{b_0-b_1\alpha_{\rm s}+ (b_1^2/b_0-b_2)\alpha_{\rm s}^2}\,.
\end{eqnarray}
The integration of the renormalization equations is facilitated by this 
modification since it turns out that the numerical solutions are stabilized
in that way.

\begin{figure}[t]
\vspace*{-1.3cm}
\centerline{\hspace*{0.6cm}
\psfig{width=11.5cm,figure=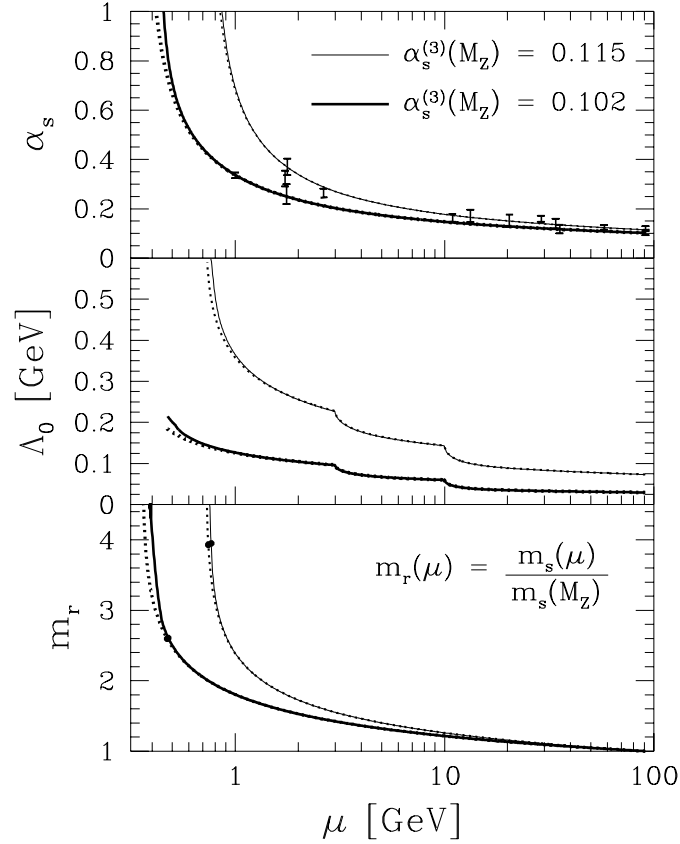}
}
\vspace*{0.cm}
\caption{ \small
$\alpha_{\rm s}(\mu)$, the $\Lambda$-parameter 
$\Lambda_0$ and $m_{\rm r}(\mu)=m(\mu)/m(M_Z)$
 as function of energy scale $\mu$. Thick lines
correspond to initial value $\alpha_{\rm s}(M_Z)\!=0.102$, thin lines
are for the initial value $\alpha_{\rm s}(M_Z)=0.115$. Dotted lines
are results obtained using the perturbative expansion for the
renormalization group functions, full lines are obtained using (0,2) Pad\'e
approximant of the $\beta$ function. Experimental results for
$\alpha_{\rm s}$ selected from 
references\protect{\protect\cite{Lan95,Vol95,Ellis96,QCD95,Zeus95}}. 
In bottom portion the dots
indicate the pair production thresholds for $m_{\rm s}(M_Z)=$~90~MeV.
\label{fig-a1}}
\end{figure}

Eq.\,(\ref{dmuda}) is numerically integrated  beginning with initial
value of $\alpha_{\rm s}(M_Z)$, using the perturbative series
(\ref{betaf}) for the $\beta$-function (dotted lines in
Fig.\,\ref{fig-a1}) or its (0,2) approximant Eq.\,(\ref{betpade})
(full lines). Thick lines correspond to  $\alpha_{\rm
s}(M_Z)=0.102$\,. This value of $\alpha_{\rm s}$ is
consistent with the precise botonium sum rule result\cite{Vol95} 
$\alpha_{\rm s}(1{\rm GeV})=0.336\pm0.011$, this extremely
precise point is shown in Fig.\,\ref{fig-a1}. We recall that
the well known  $Z$ line shape LEP data fit leads to
$\alpha_{\rm s}(M_Z)=0.123\pm 0.006$\,. However, in this fit there
is disagreement between the observed and predicted properties
of the $Z\to b\bar b$ vertex. If one proceeds with line
shape fit excluding this branching ratio{\cite{Lan95}}, 
one obtains $\alpha_{\rm s}(M_Z)=0.101\pm 0.008$. The thin solid lines
in Fig.\,\ref{fig-a1} are for the initial value $\alpha_{\rm s}(M_{{\rm
Z}})=0.115$\,, in agreement with some other experimental
results \cite{QCD95}  also shown in Fig.\,\ref{fig-a1}, including
some recent HERA data\cite{Zeus95}, as well as the recent
 measurement of the structure of 
hadronic events by the L3 detector at LEP-II\cite{L3}. 
The sum-rule study of Ellis et al.{\cite{Ellis96}} 
leads to the point at $\mu=1.7$ GeV appearing in the middle between 
thin and thick curves in the top section of Fig.\,\ref{fig-a1}.

The middle section of Fig.\,\ref{fig-a1} shows what the running of
$\alpha_{\rm s}$ implies for the value $\Lambda_0$ which is sometimes
used to characterize the variation of $\alpha_{\rm s}$ based on a first 
order result. Here $\Lambda_0(\mu)$ is defined by 
the implicit equation:
\begin{equation}\label{Lambdarun}
\alpha_{\rm s}(\mu)\equiv\frac{2b_0^{-1}(n_{\rm f})}{\ln(\mu/\Lambda_0(\mu))^2}\,.
\end{equation}
We see that $\Lambda_0(1\mbox{\,GeV})=240\pm100$ MeV, assuming that the 
solid lines provide a valid upper and lower limits on $\alpha_{\rm s}$.
However, the variation of $\Lambda_0(\mu)$ is significant for $\mu<3$~GeV, 
questioning the use of first order expressions, seen frequently in literature. 
 
With $\alpha_{\rm s}(\mu)$ from the solutions
described above Eqs.\,(\ref{dmdmu}, \ref{gamrun}) allow
to explore the quark masses. Because  Eq.\,(\ref{dmdmu}) 
is linear in $m$, it is possible to determine the universal 
multiplicative quark mass scale factor
\begin{equation}
m_{\rm r}=m(\mu)/m(\mu_0)\,.
\end{equation}
Since  $\alpha_{\rm s}$ refers to  the scale of $\mu_0=
M_Z$, it is a convenient reference point also for quark masses.
As seen in the bottom portion of Fig.\,\ref{fig-a1},
the change in the quark mass factor is highly relevant,
since it is driven by the rapidly changing
$\alpha_{\rm s}$ near to $\mu\simeq 1$~GeV.
For each of the two different functional dependences
$\alpha_{\rm s}(\mu)$ we obtain a different function
$m_{\rm r}$. Note that the difference between (0,2) approximant
result (solid lines) and perturbative expansion (dotted lines) 
in Fig.\,\ref{fig-a1}  is indeed,
at fixed $\mu\simeq 1$ GeV,  very large, however it
remains insignificant since it amounts to a slight `horizontal' shift of 
$\alpha_{\rm s}$ and $m_{\rm r}$ as function of~$\mu$.
 
Like for $\alpha_{\rm s}$, the uncertainty range in $m_{\rm r}$ due to 
the error in the initial value $\alpha_{\rm s}(M_Z)$ is
considerable. Some of this sensitivity will disappear when we consider
the cross sections, and in particular their thermal
average weighted with particle distribution.
Furthermore, the strangeness production cross section
is subject to an implicit infrared stabilization: below $\sqrt{\rm s}=1$
GeV the strange quark mass increases rapidly and the threshold mass
$m^{\rm th}_{\rm s}$ for the pair production,  defined by the solution of
the equation
\begin{equation}\label{dispersion}
m_{\rm s}^{\rm th}/m_{\rm s}(M_Z)=m_{\rm r}(2m_{\rm s}^{\rm
th})\,,
\end{equation}
is considerably greater than  $m_{\rm s}$(1 GeV). For example,
for $m_{\rm s}(M_Z)=90$ MeV:
 $m_s(1\mbox{\,GeV})$ $\simeq 160$ and 215 MeV 
for the two choices of $\alpha_{\rm s}(M_Z)$, 
within the standard range $100<m_s$(1\,GeV) $<$ 300 MeV;  
the corresponding threshold values are 470 MeV
for the smaller $\alpha_{\rm s}$ option (thick lines)
and 740 MeV for the higher option (thin line). Both values are 
indicated by the black dots in Fig.\,\ref{fig-a1}. 

We note in passing that the same effect occurs for charm quark mass:
it is running equally rapidly and we find that an appropriate value
at $\mu=M_Z$ would be 700 MeV, 
as this choice assures given $\alpha_{\rm s}(\mu)$ that 
$m_{\rm c}(1\ {\rm GeV})\simeq $ 1.5 GeV. The drop in $m_{\rm c}$ at the
production threshold, $\mu\simeq 2.5$ GeV has important 
ramifications for the rate of thermal charm production.

\begin{figure}[tb]
\vspace*{-0.7cm}
\centerline{\hspace*{-0.8cm}
\psfig{width=10cm,figure=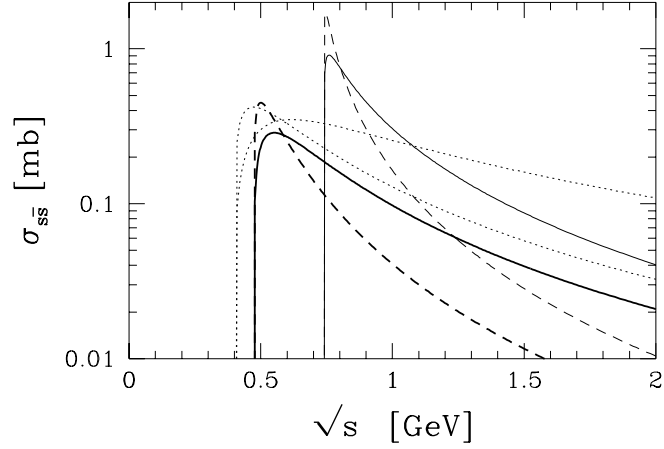}
}
\vspace*{-0.3cm}
\caption{ \small
QCD strangeness production cross sections  obtained for running
$\alpha_{\rm s}(\protect\sqrt{\rm s})$ and $m_{\rm s}(\protect\sqrt{\rm
s})$.  Thick solid line is for the small $\alpha_{\rm s}$
option, thin line for the other $\alpha_{\rm s}$-option considered in
Fig.\,\protect\ref{fig-a1}. Solid lines $gg\to  s\bar s$; 
dashed lines  $q\bar q\to s\bar s$. Dotted lines: results for fixed  
$\alpha_{\rm s}=0.6$ and $m_{\rm s}=200$ MeV. 
\label{figsigsrun}
}
\end{figure}

We can now insert into the generic production cross sections,
Eqs.\,(\ref{gl}, \ref{gk}) the running QCD parameters, identifying 
$\mu\to \sqrt{s}$. The resulting cross sections are then a sum of 
all two particle fusion contributions to strangeness production with two
particles in the final state. In Fig.\,\ref{figsigsrun}, 
the strangeness production cross sections are
shown with $m_{\rm s}(M_Z)=90$~MeV. For the two choices of the
running coupling constant considered in Fig.\,\ref{fig-a1} the
cross sections for the processes $gg\to s\bar s$ (solid
lines, upper dotted line) and $q\bar q\to s\bar s$ (dashed lines, lower
dotted line) are shown. Dotted are cross sections 
computed with fixed $\alpha_{\rm
s}=0.6$ and $m_{\rm s}=200$ MeV cross sections, shown here for comparison. We
note that the glue based flavor production dominates at high $\sqrt{s}$,
while near threshold the cross sections due to light quark heavy flavor
production dominate. We note the different thresholds for the two values
of $\alpha_{\rm s}(\mu)$ used. It is apparent that the cross
sections are `squeezed' away from small $\sqrt{s}$ as the
value of $\alpha_{\rm s}$ increases, such that the energy 
integrated cross sections ($\simeq$ rates) are little changed.
 
We note that these results justify to considerable 
extent the use of the perturbative approximation with fixed 
$\alpha_{\rm s}=0.6$ in the study of QGP based strangeness production
processes within the range $150<T<300$~MeV; not only is this value
$\alpha_{\rm s}=0.6$ in the middle of the range spanned in the
Fig.\,\ref{fig-a1}, but moreover, we see that the inelastic 
(production) cross sections shown in Fig.\,\ref{figsigsrun} 
have similar integrated strength. However the choice of the
strange quark mass impacts considerably the result we find, and
thus there is systematic uncertainty related to the relatively 
large range of permissible $m_{\rm s}(\mu)$. 
\begin{figure}[tb]
\vspace*{1.5cm}
\centerline{\hspace*{-0.5cm}
\psfig{width=9cm,figure=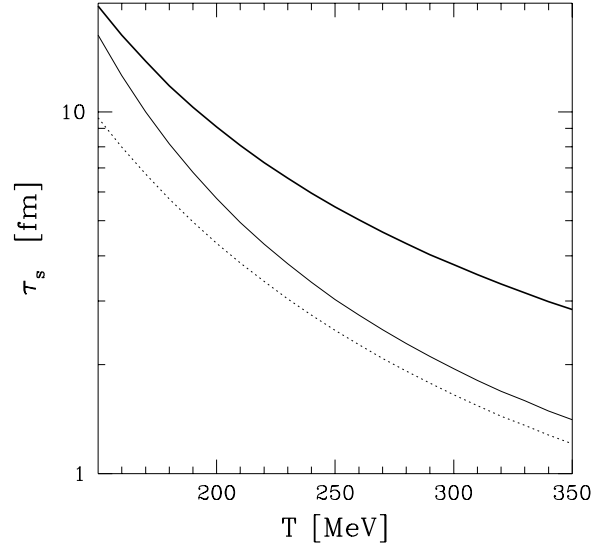}
}
\vspace*{-0.3cm}
\caption{ \small
QGP strangeness relaxation time obtained from 
the running $\alpha_{\rm s}$-cross sections shown 
in Fig.\,\protect\ref{figsigsrun}.
Thick solid line is for the small $\alpha_{\rm s}$
option, thin line for the other $\alpha_{\rm s}$-option considered in
Fig.\,\protect\ref{fig-a1}. Dotted: results for fixed  
$\alpha_{\rm s}=0.6$ and $m_{\rm s}=200$ MeV. 
\label{figTaussrun}
}\end{figure}

From this point on, given the improved cross sections, the calculation 
of thermal relaxation time constant of strangeness 
follows the pattern we described in section
\ref{strprod},   Eq.\,(\ref{qgpA}). The result, the  relaxation time 
constants $\tau_{\rm s}$ is shown in Fig.\,\ref{figTaussrun} for 
the two different choices of the strong coupling constant 
considered here. Dotted line shows,
for comparison, the result obtained using the fixed 
values $\alpha_{\rm s}=0.6$ and $m_{\rm s}=200$ MeV.
$\tau_{\rm s}$ is defined as before in Eq.\,(\ref{tauss}): 
 \begin{equation}\label{taussbis}
\tau_{\rm s}\equiv
{1\over 2}{\rho_{\rm s}^\infty\over{(A_{gg}+A_{qq}+\ldots)}}\,,
\end{equation}
but the equilibrium density which the produced particles `chase' 
require now a second thought, as it depends on the (strange) quark mass,
and the question is: at which scale $\mu$ is $m$ to be  considered. 
Recall that the equilibrium density is obtained 
in transport formulation of the evolving particle distributions in 
consequence of  
particle-particle collisions. It is thus the characteristic
energy of these interactions which determines the energy scale which enters
the determination of the mass in the density $\rho_{\rm s}^\infty$. Since
the strangeness production phenomena occur at maximum attainable 
temperature $T\ge260$ MeV, the range of values relevant in this paper 
is $m_{\rm s}\simeq 160$--$200 $ MeV.
$m_{\rm s}= 200$ MeV leads to the result shown in Fig.\,\ref{figTaussrun}.
The difference between the dotted line and the thin (or even thick line, 
provided that care is taken to choose appropriate value of
$\alpha_{\rm s}$\,) is barely significant. Thus, the gluon fusion 
motivated analytical expression  (see Eq.\,(\ref{tausg}) below) 
provides a valid description of the relaxation times in the range of 
temperatures explored in Fig.\,\ref{figTaussrun}, in particular
view of the remaining uncertainties about the radiative 
(odd-$\alpha_{\rm s}$) diagrams,  the initial value  
$\alpha_{\rm s}(M_Z)$, and the strange quark mass. 

\vspace{-3mm}
\subsection{Thermal charm production}\label{charprod}
\vspace{-1mm}
We note that since the mean energy per particle is approximately
$3T$ in the relativistic gas, rather high $\sqrt{s}$ are reached,
allowing in principle the thermal formation of charmed quark
pairs. We exploit now the above developments to obtain these results.

\begin{figure}[tb]
\vspace*{-0.7cm}
\centerline{\hspace*{-0.4cm}
\psfig{width=10cm,figure=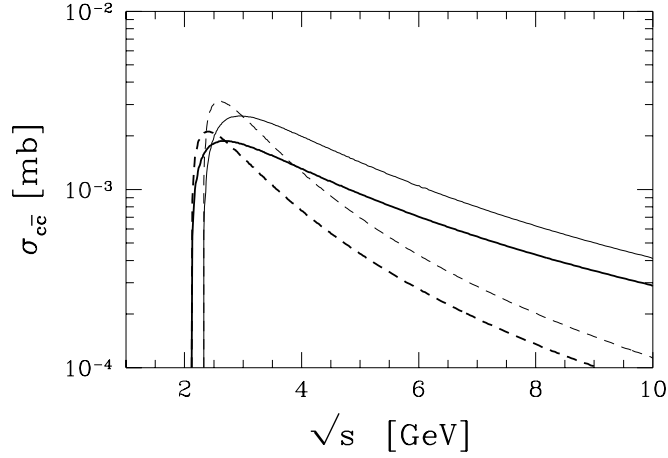}
}
\vspace*{-0.2cm}
\caption{ \small
Charm production cross sections for the two running $\alpha_{\rm s}$
(thick and thin solid lines)  and running
charmed mass  with $m_{\rm c}(M_Z) = 0.7$
MeV; dashed the light quark process,
 solid lines, $gg\to c\bar c$.}\label{sigcc}
\end{figure}

In Fig.\,\ref{sigcc} we show the gluon and 
quark-pair fusion charm production cross sections, computed for 
the two running $\alpha_{\rm s}(M_Z)=0.102\,\ =0.115$
(thick and thin solid lines)  and running
charmed mass  with $m_{\rm c}(M_Z) = 0.7$
MeV; dashed lines depict the light quark fusion process,
Noteworthy is the smallness of this cross section, due to the relatively 
large value of $\sqrt{s}$ required, given that $\sigma\propto 1/s$\,.
However, we will show that one cannot neglect the thermal charm production
in LHC or even RHIC environments, where the charm production
can lead in the end to notable phase space saturation at freeze-out.

\begin{figure}[ptb]
\vspace*{0.9cm}
\centerline{\hspace*{-.3cm}
\psfig{width=9cm,figure=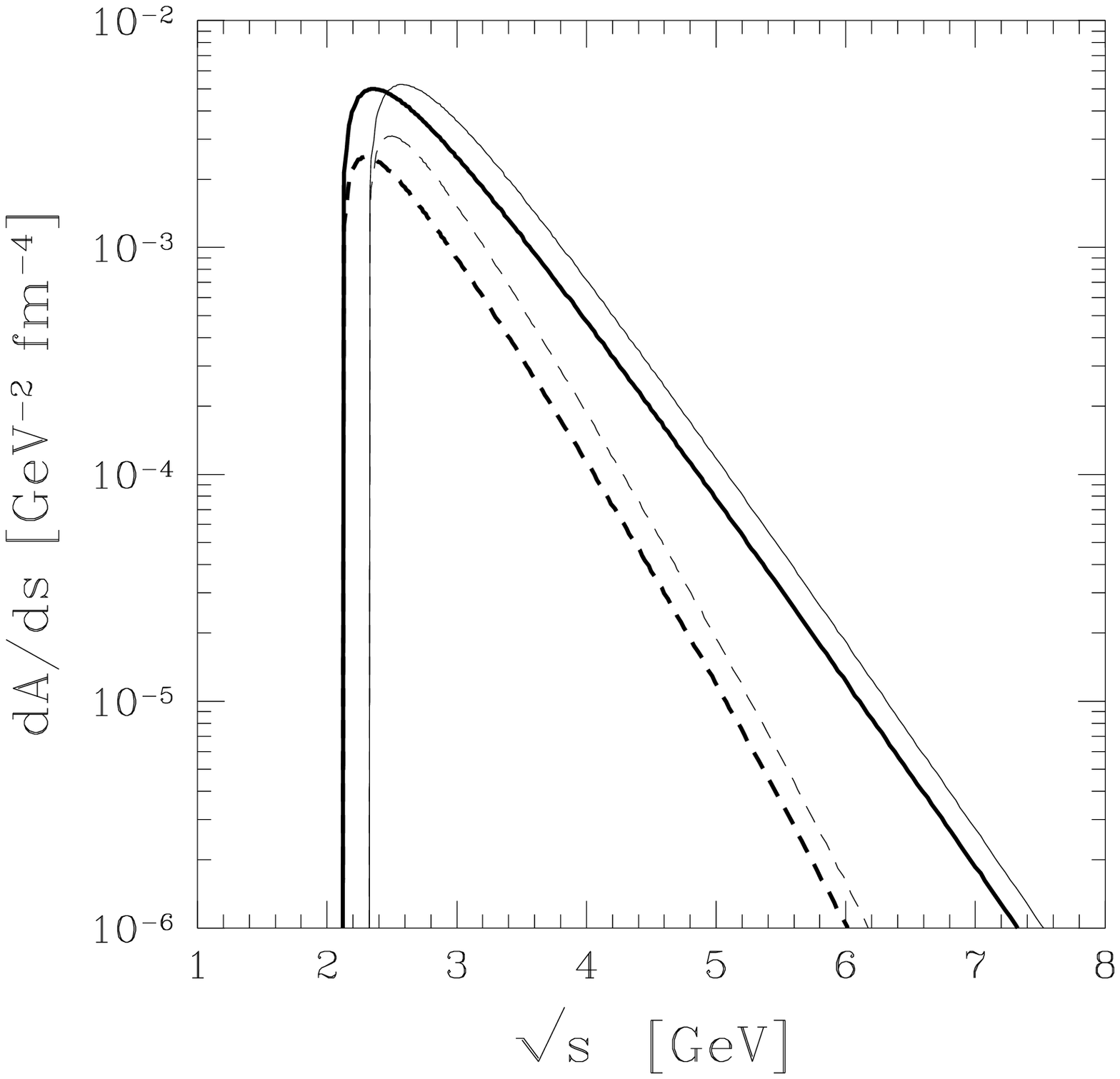}
}
\vspace*{-0.2cm}
\caption{ \small
Differential thermal charm
production rate $dA_{\rm c}/ds = P(s)\sigma(s)$, with $T$ = 500 MeV, with
$\lambda_q=1$ for gluons (solid lines) 
and $q\bar q\to c\bar c$ (dashed, includes three interacting flavors), 
for the two running $\alpha_{\rm s}$
(thick and thin solid lines)  and running
charmed mass $m_{\rm c}$\,.\label{figdAdsc}}
\vspace*{2.3cm}
\centerline{\hspace*{-.3cm}
\psfig{width=9cm,figure=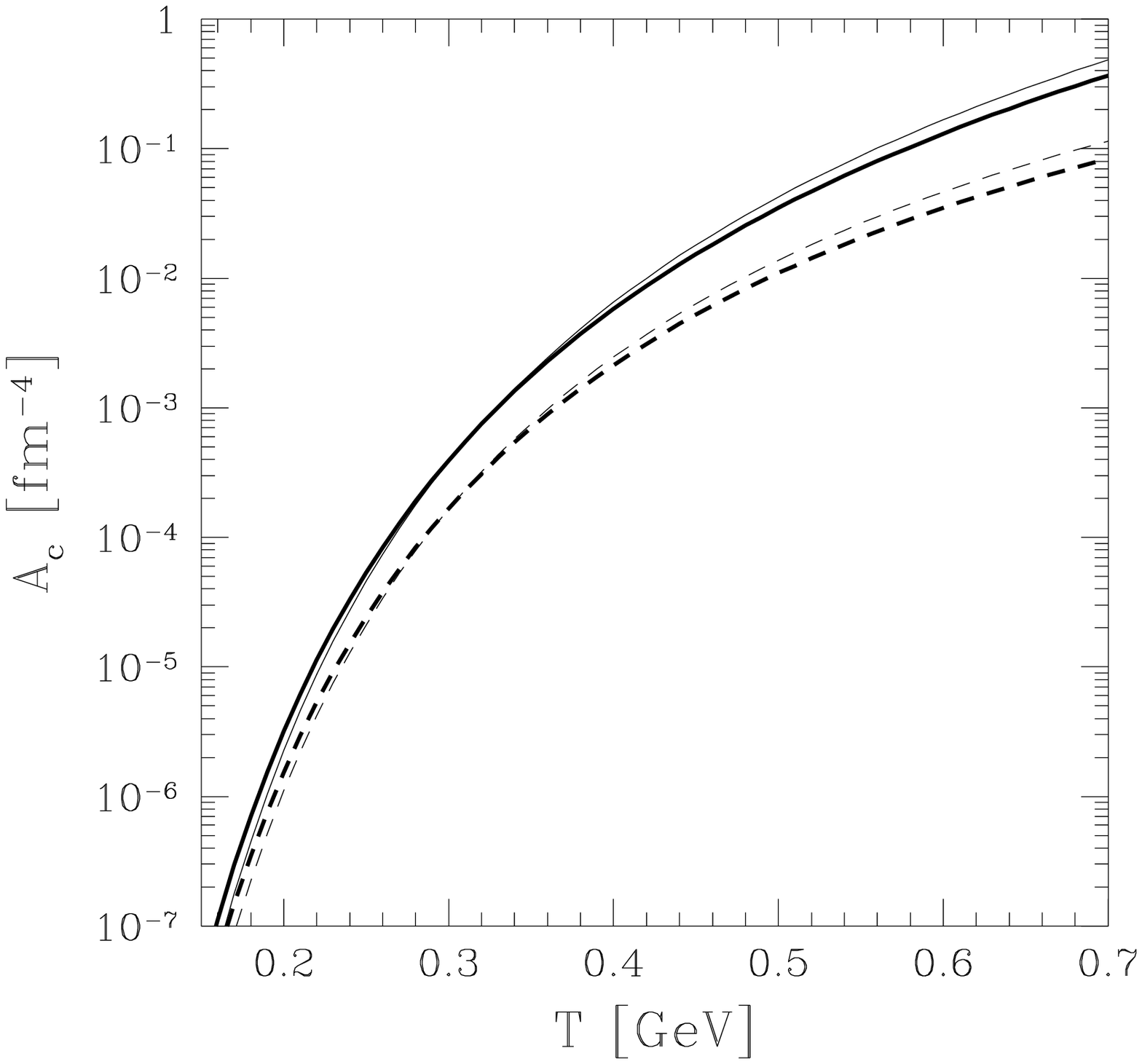}
}
\vspace*{-0.3cm}
\caption{ \small
Thermal charm
production rates $A_{\rm c}$ as function of temperature 
in QGP: total (solid line), and light quarks only (dashed line), calculated
for two cases of running-$\alpha_s\,,\ m_{\rm c}$\,.
\label{figAcc}}
\end{figure}

The resulting thermal differential production rates
$dA_i/ds$  for charm, Eq.\,(\ref{dAds}), are shown in Fig.\,\ref{figdAdsc}. 
We note that the thermal charm production peaks at
$\sqrt{s}\simeq 2.5$ GeV, near to the running mass production threshold. 

The differential production rate can  be easily integrated, and we show
the result in Fig.\,\ref{figAcc}. We see that the
charm production rate changes
by 6 orders of magnitude as the temperature 
varies between 200 and 700 MeV. This sensitivity on the 
initial temperature, while understandable due to the fact
that $m/T>1$, also implies that since the charm production rate is not 
vanishingly small, we may have found an interesting probe 
of the primordial high
temperature phase. This was also noted in a case study performed
by Levai et al. \cite{Lev95}. Note that the gluon dominance 
of the production rate is not as pronounced for charm as it is 
for strangeness because charm formation occurs near to the 
threshold, where the quark fusion cross section dominates. Only
for  $T\ge 400$ MeV we find that the glue fusion dominates the 
thermal charm production clearly. 
For charm there is the possibility that the thermal production
is overwhelmed by the direct production based on high energy parton
interactions. Calculations show \cite{SV95} that per LHC event there
may be a few directly produced charm quark pairs. However,
we have a differential production rate  
for an initial state with chemically equilibrated
gluons at $T\simeq 450$ MeV, $A_{\rm s}\simeq 10^2\,{\rm fm}^{-4}$, see
Fig.\,\ref{figAcc}, which implies that we should expect up to 
20 thermal charmed quark pairs per such event at central
rapidity, which yield is clearly dominating the reported direct production 
rate. Consequently, we continue below to evaluate 
in detail the evolution of thermal charm yield, which may 
dominate the production rate and in particular lead to rather surprising 
features in final particle yields, should the initial plasma temperature be 
sufficiently large. Moreover, the rats we find for temperatures near 250
MeV, seem to be still within the realm of the observable.

\begin{figure}[htb]
\vspace*{2.1cm}
\centerline{\hspace*{-.3cm}
\psfig{width=10cm,figure=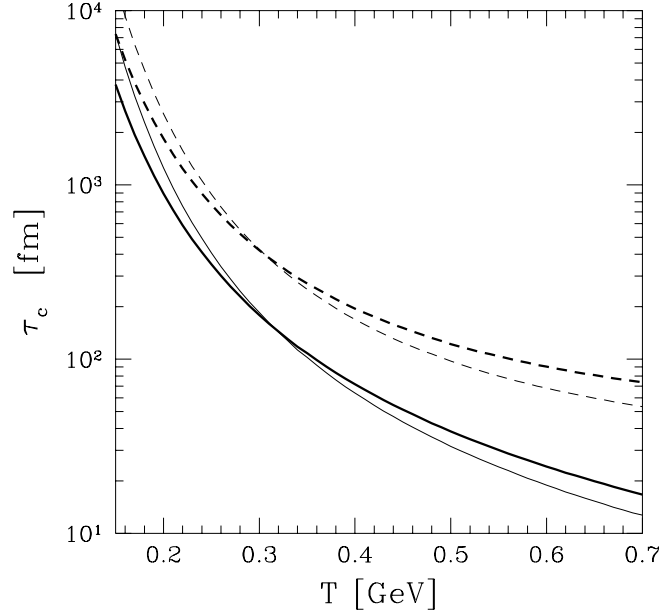}}
\vspace*{-0.2cm}
\caption{ \small
Thermal charm relaxation constant in QGP, calculated
for  $\alpha_s$ running.: same conventions and
parameters as in Fig.\,\protect\ref{figAcc}.\label{figtaucc}}
\end{figure}
In Fig.\,\ref{figtaucc}
we show the charm relaxation constant, see Eq.\,(\ref{tauss}), 
using the same conventions and parameters as in 
Fig.\,\ref{figAcc}, and using $m_{\rm c}\simeq 1.5$ GeV in order to 
establish the reference density for the approach to equilibrium. Actually,
this running value is temperature dependent, and will be slightly smaller 
at higher temperatures, since the average interaction energy is greater. 
We will return to discuss this intricate variation at another occasion. 
What it implies is that at high temperatures we are `chasing' with
the thermal rate a somewhat higher equilibrium density and thus the 
computed relaxation time underestimates slightly the correct result.

\section{Evolution of heavy quark 
Observables}\label{gammas}
\subsection{Flow model}\label{flowsec}
We now can proceed to explore two generic (strangeness) observables
as function of the impact parameter (baryon content) and collision energy:
\begin{itemize}
\vspace*{-0.3cm}\item Specific (with respect to baryon number $B$) 
strangeness yield  $N_{\rm s}/B$\\
{\it Once produced strangeness escapes, bound in diverse hadrons, from 
the evolving fireball and hence  the total abundance observed 
is characteristic for the initial extreme conditions;}
\vspace*{-0.3cm}\item Phase space occupancy $\gamma_{\rm s}$\\
{\it Strangeness freeze-out conditions at particle hadronization 
time $t_{\rm f}$, given  the initially produced abundance,
determine the final state observable phase space occupancy of strangeness 
$\gamma_{\rm s}(t_{\rm f})$.} 
\end{itemize}
\vspace*{-0.3cm}
To pursue this a more specific picture of the temporal evolution is needed: 
in the earlier discussion in section \ref{TMod} we have considered 
the chemical cooling \cite{cool} due to the strangeness production. It is
a rather complicated matter to account simultaneously for both the
chemical cooling, and the flow cooling arising from volume expansion. 
We shall concentrate here on the flow cooling which dominates the evolution
once strangeness reaches chemical equilibrium.  We therefore denote all 
initial values by the
subscript `in' in order to distinguish the here proposed schematic model
from  the earlier discussion of chemical cooling which is important in the 
early evolution stages. 

In first approximation, the particle density  in the
fireball as taken to being constant and the sharp surface of the 
volume comprising the dense matter is allowed to expand in all space
directions at most and probably near to maximal sound velocity 
$v_c\lesssim c/\sqrt{3}$. 
This value is consistent with the hydrodynamic flow studies and also 
leads to a Doppler blue-shift factor 
$F_{\rm f}=\sqrt{(1+v_c)/(1-v_c)}\lesssim 1.93$  of the 
freeze-out temperature $T_{\rm f}\simeq 140$ MeV, which is consistent
with the apparent spectral temperatures, obtained from the initial 
temperature, see table~\ref{bigtable}. Onto this collective radial 
motion there will be  superposed additional longitudinal collective 
motion related to the remainder  of the original longitudinal momentum of
the colliding particles. Furthermore, the volume and 
temperature temporal evolution constrained by the adiabatic evolution 
condition which for massless particles has the form:
\begin{equation}\label{adiaex}
V\cdot T^3=\,{\rm Const.}\,.
\end{equation}
The fireball radius grows according to
 \begin{equation}\label{R(t)}
R=R_{\rm in}+{1\over \sqrt{3}}(t-t_{\rm in})\,,
\end{equation}
and hence from the adiabatic expansion constraint
Eq.\,(\ref{adiaex}) we obtain the time dependence of temperature:
\begin{equation}\label{T(t)}
T={T_{\rm in}\over{1+{{t-t_{\rm in}}\over \sqrt{3}R_{\rm in}}}}\,.
\end{equation}

\begin{figure}[htb]
\vspace*{1.8cm}
\centerline{\hspace*{-0.4cm}
\psfig{width=7cm,figure=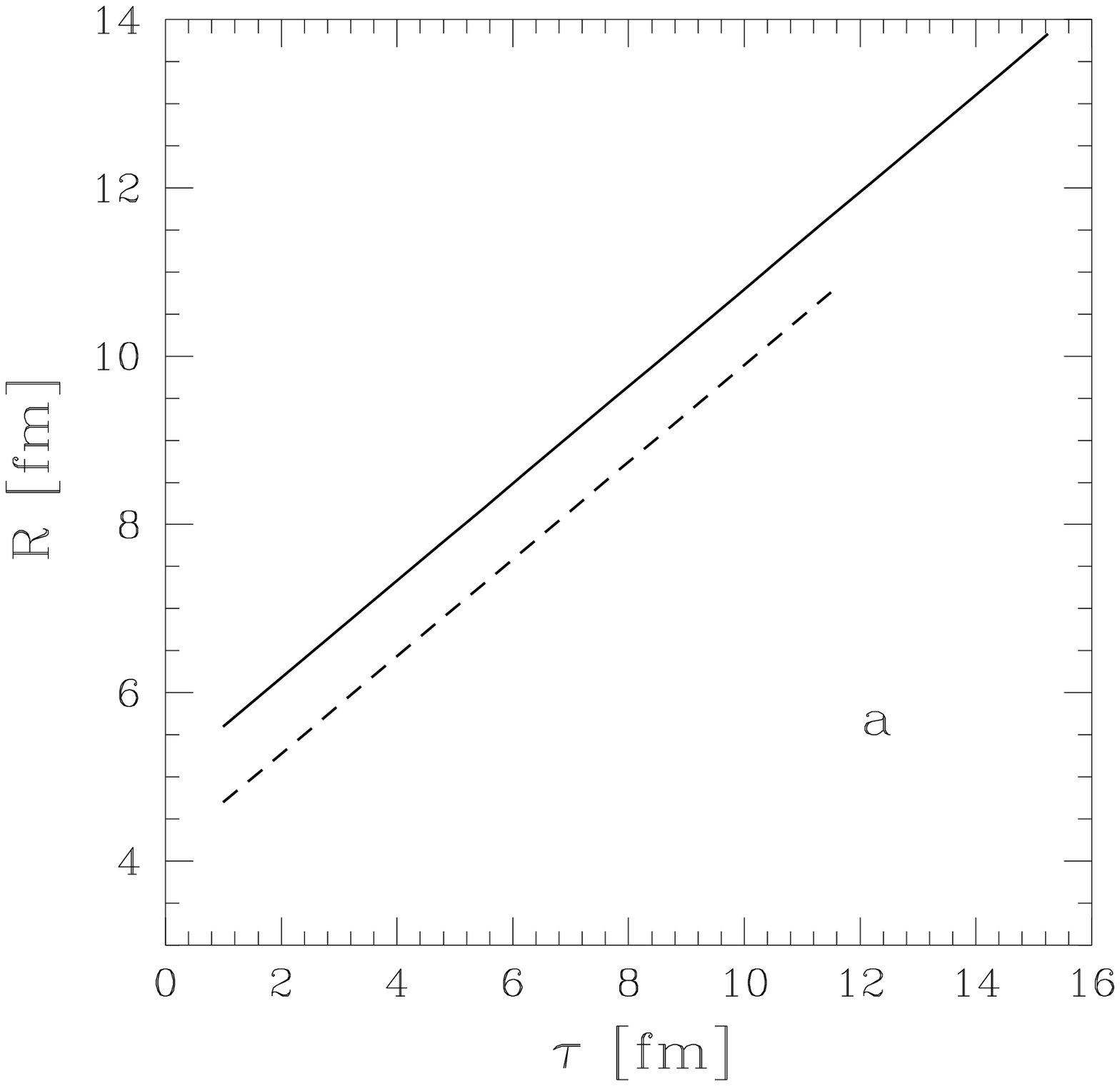}\hspace*{-1.cm}
\psfig{width=7cm,figure=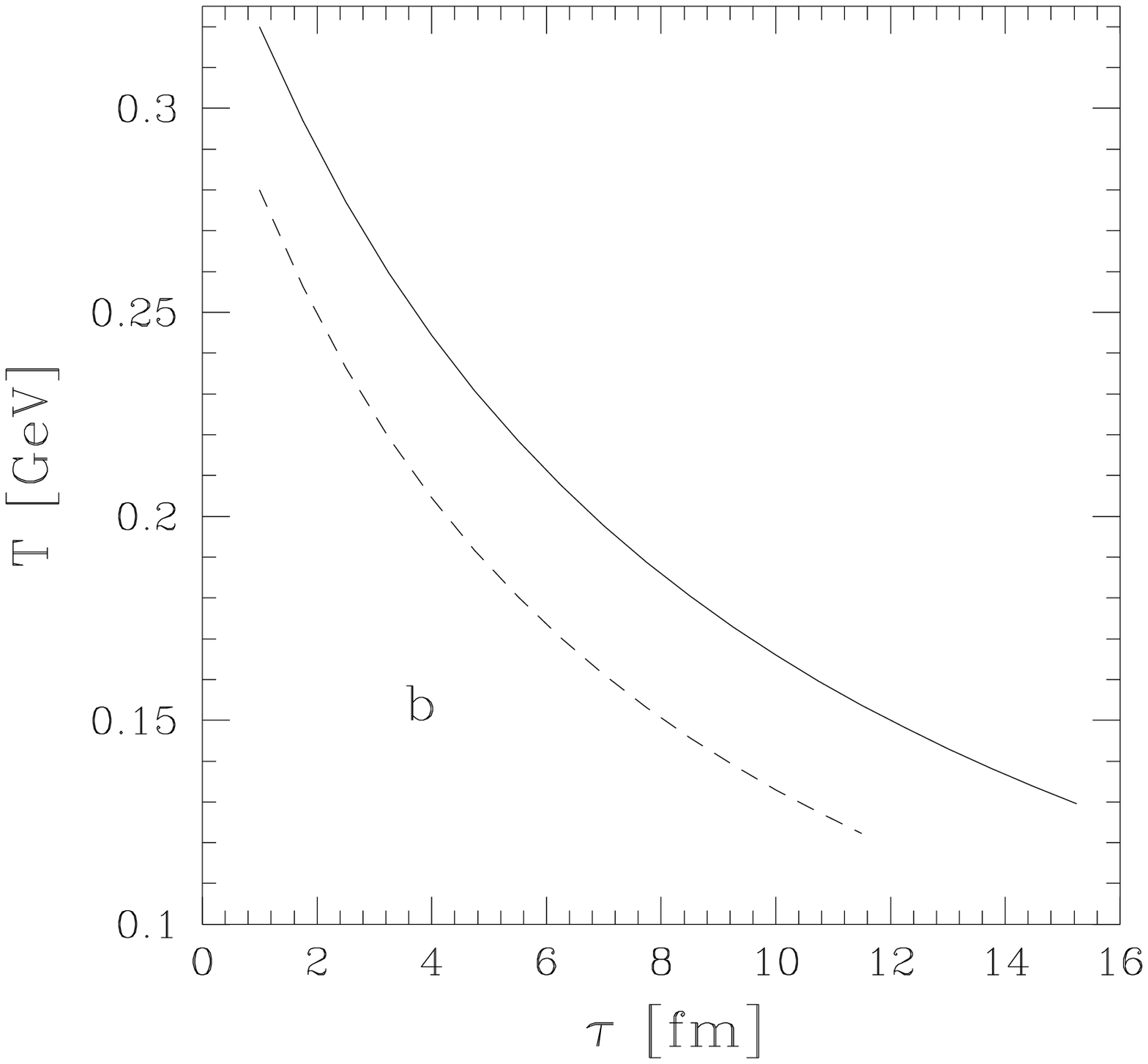}
}
\vspace*{0.2cm}
\caption{ \small
Assumed (see Eqs.\,(\protect\ref{R(t)}, \protect\ref{T(t)}) 
temporal evolution of {\bf a} the radius parameter
$R(t)$ and {\bf b} temperature $T$ of the thermal 
fireball formed in Pb--Pb collisions (solid lines) and S--W/Pb
collisions (dashed lines). \protect\label{figvarRT}}
\end{figure}
 
A set of initial conditions for the SPS experiments follows from 
the kinematic constraints\cite{dynamic}, see table~\ref{bigtable}, 
consistent with global event structure, and the hadronic 
freeze-out condition seen in HBT experiments is 
\cite{NA35HBT,NA44HBT}:
\begin{eqnarray}
T_{\rm in}=320\!&{\rm MeV;}&\! R_{\rm in}=5.6\,{\rm fm};\
        t_{\rm in}=1\,{\rm fm}/c;\
\lambda_{\rm q}=1.6;\ \mbox{\rm for\ Pb--Pb\,,}\nonumber\\
\label{initcondg}
T_{\rm in}=280\!&{\rm MeV;}&\!R_{\rm in}=4.7\,{\rm fm};
\ t_{\rm in}=1\,{\rm fm}/c;\
\lambda_{\rm q}=1.5;\ \mbox{\rm for\ S--Pb/W\,.}\nonumber
\end{eqnarray}
Here the radius $R_{\rm in}$ has been determined such that for
the QGP equations of state we employ the baryon number content
in the fireball is 380 (Pb--Pb case) and 120 (S--Pb/W case)
respectively, corresponding to zero impact parameter collisions.
The energy/baryon content at given projectile energy is computed 
assuming that same stopping governs energy and baryon number 
which we take for the current discussion to be 50\% for S--Pb/W and 
100\% for Pb--Pb. Also, we take $\gamma_s(t=t_{\rm in})=0.15$ as the 
initial strangeness abundance
after 1 fm/$c$. The chosen values of $\lambda_{\rm q}$ are of minimal
importance, as they enter marginally into the quark-fusion rate, 
which is a minor contribution to the strangeness production rate. 
However,  $\lambda_{\rm q}$ has
indirect importance as it determines the initial fireball size
for given $B$ and impacts greatly the strange particle
(baryon/antibaryon)  ratios. The most important parameter, which
follows from the theoretical model developed for the collision, is the
temperature $T_{\rm in}$. We use in the calculations 
$\gamma_{\rm s}(t_{\rm in})=0.15$ as the initial
strangeness occupancy factor after 1 fm/$c$\cite{RM82}. 

From Eqs.\,(\ref{R(t)}, \ref{T(t)}) it follows  
that the temperature drops to the commonly
accepted phase transition value  $T_{\rm f}\simeq 140$ MeV \cite{lattice}
at $t=9.2$ and 13.5 fm/$c$ for S--W/Pb and Pb--Pb systems, respectively.
At these instants the size of the fireball has reached 9.4 
and 12.8 fm, respectively.

Note that some of the  results shown below here were 
obtained with QCD parameters, 
$\alpha_{\rm s}=0.6$ and $m_{\rm s}=160$  MeV \cite{dynamic}, 
other were computed with running $\alpha_{\rm s}$ with initial 
values as discussed, see Fig.\,\ref{fig-a1} \cite{impact}.
While these results are nearly consistent with each other,
the remaining differences illustrate some of the uncertainties in 
our theoretical modeling of the heavy flavor production.

\vspace{-3mm}
\subsection{Dynamical description of observables 
in the fireball}\label{dynfbsec}
\vspace{-1mm}

As we have discussed in section 
\ref{secanalysis} it is rather straightforward to 
extract from the strange antibaryon experimental particle yields
the value of strangeness phase space occupancy $\gamma_{\rm s}(t=t_{\rm f})$,
since the hadronization value of $\gamma_{\rm s}(t_{\rm f})$ governs the
 ratios of particles with different strangeness content.
We recall that since the thermal equilibrium is by hypothesis 
established within a considerably shorter time scale than the (absolute)
heavy flavor chemical equilibration, we can characterize the saturation
of the  phase space by an average over the momentum
distribution, see also Eq.\,(\ref{rhos(t)}):
\begin{equation}\label{gamth}
\gamma_{\rm s,c}(t)\equiv {
           \int d^3\!p\,d^3\!x\,n_{\rm s,c}(\vec p,\vec x;t)\over 
     \int d^3\!p\,d^3\!xn_{\rm s,c}^{\infty}(\vec p,\vec x)}\, ,
\end{equation}
where $n_{\rm s,c}$ is the sum over all heavy flavor containing
 particle densities, and
should multistrange/charmed objects be present, this sum 
 contains the associated 
weight. $n_{\rm s,c}^\infty$ is the same, but for the equilibrium particle 
densities. In QGP deconfined state, of course we have just the free quarks.
When we assume that the fireball is homogeneous in  $T$ and $\lambda_{\rm q}$ 
we can write:
\begin{equation}\label{gamdef2}
n_{\rm s}(\vec p;t)=\gamma_{\rm s}n_{\rm s}^\infty
     (\vec p;T,\mu_{\rm s})\,.
\end{equation}

Since  $\tau_{\rm s}$ is just of the magnitude of the 
life span of the deconfined state, see Fig.\,\ref{figtauss}, 
strangeness will be close to fully
saturate the final state phase-space in the QGP fireball. 
However, this accidental similarity of the life span of the
fireball and the relaxation time of strangeness implies that 
changes in the collision conditions should lead to measurable
changes of $\gamma_{\rm s}$. This would be a highly desirable situation,
allowing a test of the theoretical predictions. It can be expected
that in the near future  $\gamma_{\rm s}$ will  be
studied varying a number of parameters of the collision, such as the
volume occupied by the fireball (varying size of the colliding nuclei and
impact parameter), the trigger condition (e.g., the inelasticity), the
energy of colliding nuclei when searching for the threshold energy of
abundant strangeness formation. We thus develop in this section a 
more precise understanding of the time evolution of the observed value of 
$\gamma_{\rm s}$, as function of the collision parameters. This variable
comprises as the dynamic element the specific strangeness yield, 
indeed we can easily see that:
the ratio of the observed phase space occupancy $\gamma_{\rm s}(t_{\rm f})$
to the specific yield $N_{\rm s}/B$ is {\it independent of the 
initial conditions and only dependent on the freeze-out}:
\begin{eqnarray}
\frac{\gamma_{\rm s}(t_{\rm f})}{N_{\rm s}/B}= \frac{B}{N_{\rm s}^\infty}\,,
\end{eqnarray}
where 
$N_{\rm s}^\infty=\rho_{\rm s}^\infty(T_{\rm f}) 
		V_{\rm f}$ 
is the equilibrium abundance of 
strangeness in the fireball at  dissociation/freeze-out.
We thus see that if we can interpret both observables 
$\gamma_{\rm s}(t_{\rm f})$ and $N_{\rm s}/B$ successfully, we have 
in all likelihood obtained a valid model both of the initial 
and freeze-out conditions. 

Similarly, as alluded to above in section \ref{charprod}, 
the thermal charm production is 
sensitive to the initial temperature, but clearly the
production of charmed particles will not saturate the initially
available phase space. However, it is interesting to 
see what values of $\gamma_{\rm c}$ would be found in the final
state, since the equilibrium density of charm at hadronization
is very low. Also here we need to consider in some
more detail the temporal evolution with the plasma expansion
of the off-equilibrium parameter $\gamma_{\rm c}$.

The general expression for strangeness production is given by 
Eq.\,(\ref{drho/dt3}). Taking the particle density everywhere in the
fireball as constant, we have:
\begin{equation}\label{dNsdt}
{1\over V}{{dN_{\rm s}(t)}\over {dt}} = 
A\left[1-\gamma_{\rm s}^2\right]\,.
\end{equation}
The $\gamma_{\rm s}^2$ term arises when the back reaction, 
e.g., $s\bar s\to gg$ is considered in the Boltzmann 
approximation, and the unsaturated \cite{cool} thermally 
equilibrated quantum Bose/Fermi phase space  distributions
are expanded:
 \begin{equation}\label{qdist}
n^{\rm B,F}_i=
{1\over{\gamma^{-1}_i\lambda_i{\rm e}^{\beta\epsilon_i}\mp 1}}\,
	\to \gamma_i\lambda_i^{-1}{\rm e}^{-\beta\epsilon_i}\,.
\end{equation}
 Evaluation of relaxation
constants with complete  quantum phase space \cite{SM86} has not 
revealed any  significant effects, thus Boltzmann terms provide  
here a very  good approximation. Recall that the subtle difference between 
$\gamma$ and 
\baselineskip=13.9pt
$\lambda$ is that while the latter is conjugated between
particles and antiparticles, see Eq.\,(\ref{lam}), $\gamma$ is the 
same for particles and antiparticles. Equivalently, one can introduce 
different chemical potentials for particles and antiparticles, as was
the case in Ref.\cite{SM86}
 
It is common practice to illustrate the impact of volume expansion 
dilution 
to write $N=\rho V$ which when inserted on the left hand side of 
Eq.\,(\ref{dNsdt}) leads to:
\begin{equation}\label{dNsdtbis}
{d\,\rho_{\rm s}\over{dt}}+\rho_{\rm s}{1\over V}{{dV}\over {dt}} = 
A\left[1-\gamma_{\rm s}^2\right]\,.
\end{equation}
The second term on the left hand side is referred to as the volume
dilution term.
 
In order to obtain a dilution equation for $\gamma_{\rm s}$, 
let us instead proceed, using in Eq.\,(\ref{dNsdt}) the definition of
$\gamma_{\rm s}$ in the form:
\begin{equation}\label{Ndef}
N_{\rm s}(t)=\gamma_{\rm s}(t)N_{\rm s}^\infty(T(t))\,.
\end{equation}
Note that when dividing Eq.\,(\ref{Ndef}) by $V(t)$ we recover our earlier 
definitions of $\gamma_{\rm s}$, see Eqs.\,(\ref{rhos(t)}, \ref{gamdef2}).

Inserting Eq.\,(\ref{Ndef}) into Eq.\,(\ref{dNsdt}) we obtain:
\begin{equation}\label{dNsdtinf}
2\tau_{\rm s}\left({{d\gamma_{\rm s}}\over{dt}}
+\gamma_{\rm s}{d\over{dt}}\ln N^\infty\right)
=1-\gamma_{\rm s}^2\,.
\end{equation}
It is noteworthy that $N^\infty$, the final total abundance 
of particles, as given in
Eq.\,(\ref{Nsinfty}), changes only slowly in time when the volume and
temperature temporal evolution is governed by the adiabatic 
evolution condition, Eq.\,(\ref{adiaex}).
Thus the logarithmic derivative in the dilution term in
Eq.\,(\ref{dNsdtinf}) is in many cases very small since:
\begin{equation}\label{dilute}
{d\over{dt}}\ln N^\infty
={d\over{dt}}\ln \left(x^2{k}_2(x)\right)\,;\quad x={m\over T(t)}\,.
\end{equation}
What we see happening is that the volume dilution seen in Eq.\,(\ref{dNsdtbis})
is  nearly completely compensated by the dilution of the value of
$\rho^\infty(T)$  in presence of adiabatic cooling. 

\baselineskip=13.5pt
In many cases it is sufficient to study an approximate solution 
of Eq.\,(\ref{dNsdtinf}). 
For $m_{\rm s}/T=x<1$ we have $x^2k_2(x)\simeq$\,Const., 
and hence we have the analytical solution:
\begin{equation}\label{dgamsmallx}
\gamma_{\rm s}\simeq \tanh \left(\int_0^{t_{\rm freeze}} 
{{dt}\over {2\tau_{\rm s}(T(t))}}\right)\,<\,1\quad m_{\rm s}/T<1\,,
\end{equation}
where the semi-convergent approximation  for the dominant
gluon fusion term has been used in the past\cite{RM82}:
\begin{equation}\label{tausg}
\tau_{\rm s}^gm_{\rm s}=\alpha_s^{-2}\,{{9}\over 7}\sqrt{\pi\over 2}\, 
{{x^{5/2}}\over {{\rm e}^{-x}(x+99/56+\ldots)}}\,
\end{equation}
in order to argue that the value of $\gamma_{\rm s}$ in many cases of 
interest approaches unity. 

But the approximate solution, Eq\,.(\ref{dgamsmallx}), presumes that the 
final freeze-out occurs such that $m_{\rm s}/T<1$, which condition
is not fulfilled if the plasma hadronizes at temperatures of 
the magnitude $T=140$ MeV as seems to be the case today for 
the baryon rich plasma, see our discussion in section
\ref{results} and the results of lattice gauge simulations of
QCD \cite{lattice}. We will now show 
numerically that major deviations from the approximate solution
arise and in particular $\gamma_{\rm s}$ can easily become much
greater than unity, depending on the precise value of the
freeze-out temperature. To see this note that 
a slight rearrangement of Eq.\,(\ref{dNsdtinf}) leads to the form:
\begin{equation}\label{dgdtf}
{{d\gamma_{\rm s}}\over{dt}}=
\left(\gamma_{\rm s}{{\dot T m_{\rm s}}\over T^2}
	{d\over{dx}}\ln x^2k_2(x)+
{1\over 2\tau_{\rm s}}\left[1-\gamma_{\rm s}^2\right]\right)\,,
\end{equation}
which shows that even when $1-\gamma_{\rm s}^2<1$ we still can have 
a positive derivative of $\gamma_{\rm s}$, since the first term
on the right hand side of Eq.\,(\ref{dgdtf}) is always positive,
both $\dot T$ and $d/dx(x^2k_2)$ being always negative. 
Note that $1/\tau$ becomes small when $T$ drops below 
$m_{\rm s}$ and whence the dilution term dominates the evolution of
$\gamma_{\rm s}$. 

\vspace{-3mm}
\subsection{Strangeness and charm in final state}\label{finscsec}
\vspace{-1mm}

\begin{figure}[htb]
\vspace*{1.3cm}
\centerline{\hspace*{-.5cm}
\psfig{width=7cm,figure=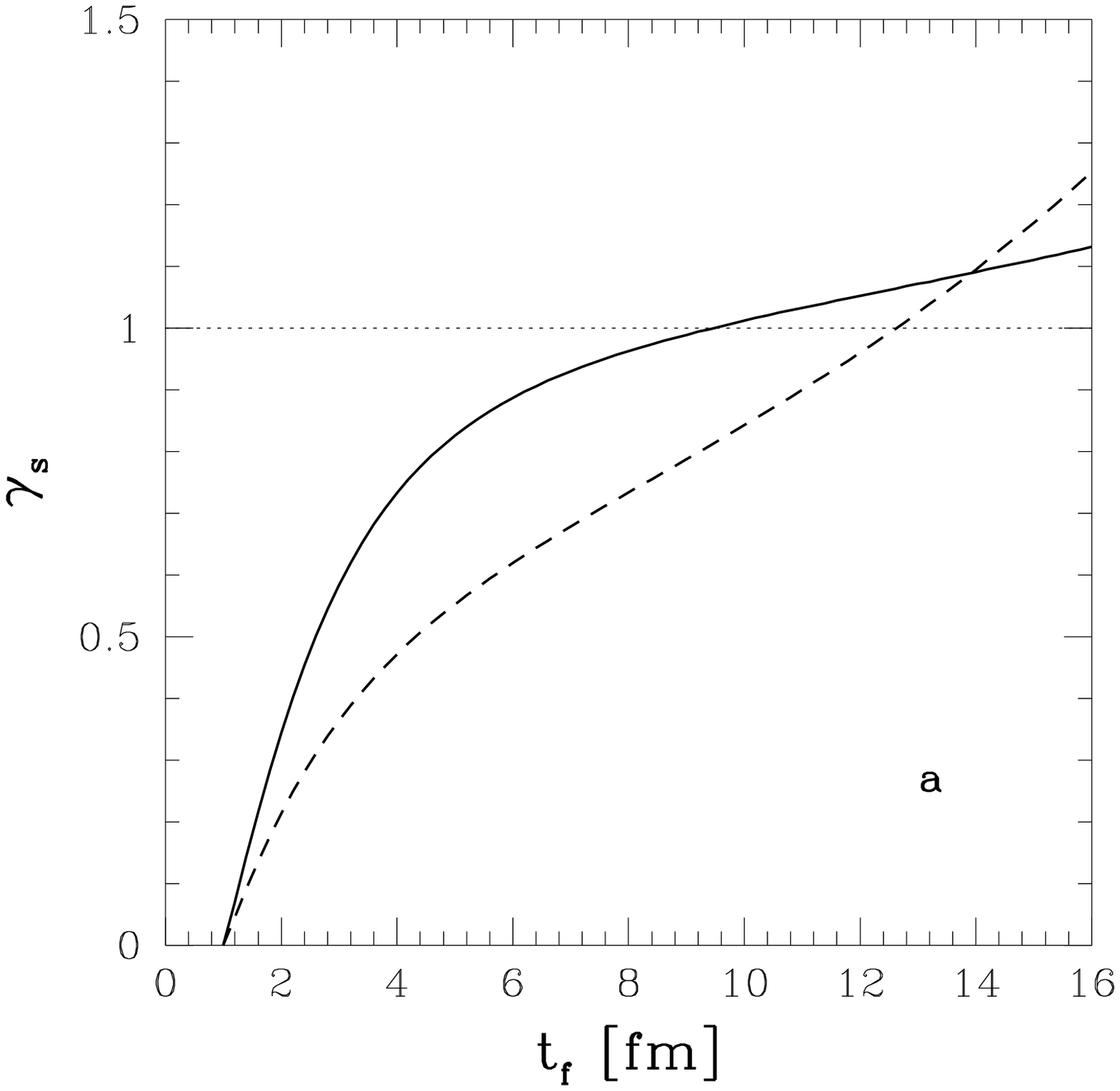}\hspace*{-.9cm}
\psfig{width=7cm,figure=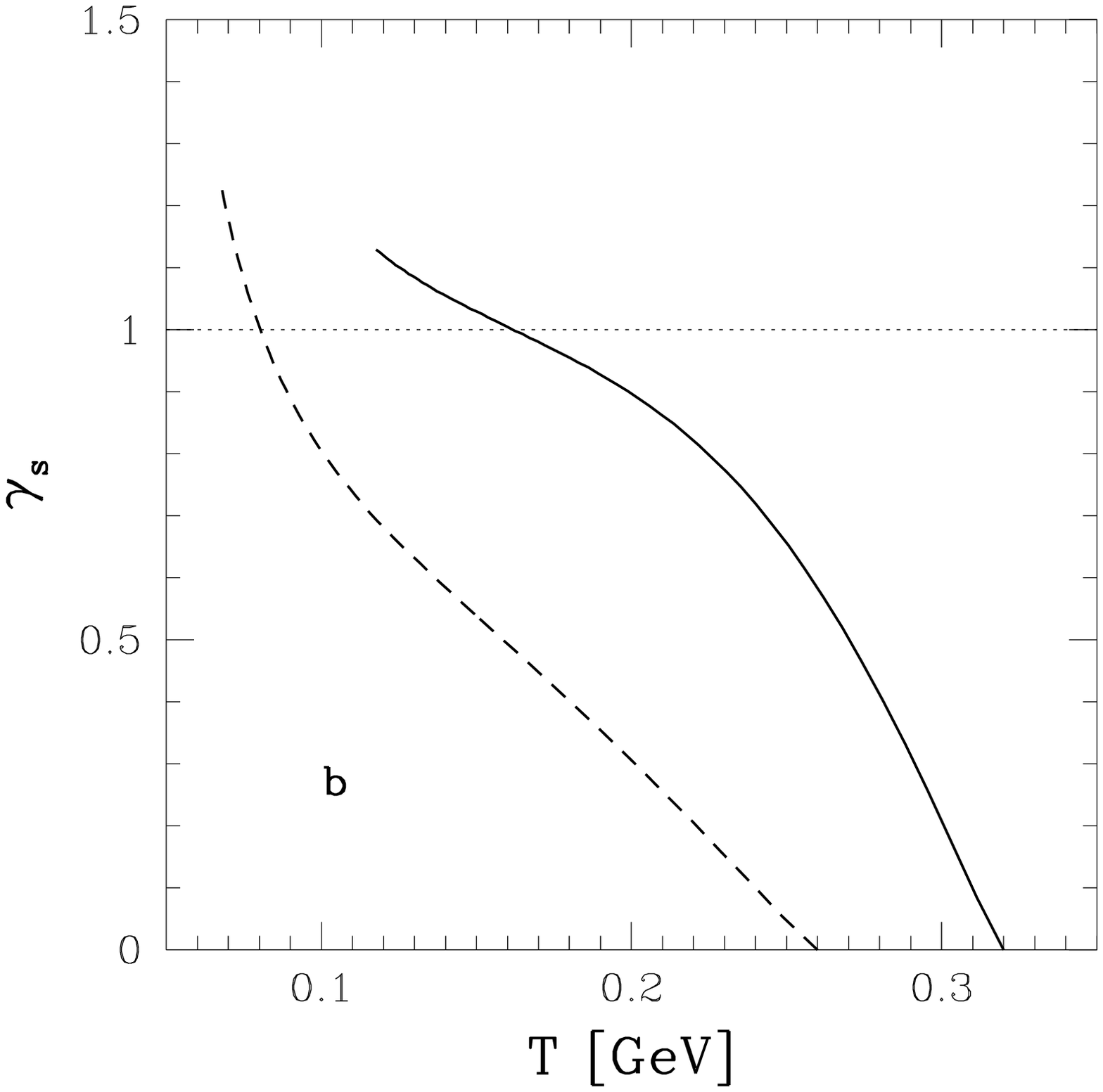}
}
\vspace*{-0.2cm}
\caption{ \small
QGP-phase strangeness phase space occupancy $\gamma_{\rm s}$ 
{\bf a}) as function of  time and {\bf b}) as function
of temperature, for 
$\alpha_s=0.6$ and $m_{\rm s}=160$ MeV, for initial conditions
pertinent to maximum SPS energies (200A GeV S beam and 158A GeV
Pb beam). Solid lines: conditions relevant to central Pb--Pb 
interactions, dashed lines: conditions relevant to S--W/Pb
interactions (see table~\protect\ref{bigtable}).\label{figvarsgam}} 
\end{figure}

The numerical integration of Eq.\,(\ref{dgdtf}) is now possible, up to the
point at which the plasma phase ceases to exist or/and the final state
strange particles are emitted. According to our hypothesis, which
leads to a successful interpretation of the experimental data, the
abundances of rarely produced strange (anti)baryons is not further
affected by subsequent evolution. We present $\gamma_s$ for the
case of S--W/Pb
collisions (dashed lines) and Pb--Pb collisions (solid lines),
in Fig.\,\ref{figvarsgam}{\bf a} as function of final time and in 
Fig.\,\ref{figvarsgam}{\bf b} as function of final temperature. We note  
that for 8 fm/c we obtain the observed  value
$\gamma_{\rm s}\simeq 0.75$ for the S--W/Pb collisions. However,
this time is associated with a low final temperature of $T=110$
MeV, as can be deduced from the result shown in
Fig.\,\ref{figvarsgam}{\bf b}. Taking the final temperature
value to be $T\simeq 140$ MeV
for the S--W/Pb case, one arrives at $\gamma_{\rm s}\simeq
0.57$. This is slightly less than the experimental result 
$\gamma_{\rm s}\simeq 0.75$ which suggests that our ideal flow
temporal evolution model may be leading to a too fast cooling
or/and that the perturbative estimate of the strangeness
production rate is a bit too low ---  to reach
exact agreement between experiment and theory we would need a
cumulative change in these two here relevant quantities (flow
velocity and QGP-strangeness production rate) of magnitude
20\%. There is clearly plenty of room for an improvement of this
magnitude in both these quantities.

When considering  the Pb--Pb collisions we are primarily interested to
find if it is likely that we reach $\gamma_{\rm s}=1$. 
For this to occur, our results, see Fig.\,\ref{figvarsgam}, 
suggest that the final QGP fireball 
temperature should be lower than 160 MeV. 
Note that allowing for the above discussed likely further
increase in production rate and/or reduction in flow, pushes
this temperature limit to 210 MeV.
We thus can be practically certain that in Pb--Pb collisions at
158A GeV one observes $\gamma_{\rm s}\ge 1$ with the associated
interesting consequences for strange particle abundances (see
section \ref{results}). 

\begin{figure}[htb]
\vspace*{1.3cm}
\centerline{\hspace*{-.5cm}
\psfig{width=7cm,figure=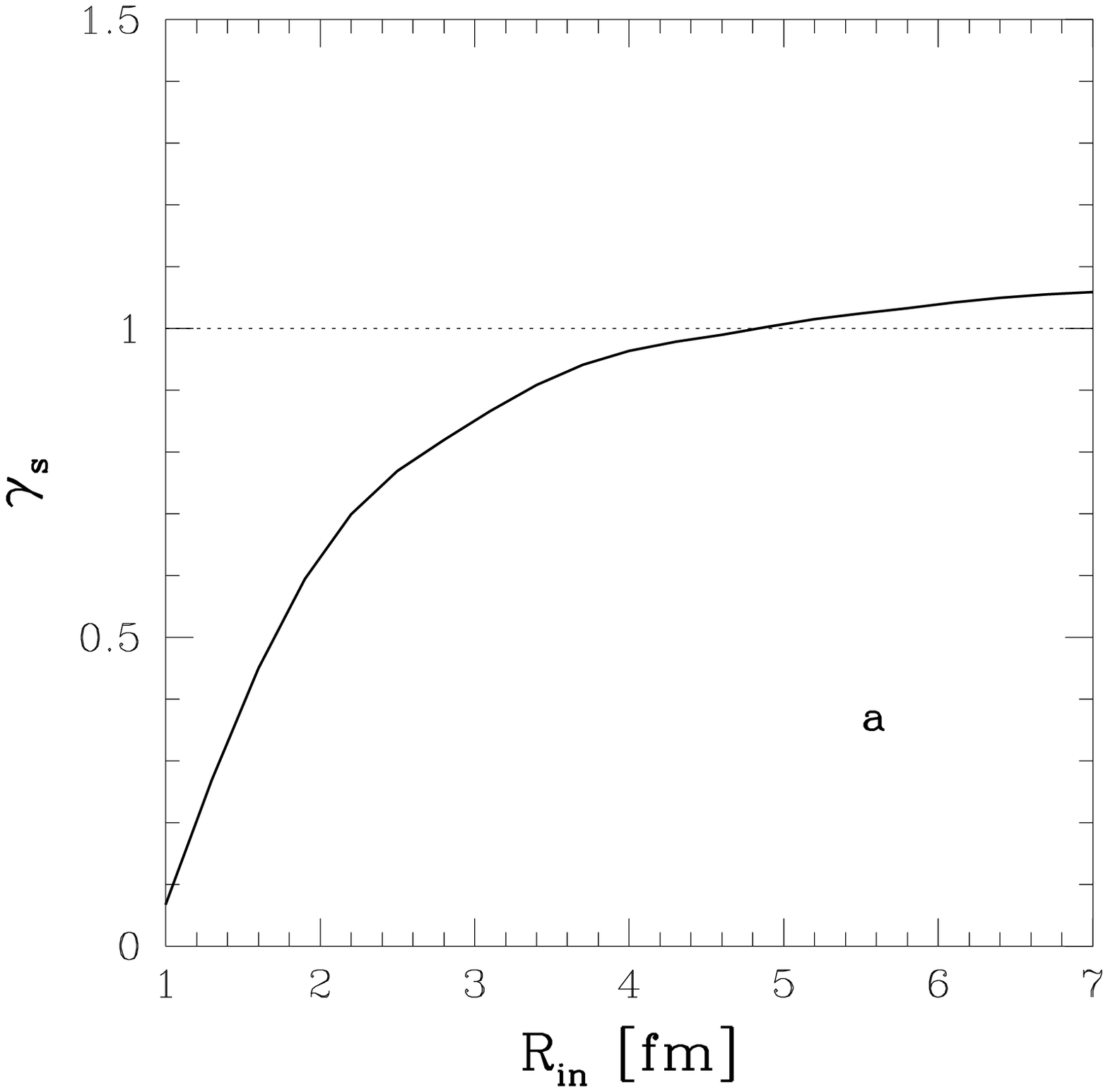}\hspace*{-.9cm}
\psfig{width=7cm,figure=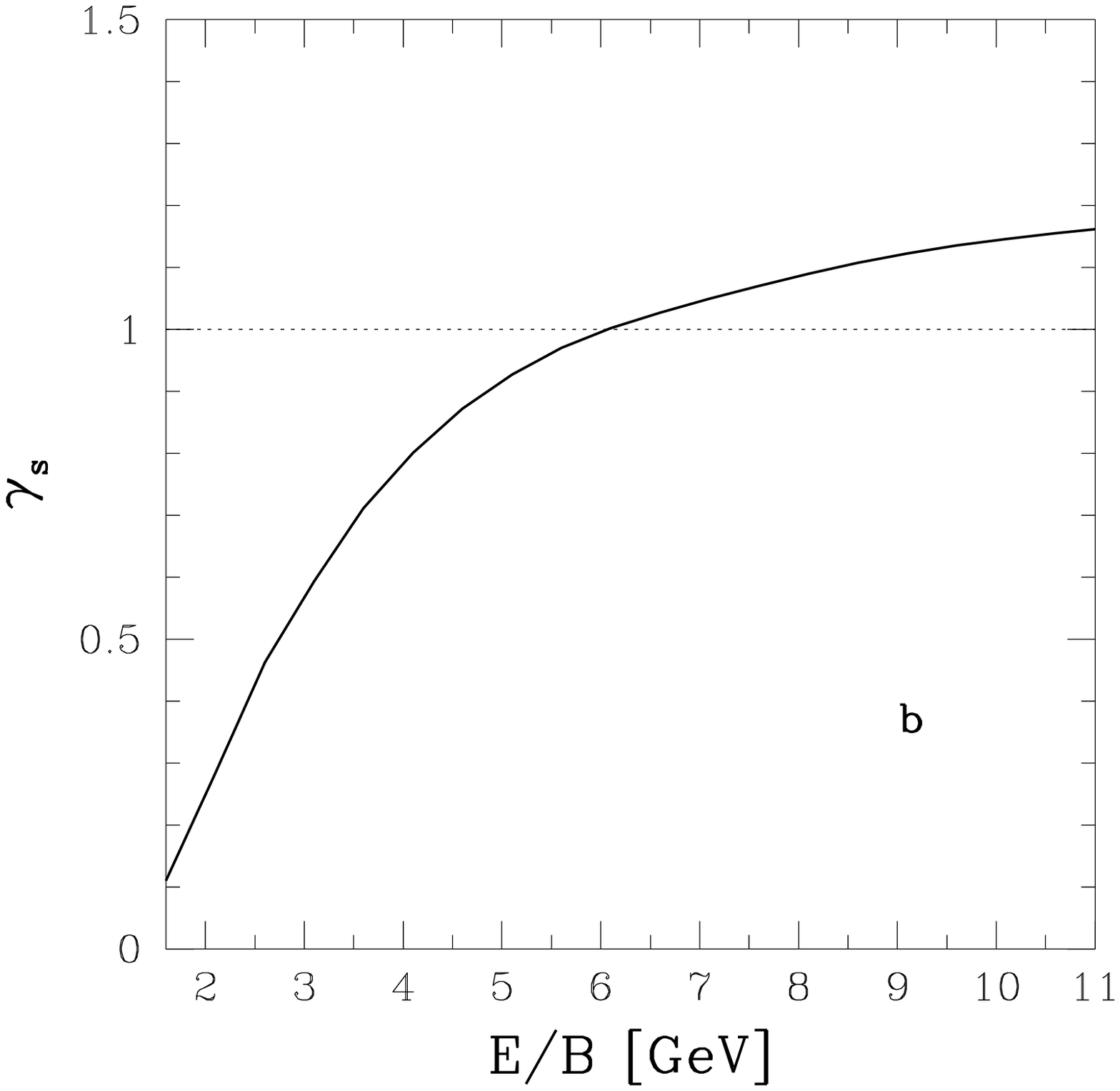}
}
\vspace*{-0.2cm}
\caption{ \small
Strangeness phase space occupancy {\bf a}) as function of initial fireball
size $R_{\rm in}$ assuming initial conditions
of the zero impact parameter 158A GeV Pb--Pb collisions; {\bf b}) as
function of the CM-specific ion collision energy content, assuming
a $R_{\rm in}=5$ fm initial fireball size and freeze-out at 140 MeV.
\label{gammasr}}
\end{figure}

To study the dependence on the impact parameter on strangeness
saturation, we vary the magnitude of the initial fireball size 
$R_0$. From geometric considerations one finds roughly the 
relation between the impact parameter in Pb--Pb collisions, $b$
and $R_{\rm in}$ to be $R_{\rm in}\simeq 6-b/2>0$ fm; for
small impact parameters $0<b<2$ fm we assume here formation of a
`standard' fireball of 5 fm radius.  A further 
assumption is needed regarding initial temperature of the
fireball: we will not vary this parameter, leaving it for the Pb--Pb
collisions at $T=320$ MeV for all fireball sizes. However, for larger
impact parameters (small fireball sizes) the actual momentum 
stopping is reduced and thus the heating and compression of the
fireball is less than we have implicitly assumed using a
constant value for $T_{\rm in}$ for different initial fireball
volumes. It is
impossible for us to improve on this hypothesis here, since this 
requires the understanding of the hadronic matter stopping as
function of the amount of hadronic matter involved.
With this set of initial conditions we integrate the dynamical
equation (\ref{dgdtf}) for $\gamma_{\rm s}$ up to final
temperature $T\simeq 140$ MeV (see discussion below in section
\ref{hadromod})  for the 200A GeV S--W/Pb collisions. We find that full
strangeness phase space saturation occurs for fireballs 
with a radius $R_{\rm in}>4$ fm, which
includes impacts parameters $b$ up to about 3-4 fm. This result
suggests that there is no need to trigger onto very central
collisions in order to observe $\gamma_{\rm s}\simeq 1$. Moreover,
The relatively sudden onset of the phase space saturation seen in 
Fig.\,\ref{gammasr}{\bf a} as function of fireball size is 
very probably even more sudden, had we incorporated the changing
stopping related to the change of volume. 

It is also most interesting to study how $\gamma_{\rm s}$ 
depends on the Pb--Pb collision energy. We
obtain this result by varying the initial fireball temperature
$T_{\rm in}$ and relating this value to the specific energy
content in the fireball by the results given
in Fig.\,\ref{fig1S95} --- we take here the result we obtained 
for full stopping $\eta=1$\,. Recall that baryon and energy
stopping being equal, 
\baselineskip=12.9pt
$E/B=8.6$ GeV corresponds to 158A
GeV Pb--Pb collisions. For each initial temperature $T_{\rm
ch}$ we assume that the initial value of $\gamma_{\rm s}$ is
0.15, and integrate the temporal evolution of $\gamma_{\rm s}$,
Eq.\,(\ref{dgdtf}), till the final temperature which is taken 
for all collision energies to be at $T=140 $ MeV. 
As shown in Fig.\,\ref{gammasr}{\bf b} for the full SPS range 
$4.3<E/B<8.6$ GeV we find as expected fully
saturated phase space, with $0.8<\gamma_{\rm s}<1.1$\,. 
Between the AGS $E/B=2.6$ GeV and CERN energies $\gamma_{\rm s}$
increases from $0.45$ to 0.85. We recall that our study of the
S--Pb collision system suggests somewhat more effective chemical
equilibration, thus the small variation of $\gamma_{\rm s}$ 
with energy reported here may be even less pronounced. On the
other hand this small variation impacts the
final particle yields as we shall see in section \ref{results},
in that it makes relative yields of strange antibaryons such as
$\overline{\Lambda}/\bar p,\ \overline{\Xi}/\overline{\Lambda}$
nearly independent of collision energy. 

We can now briefly return to the discussion of the result of the
NA35 collaboration\cite{NA35pbar} shown in Fig.\,\ref{lbpbNA35}:
despite the large error bar it is noticeable that there is a
tendency for the $\overline{\Lambda}/\bar p$-ratio to increase 
as the  collision system becomes smaller.
This can be interpreted in terms of $\gamma_{\rm s}$ and one
finds the normally unexpected result that while S--Au collisions
lead to $\gamma_{\rm s}\simeq 0.8\pm0.2$, the S--S collisions
may require a greater value $\gamma_{\rm s}\simeq 1.2\pm0.3$\,. 
In the earlier analysis\cite{analyze} of S--S data (excluding 
$\bar p$) this tendency towards $\gamma_{\rm s}\simeq 1$ was
also found, while the
S--W/Pb results always invariably lead to $\gamma_{\rm s}\simeq
0.75$\,. In light of the model calculations done above,
it is not anymore impossible to imagine 
that the combination of initial and disintegration
conditions of these two systems reverses the naive expectations
regarding the final observable values of $\gamma_{\rm s}$, leading 
to a greater value for the smaller system. 
 
\begin{figure}[htb]
\vspace*{1.3cm}
\centerline{\hspace*{-.3cm}
\psfig{width=7cm,figure=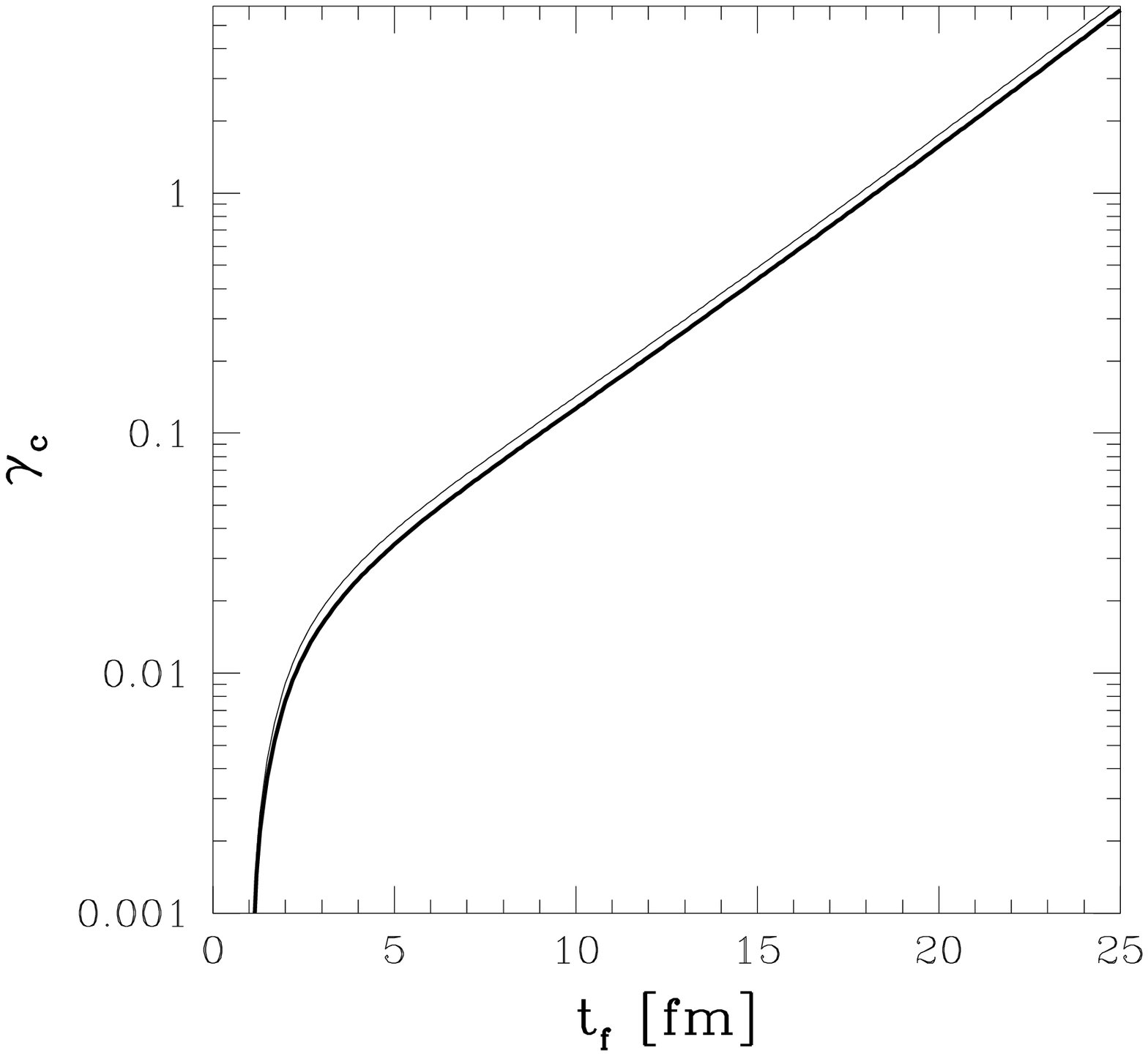}\hspace*{-1.1cm}
\psfig{width=7cm,figure=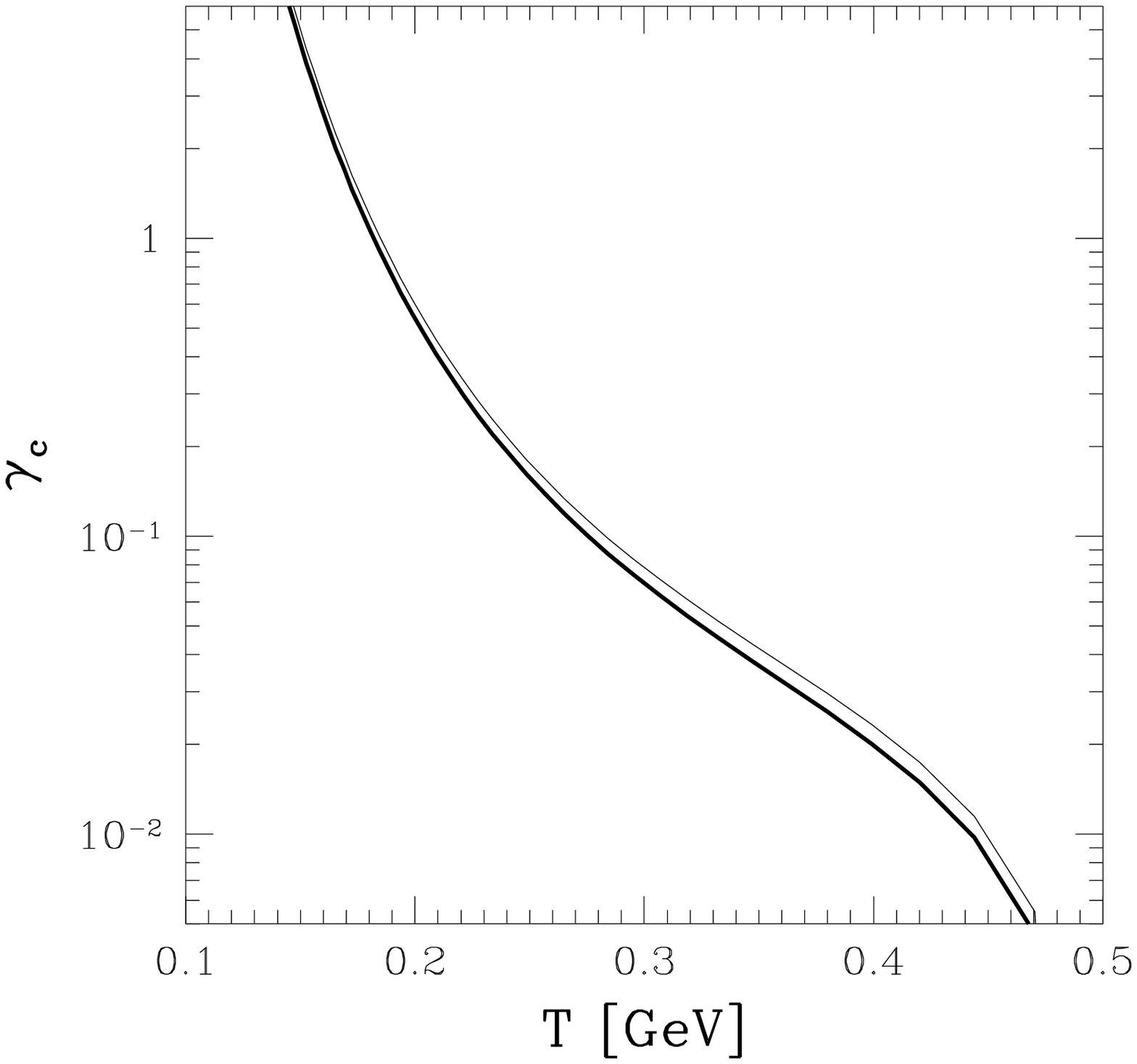}
}
\vspace*{-0.2cm}
\caption{ \small
QGP phase charm phase space occupancy $\gamma_{\rm c}$ in
central Pb--Pb interactions: {\bf a}) as function time and 
{\bf b}) as function of temperature, for running $\alpha_s$\,.
\label{figvarcgam}}
\vspace*{-0.2cm}
\centerline{\hspace*{-.7cm}
\psfig{width=11.5cm,figure=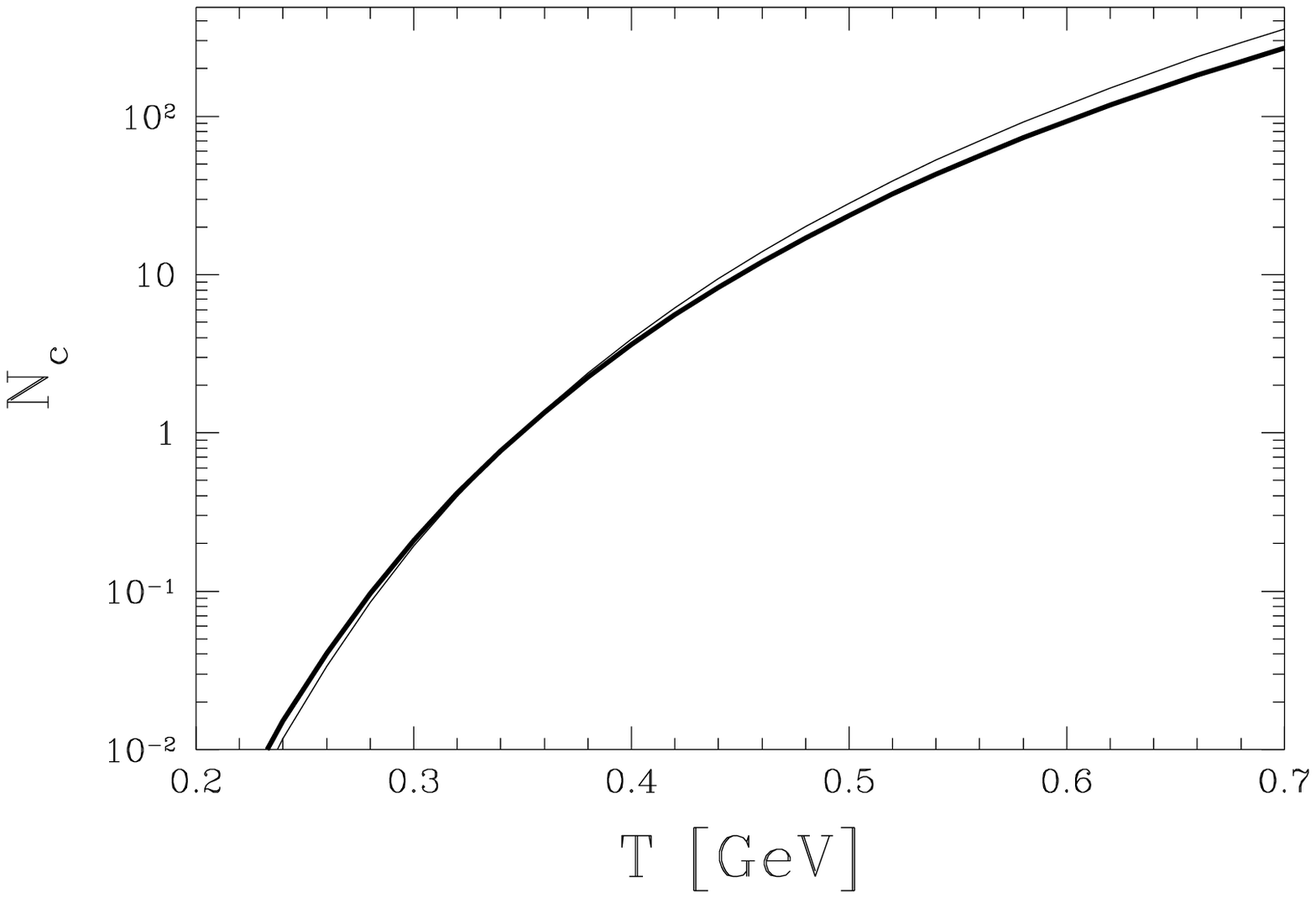}
}
\vspace*{-.6cm}
\caption{ \small 
Total charm production yield in longitudinally expanding QGP as 
function of initial temperature. 
\protect\label{charmtot}}
\end{figure}

We now explore the saturation of the charmed quark phase space
in conditions sensible for the forthcoming RHIC and LHC environments. We
consider the temporal evolution for the initial temperature 500
MeV. Due the to likely dominance of the expansion by the longitudinal flow
we take for the adiabatic condition the relation $LT^3=$ Const.
We take that $L$ expands with light velocity. As can be seen in
the results shown in Fig.\,\ref{figvarcgam} thermal production
of charm is small, being very slow, but because the freeze-out temperature
should here also be taken in the vicinity of 150 MeV, the phase
space occupancy reaches a stunning value {\em exceeding unity}. 
Still greater
values result for higher initial temperatures and/or lower
freeze-out temperatures. We note that the discussion of thermal
charm production in this language makes only sense if the number
of charmed quark pairs produced in the initial moments is 
considerably greater than unity. We thus present in 
Fig.\,\ref{charmtot} the pair yield as function of initial temperature
(assuming small impact parameter collisions). We note that our
calculations apply to initial temperatures above 300 MeV and
that for $T_{\rm in}=500$ MeV we would be reaching a yield of
twenty charm quark pairs per event. 
We also note that the difference between the thick and thin lines in 
Figs.\,\ref{figvarcgam}, \ref{charmtot}, indicating that 
at the energy scale of charm production the uncertainty arising from the
error in $\alpha_{\rm s}(M_Z)$ is negligible. There remains considerable 
uncertainty in the evolution model of the fireball, which in the here 
presented scenario would evolve for up to 25 fm/c.
 
\begin{figure}[ptb]
\vspace*{1.7cm}
\centerline{\hspace*{-.5cm}
\psfig{width=9cm,figure=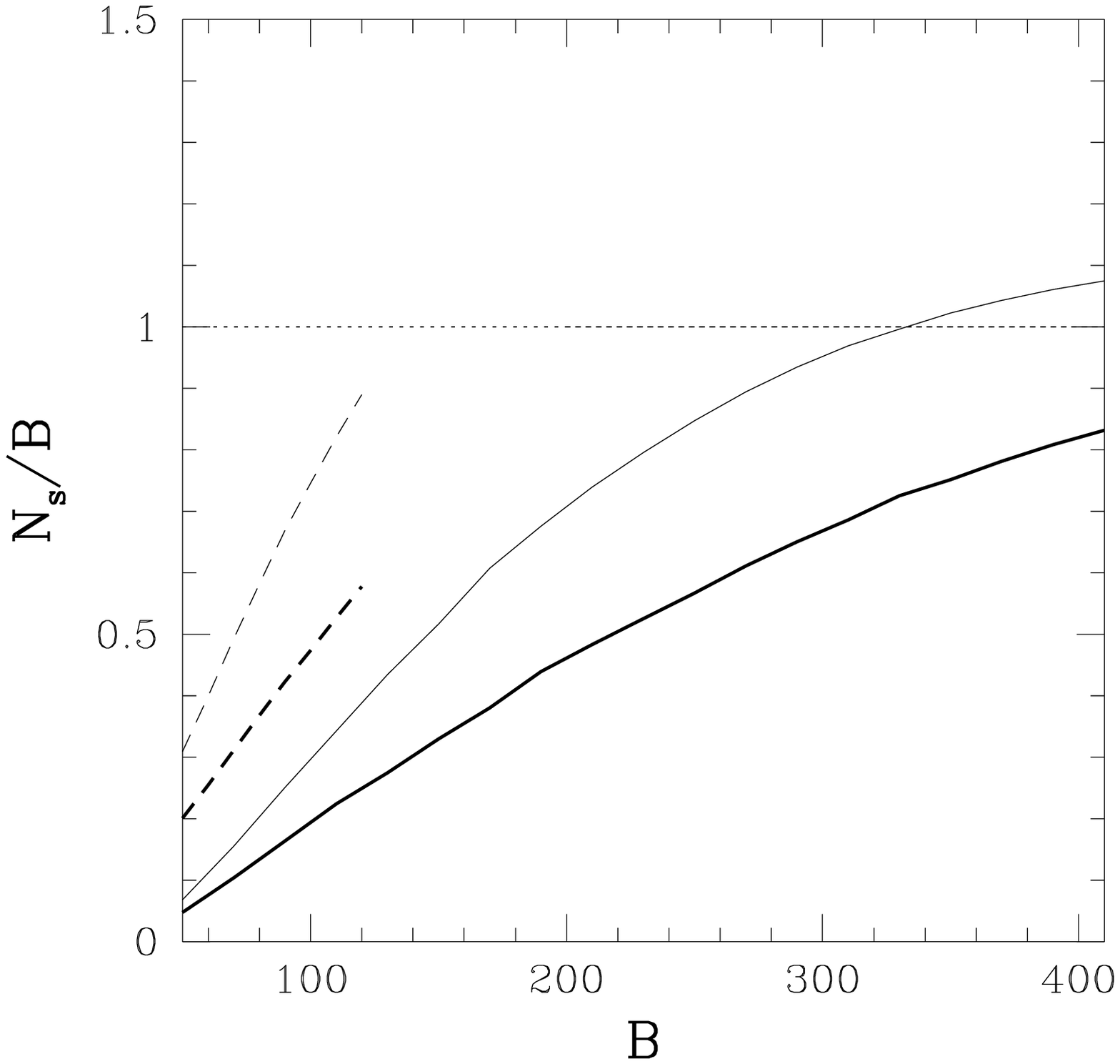}
}
\vspace*{-0.3cm}
 \caption{ \small
QGP fireball specific strangeness abundance  as function of
number of participants (solid lines for Pb--Pb, dashed lines for S--W) 
for running  $\alpha_{\rm s}$ and $m_{\rm s}$.
The freeze-out point is fixed at $T_{\rm f}=140$ MeV. 
Thick lines for the smaller
$\alpha_{\rm s}$ option, thin lines for the larger option.
\label{figsbrunB}}
\vspace*{2.7cm}
\centerline{\hspace*{-.5cm}
\psfig{width=9cm,figure=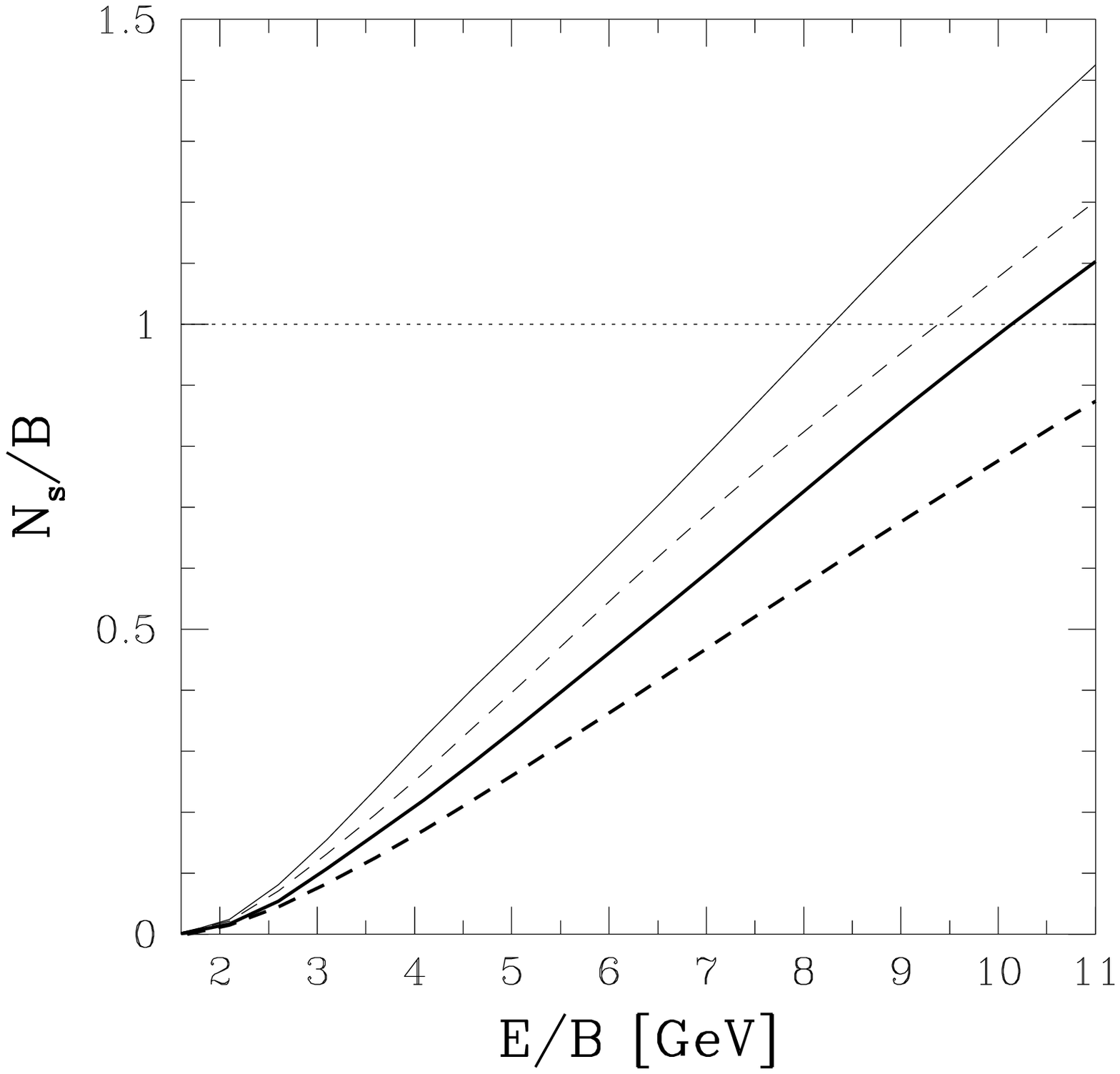}
}
\vspace*{-0.2cm}
 \caption{ \small
QGP fireball specific strangeness abundance  as function of energy per baryon
in the fireball $E/B$. See Fig.\,\protect\ref{figsbrunB} for
definitions.
\label{figsbrunE}}
\end{figure}

Let us consider next the running-QCD case and present a series of 
results for the two strangeness observables $N_{\rm s}/B$ and
$\gamma_{\rm s}(t_{\rm f}$ as function of the energy and impact 
parameter \cite{impact}. 
We present in Fig.\,\ref{figsbrunB} the ratio
$N_{\rm s}/B$ as function of $B$ and in  Fig.\,\ref{figsbrunE} 
as function of $E/B$. Solid lines give results for
the Pb--Pb collisions, dashed lines correspond to the
S--Pb/W case. Thick/thin lines as before refer to small/large
$\alpha_{\rm s}$ options. In Fig.\,\ref{figsbrunB} we see
that the specific yield is expected to be 20--40\% higher in 
{\it central\/} Pb--Pb collisions than 
in {\it central\/} S--Pb/W collisions --- the effect is smaller for the
`large' $\alpha_{\rm s}$ case since we are closer to the
saturation of the phase space in the early collision stage. As
Fig.\,\ref{figsbrunE} shows, the specific yield of
strangeness is in a wide range of CM-energies, comprising all
the accessible SPS-range, nearly proportional to $E/B$. This
pattern arises from a number of factors, such as the change in
initial temperature with collision energy, the change of
strangeness production with temperature, etc. It would be quite
surprising to us, if other reaction models without QGP would find this
linear behavior with similar coefficients. We therefore believe that this
result is an interesting characteristic feature of our QGP thermal
fireball model.

\begin{figure}[h]
\vspace*{1.3cm}
\centerline{\hspace*{-.5cm}
\psfig{width=7cm,figure=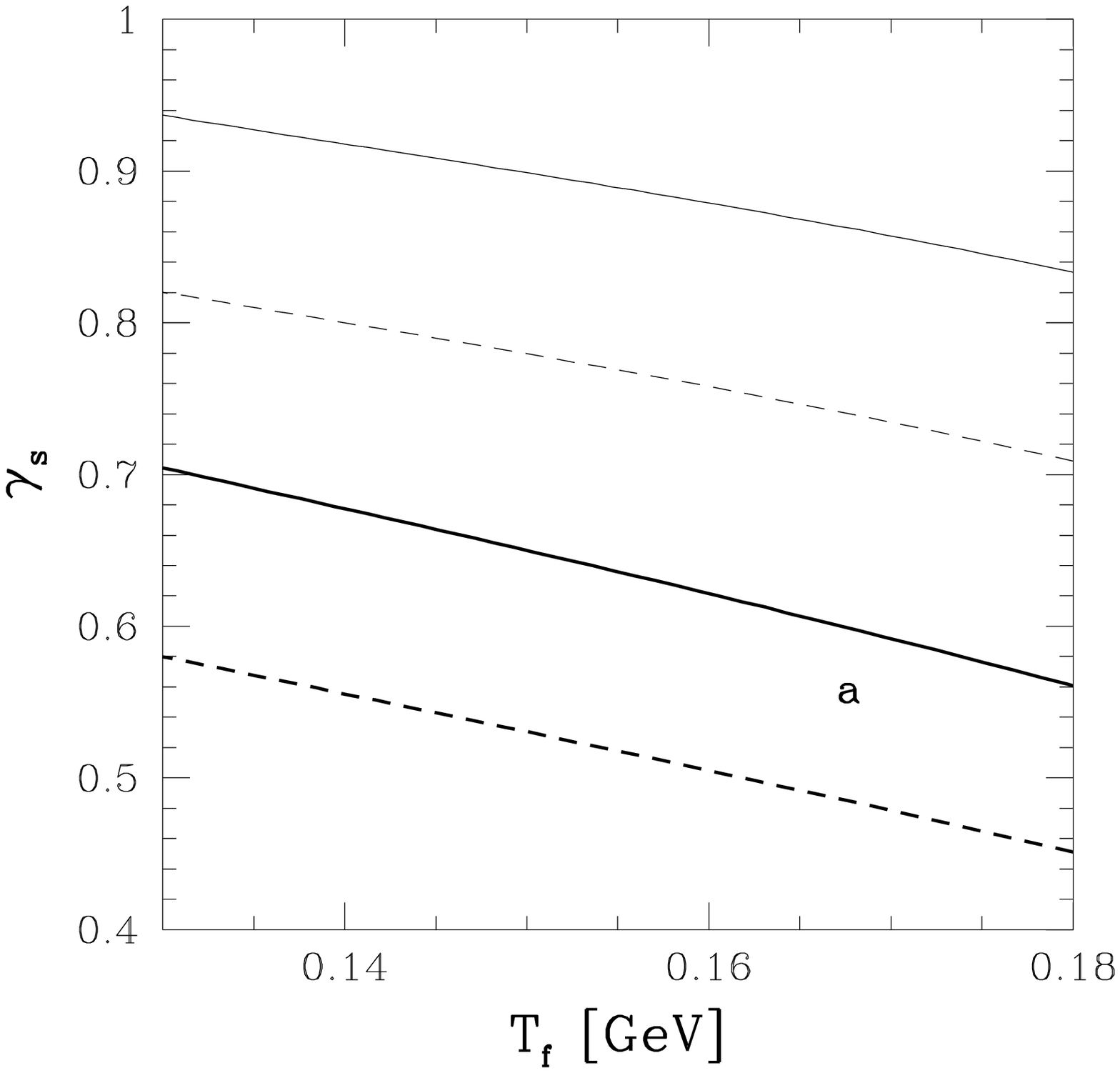}\hspace*{-.7cm}
\psfig{width=7cm,figure=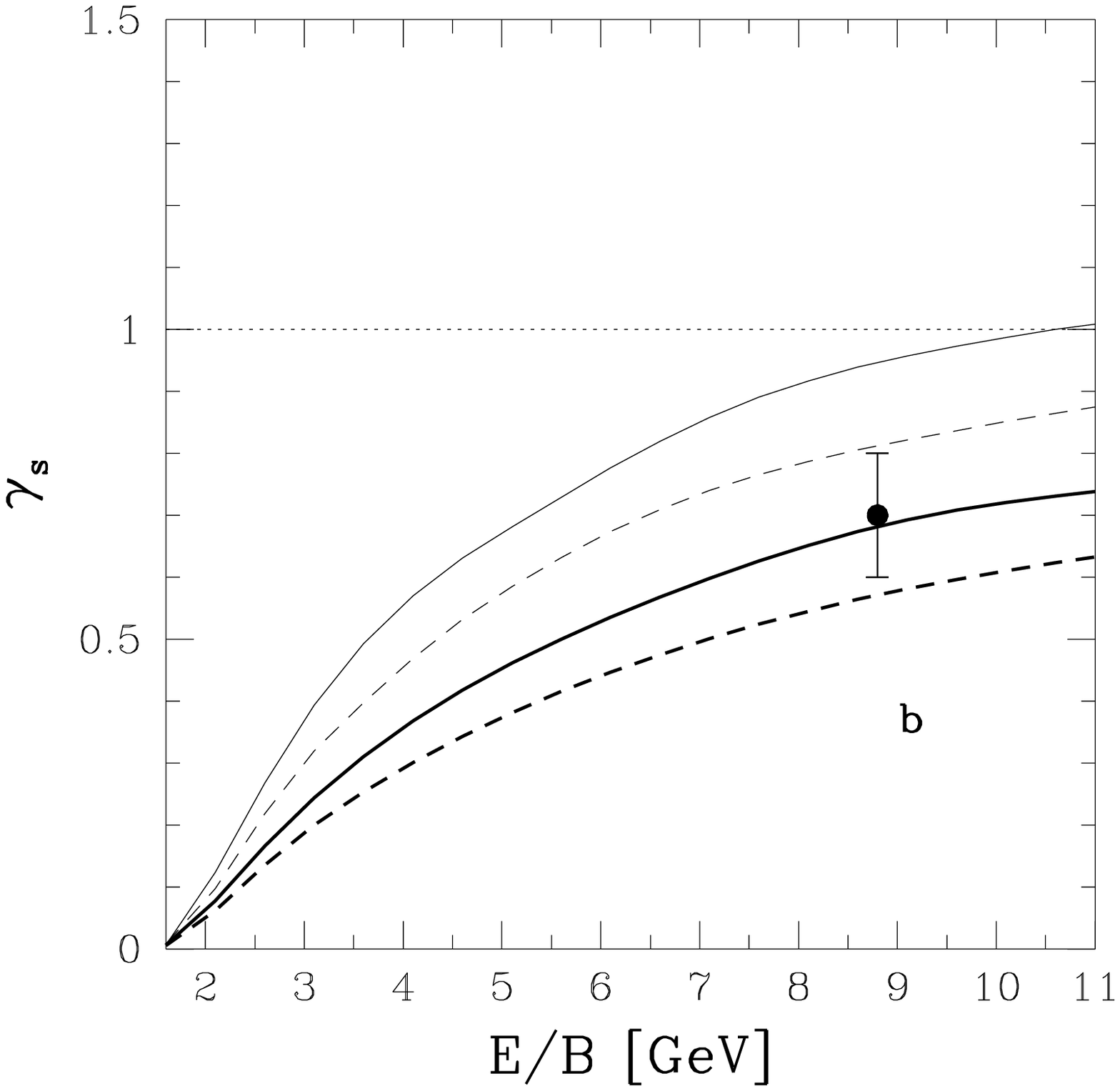}
}
\vspace*{0cm}
\caption{ \small
QGP strangeness phase space occupancy $\gamma_{\rm s}$
{\bf a)} as function of  freeze-out temperature $T_{\rm f}$
and {\bf b)} as function of energy per baryon 
$E/B$ in the fireball. 
Lines according to convention in Fig.\,\protect\ref{figsbrunB}.
Initial conditions in {\bf a)} chosen to correspond to Pb--Pb collisions
at maximum SPS energies with $E/B=8.6$ GeV and $B\simeq 380$; in {\bf b)}
we keep the number of constituents at $B=380$  in  Pb--Pb 
and $B=120$ in S--W and fixed freeze-out
point at $T_{\rm f}=140$ MeV. The vertical bar corresponds to the value of 
$\gamma_{\rm s}$  obtain in S--W data analysis \protect\cite{analyze}. 
\label{figvarefgam}}
\end{figure}

The value of $\gamma_{\rm s}$ observed by studying the final
state hadrons, reflects on the initial production and the enrichment of
the strange phase space occupancy by the dilution effect. Consequently it
depends, in addition to the dependences we saw in strangeness abundance,
sensitively on the freeze-out temperature. High values of
$\gamma_{\rm s}$ could accompany low freeze-out temperature,
provided that there has been extreme initial conditions allowing
to reach strangeness phase space saturation long before hadronization.
When considering the Pb--Pb collisions at presently explored
energy of 158A GeV ($E/B\simeq 8.6$ GeV), we are primarily interested to
know if it is likely that we reach $\gamma_{\rm s}=1$. 
In Fig.\,\ref{figvarefgam}{\bf a} we present
the evolution of $\gamma_{\rm s}$ as function of freeze-out
temperature in the `reasonable' range $130<T_{\rm f}<180$ MeV. 
We assumed here (solid lines) that the fireball was formed in
Pb--Pb collisions at nearly zero impact parameter, hence it
comprises $B\simeq 380$ participants and that the
energy available for collisions of projectiles with 158A GeV
is $8.6$ GeV, as is the case should the energy
and baryon number stopping be the same. Dashed lines correspond
to the 200A GeV S--Pb/W collisions leading to a smaller baryon
content $B=120$ in central collisions, thus also to smaller
$\gamma_{\rm s}$ at the same freeze-out temperature. As before,
thin lines are for the `large' $\alpha_{\rm s}$ case, and thus
the presented results are systematically ($\Delta \gamma_{\rm
s}\simeq 0.25$) greater than the
thick line results. Considering that the freeze-out temperature
is not cut in stone as well, this is the typical magnitude of
the theoretical uncertainty of the present theoretical evaluation of
$\gamma_{\rm s}$. 
 
In Fig.\,\ref{figvarefgam}{\bf b} 
we reversed the roles of the variables $E/B$ and $T_{\rm
f}$: we set freeze-out temperature $T_{\rm f}=140$ MeV and consider a zero
impact parameter collision with $B=380$ (solid lines) and $B=120$ (dashed
lines) as function of energy content $E/B$. The 
result (dot with vertical bar) we give $(0.75\pm 0.1)$ arises
from our analysis \cite{analyze} of experimental data for 200A
GeV S--W/Pb collisions. It is remarkably consistent with our
model results. It may be unwise to view its better agreement with
the `large' $\alpha_{\rm s}$ option as an effective measurement
of $\alpha_{\rm s}$: there are too many uncertainties to be
refined before we could use strangeness as a measure of the QCD
coupling strength. However, our results clearly show that we
have the required sensitivity to relate properties of QCD and
the perturbative vacuum to the particle yields observed in
relativistic nuclear collisions. 

\section{QGP hadronization} 
\label{hadromod}
\subsection{Hadronization constraints}
\baselineskip=13.4pt
It is easy to imagine QGP-hadronization mechanisms that would
largely erase memory of the transient deconfined phase. We will not 
discuss such {\it re-equilibrating} hadronization models
\cite{reequilib} of strange particles which are not observed at
least at SPS energies \cite{analyze}. 
Instead, we shall focus our attention on the alternative that
the particles emerge directly and without re-equilibration 
from the deconfined phase. 

Our approach to hadronization and particle production is
schematic and does not involve development of a 
dynamical model \cite{sudden}.
Instead, we introduce two parameters which describe how far are 
from the hadronic gas equilibrium the produced meson and baryon
abundances. To justify the introduction of these
parameters we note that there is no reason whatsoever to expect that the 
rapid disintegration
of the deconfined state will lead to particle abundances that are
associated with full chemical equilibrium of any individual particle 
species in the final state. These hadron nonequilibrium
constants $C_i$  are in principle different for each 
particle species, but if we presume that the mechanisms that lead to
particle production are similar for all mesons ($i=$ M), and all baryons
($i=$ B), we can group the hadronic particles into these two families,
keeping just two unknown quantities. Note that also the
{\it relative} abundances of mesons and baryons emerging from hadronizing QGP
are difficult to equilibrate, because processes which convert meson into
baryon---antibaryon pairs are relatively slow. The magnitude of
these abundance coefficients $C_i$ is determined theoretically
by the need to conserve or
increase entropy, conserve baryon number and strangeness in the 
hadronization process.

We now consider strangeness conservation in the final state:
the abundances of the final state strange particles can be gauged by
considering the Laplace transform of the phase space distribution
which leads to a partition function {\it like} expression ${\cal Z}_{\rm
s}$.  The individual components comprise aside of the chemical
factors $\lambda_{\rm q},\,\lambda_{\rm s}$ the non-equilibrium 
coefficients $\gamma_{\rm s}$ and  $ C_{\rm B}^{\rm s}$, 
$ C_{\rm M}^{\rm s}$ (we have added here the superscript `s' to
the factors $C$ since at present we look only at strange particles):  
\begin{eqnarray}
\ln{\cal Z}_{\rm s} = { {V T^3} \over {2\pi^2} }
\left\{(\lambda_{\rm s} \lambda_{\rm q}^{-1} +
\lambda_{\rm s}^{-1} \lambda_{\rm q}) \gamma_{\rm s} C^{\rm s}_{\rm M}
F_K +(\lambda_{\rm s} \lambda_{\rm q}^{2} +
\lambda_{\rm s}^{-1} \lambda_{\rm q}^{-2}) \gamma_{\rm s} C^{\rm
s}_{\rm B} F_Y \right.\nonumber \\ 
\left.+ (\lambda_{\rm s}^2 \lambda_{\rm q} +
\lambda_{\rm s}^{-2} \lambda_{\rm q}^{-1}) \gamma_{\rm s}^2 C^{\rm
s}_{\rm B}  F_\Xi + (\lambda_{\rm s}^{3} + \lambda_{\rm s}^{-3})
\gamma_{\rm s}^3 C^{\rm s}_{\rm B} F_\Omega\right\}\, , 
\label{4a}
\end{eqnarray}
where the kaon, hyperon, cascade and omega 
degrees of freedom are included. Here $T$ is the freeze-out temperature.
The phase space factors $F_i$ of the strange particles are (with $g_i$
describing the statistical degeneracy):
\begin{eqnarray}
F_i&=&\sum_j g_{i_j} W(m_{i_j}/T)
\, .
\label{FSTR}
\end{eqnarray}
In the resonance sums $\sum_j$ all known strange hadrons should be
counted.  
\baselineskip=12.9pt
The function $W(x)$ arises from the phase-space integral of the
different particle distributions $f(\vec p)$. For the Boltzmann particle
phase space (appropriate when the final state mass is equal or greater than 
the temperature of the source) and when the integral includes the entire
momentum range, we have
\begin{equation}\label{therspec}
W(x)\equiv (4\pi)^{-1}\int d^3(p/T)f(\vec p)=x^2{\rm K}_2(x)\,,
\end{equation}
where as before $x=m/T$ and ${\rm K}_2(x)$ is the modified Bessel function.

There is a strong constraint between the two fugacities
$\lambda_{\rm q}$, and $\lambda_{\rm s}$ arising from the requirement of
strangeness conservation among the final state particles, 
which was discussed at length recently \cite{analyze}.
The non-trivial 
relations between the parameters characterizing the
final state are in general difficult to satisfy and the resulting
particle
distributions are constrained in a way which  differs considerably
between different reaction scenarios which we have considered in detail:
the rapidly disintegrating QGP or the equilibrated HG phase. These two
alternatives differ in particular by the value of the
strange quark chemical potential $\mu_{\rm s}$:
\begin{itemize}
\vskip -0.3cm 
\item[1.] In a strangeness neutral QGP fireball $\mu_{\rm s}$
is always exactly zero, independent of the prevailing temperature and
baryon
density, since both $s$ and $\bar s$ quarks have the same phase-space
size. 
\vskip -0.3cm 
\item[2.] In any state consisting of locally confined hadronic
clusters, $\mu_{\rm s}$ is generally different from zero at finite baryon
density, in order to correct the asymmetry introduced in the phase-space
size by a finite baryon content.
\end{itemize}
At non-zero baryon density, that is for $\mu_{\rm B}\equiv 3\mu_{\rm
q}\ne 0$,  there is just one (or perhaps at most a few) special value
$\mu_{\rm B}^0(T)$ for which $\langle s \rangle = \langle \bar s \rangle$
at $\mu_{\rm s}^{\rm HG}=0$, which condition mimics the QGP. We 
have studied these values carefully \cite{analyze} for the final
state described by  Eq.\,(\ref{4a}): here the condition of 
strangeness conservation takes the simple 
analytical form \cite{analyze,entropy}:
\begin{eqnarray}
\mu_{\rm q}^0=T{\cosh}^{-1}\left(R^{\rm s}_{\rm C}{F_{\rm K}\over
2F_{\rm Y}} -\gamma_{\rm s} {F_{\Xi}\over F_Y}\right),\quad \mbox{for}\
\mu_{\rm s}^{\rm HG}=0\, . \label{zero}
\end{eqnarray}
Here, and when we consider relative abundance of particles, only the ratio
\begin{equation}\label{RsC}
R^{\rm s}_{\rm C}=C^{\rm s}_{\rm M}/C^{\rm s}_{\rm B}
\end{equation} appears. We note that there is at most one
non-trivial real  solution of Eq.\,(\ref{zero}) for monotonous arguments 
of $\cosh^{-1}$, and only when this argument is greater than unity.   
 
Clearly, the observation \cite{analyze,Heinzy} of $\lambda_{\rm s}=1\ 
(\mu_{\rm s}=0)$  is, in view of the accidental nature of this value  in
the confined phase, a
rather strong indication for the direct formation of final state hadrons 
from a deconfined phase. In such a process the particle
abundances retain memory of the chemical (fugacity) parameters,
the conservation of strangeness and other properties is assured
by the (non-equilibrium hadronic gas) abundance numbers of the 
particles produced. For example the number
of baryons emitted even at very low temperatures must remain
conserved and thus cannot be tiny despite the thermal
suppression factor $e^{-m/T}$ --- a big change in chemical
potentials would require lengthy reequilibration. 
These effects are absent since $\lambda_{\rm s}=1$, 
at least in the strangeness
chemical potential: for
the S--W/Pb collisions at 200A GeV this was found already in the first data
analysis\cite{Raf91} and this remarkable result was corroborated
by an  extensive study of the resonance decays and flow 
effects\cite{analyze}. For the S--S collisions at 200A GeV  
a further refinement  \cite{Heinzy} which
allows for a rapidity dependence of $\lambda_{\rm q}$ due to flow 
further underpins the finding $\lambda_{\rm s}=1$.

We can thus safely conclude that strange particles 
produced in 200A GeV Sulphur interactions with diverse targets indicate a
particle source which displays a symmetry in phase space size of strange
and antistrange particles, which fact is more than just an accident of
parameters considering that it appears for two widely different
collision systems, S--S and S--W/Pb. A natural explanation is
that such  a source
is deconfined, and that it disintegrates so rapidly, that its
properties remain preserved in emitted strange particles. It
will be very interesting to see,
if this behavior will be confirmed in the Pb--Pb system, with
present experiments operating at 158A GeV and possibly later at different
collision energies.
 
\begin{figure}[ptb]
\vspace*{1.7cm}
\centerline{\hspace*{-.3cm}
\psfig{width=10cm,figure=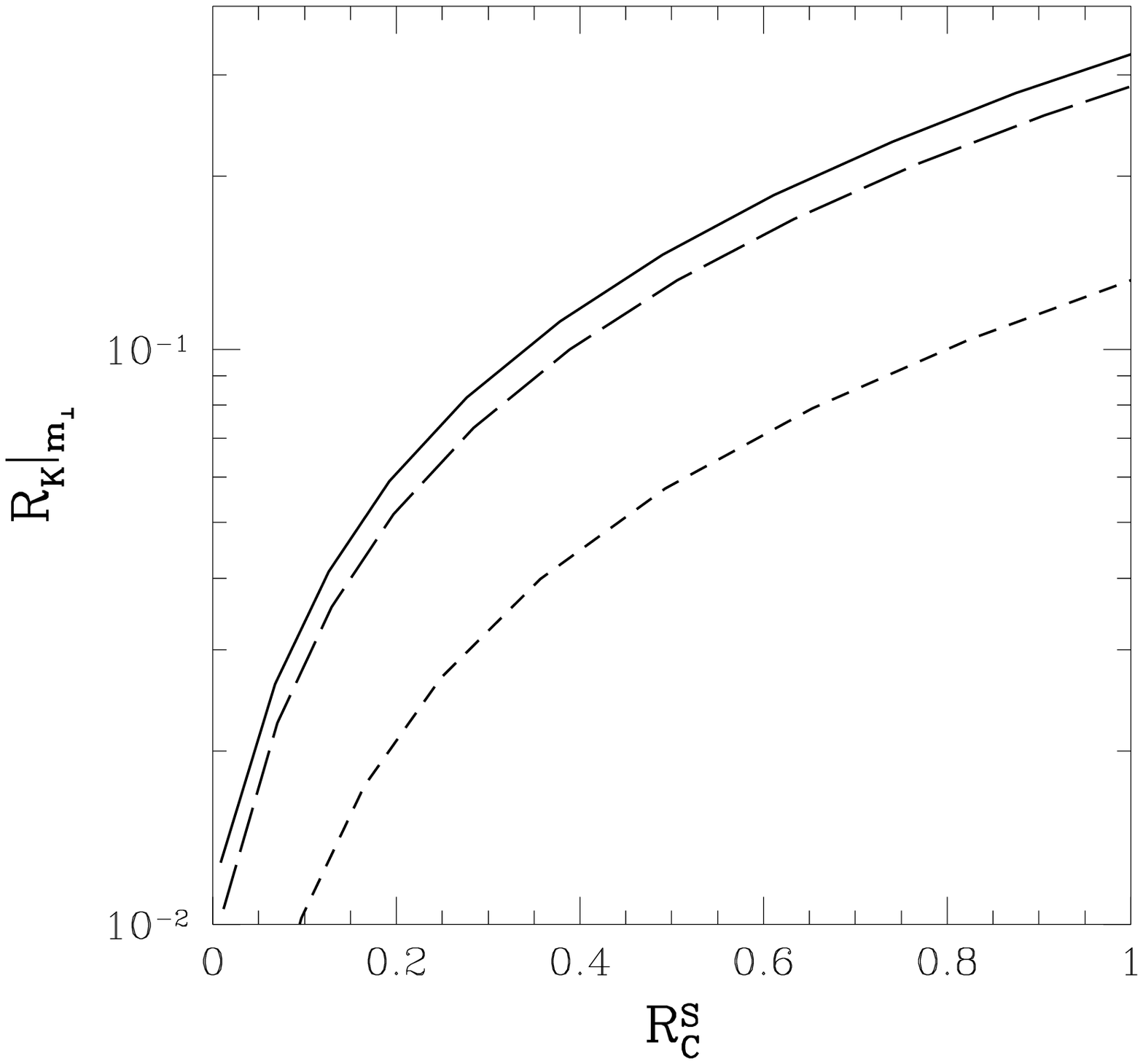}
}
\vspace*{-.2cm}
\caption{ \small 
$R_{\rm K}\vert_{m_\bot}$ as function of 
$R_{\rm C}^{\rm s}$ for $\lambda_{\rm q}=1.5$ (solid line), 
$\lambda_{\rm q}=1.6 $ (long-dashed line) and 
$\lambda_{\rm q}=2.5$ (short-dashed line).
\protect\label{Rcfreeze}}
\vspace*{2.3cm}
\centerline{\hspace*{-.3cm}
\psfig{width=10cm,figure=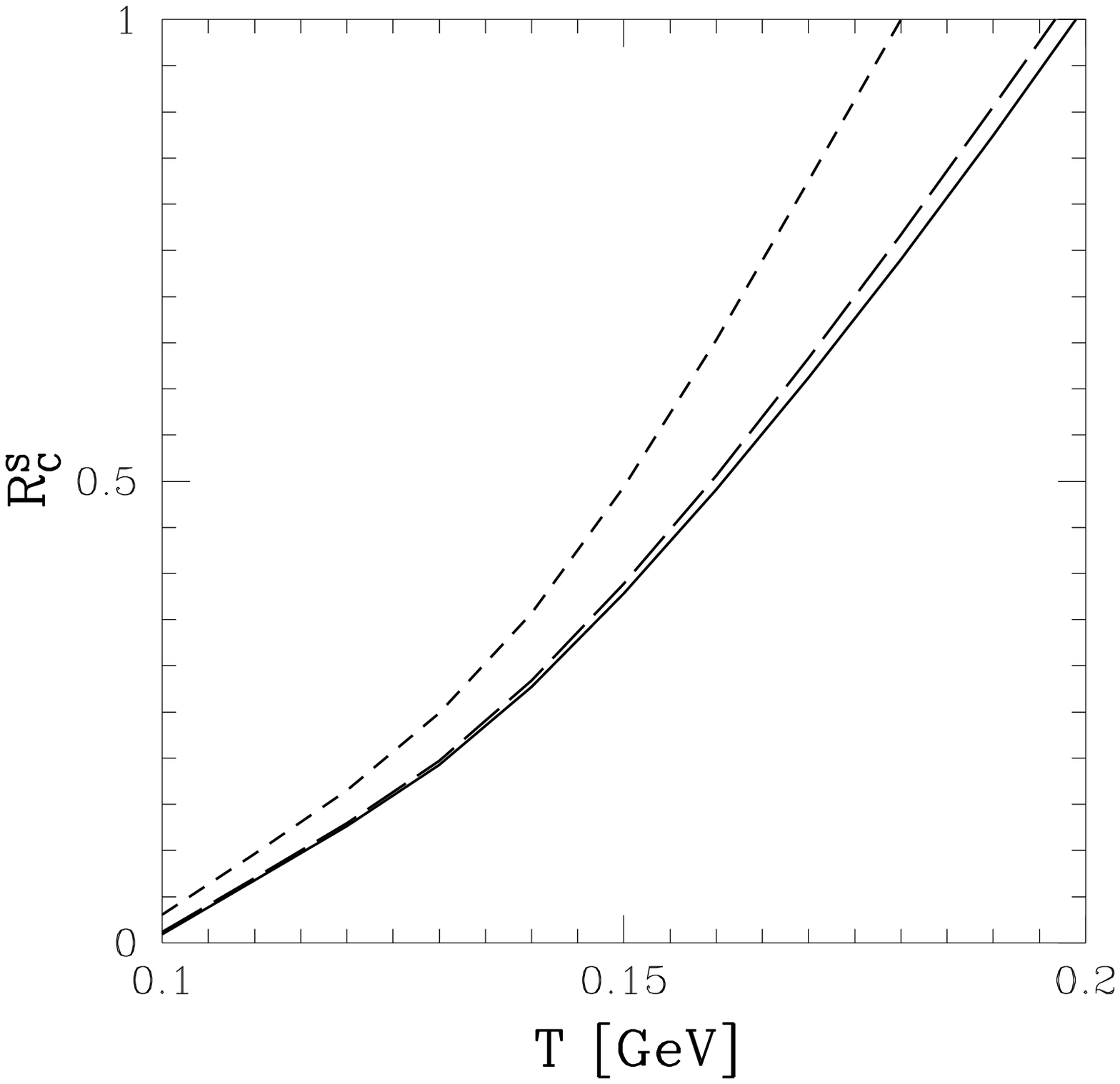}
}
\vspace*{-.2cm}
\caption{ \small 
Strangeness neutrality line: $R_{\rm C}^{\rm s}$ versus freeze-out 
temperature for $\lambda_{\rm
q}=1.5$ (solid line), $\lambda_{\rm q}=1.6 $ (long-dashed line) and 
$\lambda_{\rm q}=2.5$ (short-dashed line).
 \protect\label{F2freeze}}
\end{figure}

We now explore the values of the parameter $R_{\rm C}^{\rm s}$.
We consider the constraint imposed by Eq.\,(\ref{zero}), taking  
$\gamma_{\rm s}=0.7$ (the deviation from unity is of little
numerical importance), $\lambda_{\rm s}=1$. For $\lambda_{\rm
q}$ we take three values in Fig.\,\ref{F2freeze}: the solid line
is for $\lambda_{\rm q}=1.5$, choice motivated by the case of S--W/Pb
collisions at 200A GeV, the long-dashed line is for 
$\lambda_{\rm q}=1.6$ suitable for the case of Pb--Pb 160A GeV 
collisions; the short-dashed curve is for
$\lambda_{\rm q}=2.5$, the value which our model calculations 
suggest for the 40A GeV collisions (see table~\ref{bigtable}). The
value  $R_{\rm C}^{\rm s}=1$ is found for $T\le 200$ MeV at 
$\lambda_{\rm q}\simeq1.48$--$1.6$. For lower disintegration 
temperatures we would have  $R_{\rm C}^{\rm s}<1$, as shown 
in Fig.\,\ref{F2freeze}.

The physical observable which we find to be primarily 
sensitive to the parameter $R_{\rm C}^{\rm s}$, and to a lesser degree to
the other thermal model parameters, is the  kaon to
hyperon abundance ratio at fixed $m_\bot$:
\begin{eqnarray}
R_{\rm K}\vert_{m_\bot}\equiv{{{\rm K}_{\rm S}^0}\over
	{\Lambda+\Sigma^0}}\,.
\label{RK0}
\end{eqnarray}
When computing this ratio, we have incorporated the decay
pattern of all listed resonances numerically and included the 
descendants of strong and weak decays in order to
facilitate comparison below with experimental data. 
In 
Fig.\,\ref{Rcfreeze} we show $R_{\rm K}\vert_{m_\bot}$  as
function of $R_{\rm C}^{\rm s}$ for $\lambda_{\rm
q}=1.5,\,1.6,\,2.5$\,, with the same line conventions as in the 
Fig.\,\ref{F2freeze}. We assumed that the distribution of parent
particles for kaons  and hyperons is according to the thermal 
equilibrium condition evaluated at temperature implied by 
Fig.\,\ref{F2freeze}. 

There is no officially reported value for the $R_{\rm K}$ ratio. However,
WA85 collabora\-tion\cite{WA85T} has presented results for the yields of 
$\Lambda,\,\overline{\Lambda}$ and ${\rm K}_{\rm S}$ obtained in
S--W collisions at 200A GeV, shown here in Fig.\,\ref{specWA85},
in the interval  $1.1<m_\bot<2.6$ GeV for the central
rapidity region $2.5<y<3$\,.  No cascading corrections were applied 
to these experimental results. From these results we obtain
$R_{\rm K}\vert_{m_\bot}=0.11\pm0.02$. This implies 
a far off-HG-equilibrium result $R_{\rm C}^{\rm s}=0.38$ as can
be seen in Fig.\,\ref{Rcfreeze}, which according to
Fig.\,\ref{F2freeze} leads to a freeze-out temperature 
$T_{\rm f}\simeq 145$ MeV. The equilibrium HG source with  
$R_{\rm C}^{\rm s}\simeq 1$ ($R_{\rm K}\vert_{m_\bot}\simeq0.3$)
is experimentally  completely excluded. The factor $R_{\rm C}^{\rm s}\ne
1$ confirms the expectation that these strange particles
are produced in non-equilibrium processes --- in our model 
they originate from directly disintegrating QGP fireball. Strangeness
conservation constraint fixes the freeze-out condition at
$T\simeq 145$~MeV.

The final issue is how, from the value  $R_{\rm C}^{\rm s}\simeq0.4$, we can 
infer the values of the abundance constants $C^{\rm s}_{\rm M}$ and
$C^{\rm s}_{\rm B}$ which (see Eqs.\,(\ref{4a}, \,\ref{RsC})) express the 
relative strange meson and baryon production abundance to the thermal 
equilibrium values. If we argue that the strange meson abundance, akin 
to total meson abundance is enhanced by factor two (i.e., 
$C^{\rm s}_{\rm M}=2$) as we found studying the entropy
enhancement \cite{entropy}, 
then the conclusion would be that the strange baryons are enhanced 
(against their tiny HG equilibrium abundance at $T_{\rm f}\simeq
145$ MeV) by the factor $C^{\rm s}_{\rm B}=5$.

We thus see that the hadronization abundance of mesons is
 enhanced by factor 2 and that of baryons by factor $2\cdot 2.5=5$ 
compared  to the yields that would be expected from a 
chemically equilibrated HG phase dissociating at about $T=145$ MeV.
Clearly, one of the important aspects of this result is the 
relation of the meson enhancement factor to entropy production in
heavy ion collision, and we briefly recapitulate the situation
which lead to the expected enhancement of meson yield.

\vspace{-3mm}
\subsection{Entropy content of heavy ion collisions}
\label{entropysec}
\vspace{-1mm}
\baselineskip=12.7pt

One of the fundamental differences  between the QGP and the
conventional Hagedorn type hadron gas (HG) structure of the fireball, is
the specific entropy content per baryon ($S/B$) evaluated at some given
(measured) values of statistical parameters. This entropy content can be
determined in terms of the final state particle multiplicity.
However, in the central rapidity region only few
experiments have been able to obtain a full phase space coverage. Because
of the need to observe relatively small momentum particles, this is a
particularly difficult experimental task and at present the best
experimental access to this issue is by means of emulsion techniques. We
present here one set of preliminary experimental results which give
already a pretty good representation of the specific entropy content in
the fireball.
 
\begin{figure}[ptb]
\vspace*{-0.cm}
\centerline{\hspace*{4.cm}\psfig{height=11.4cm,figure=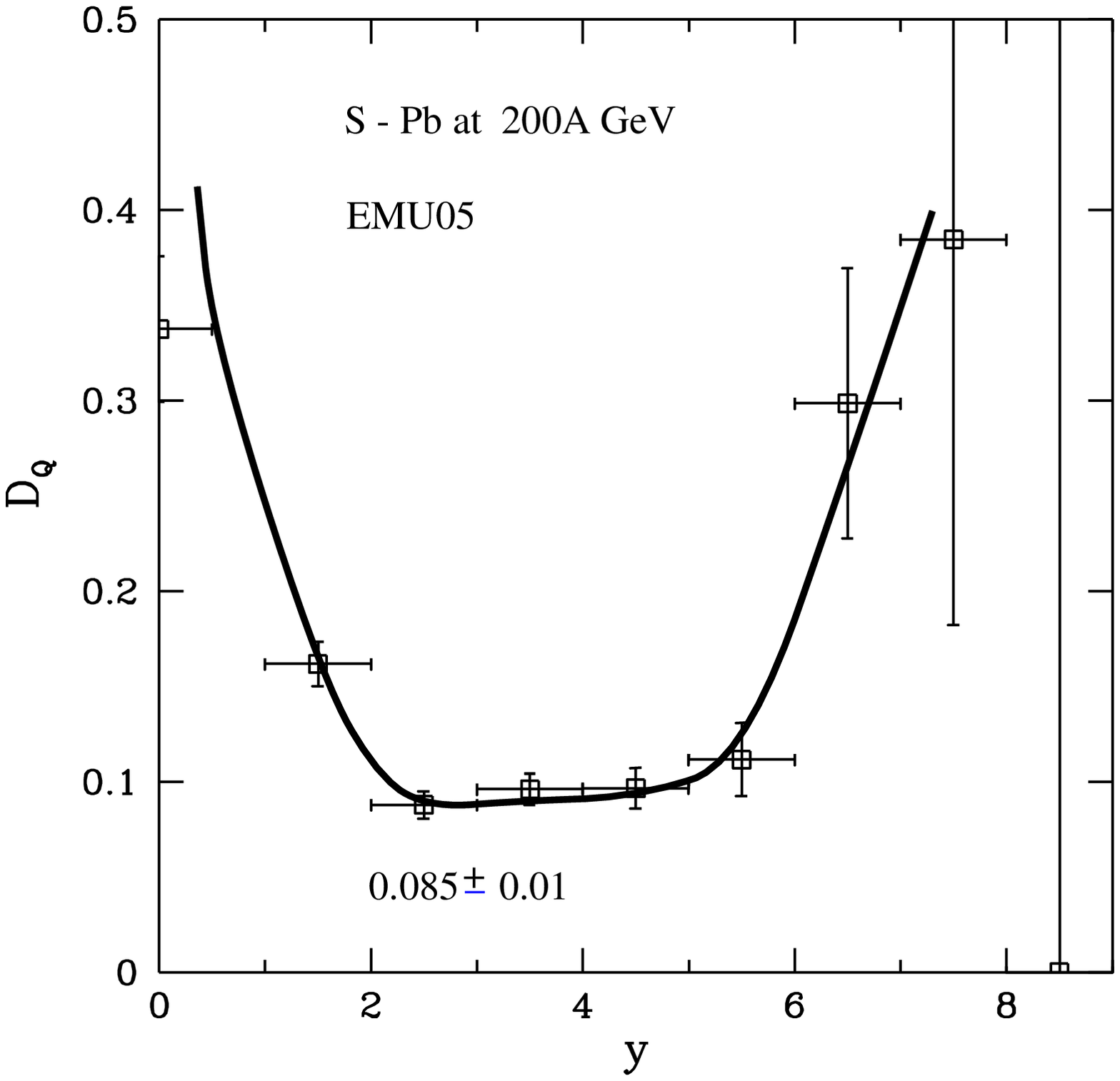}}
\vspace*{-4.1cm}
\caption{ \small 
Emulsion data \protect\cite{EMU05} for the charged particle multiplicity
from central S--Pb collisions at 200A GeV as a function of rapidity: the
difference of positively and negatively charged particles normalized by
the sum of both polarities. 
\protect\label{fig1entro}}
\vspace*{0.5cm}
\centerline{\hspace*{4.cm}\psfig{height=12cm,figure=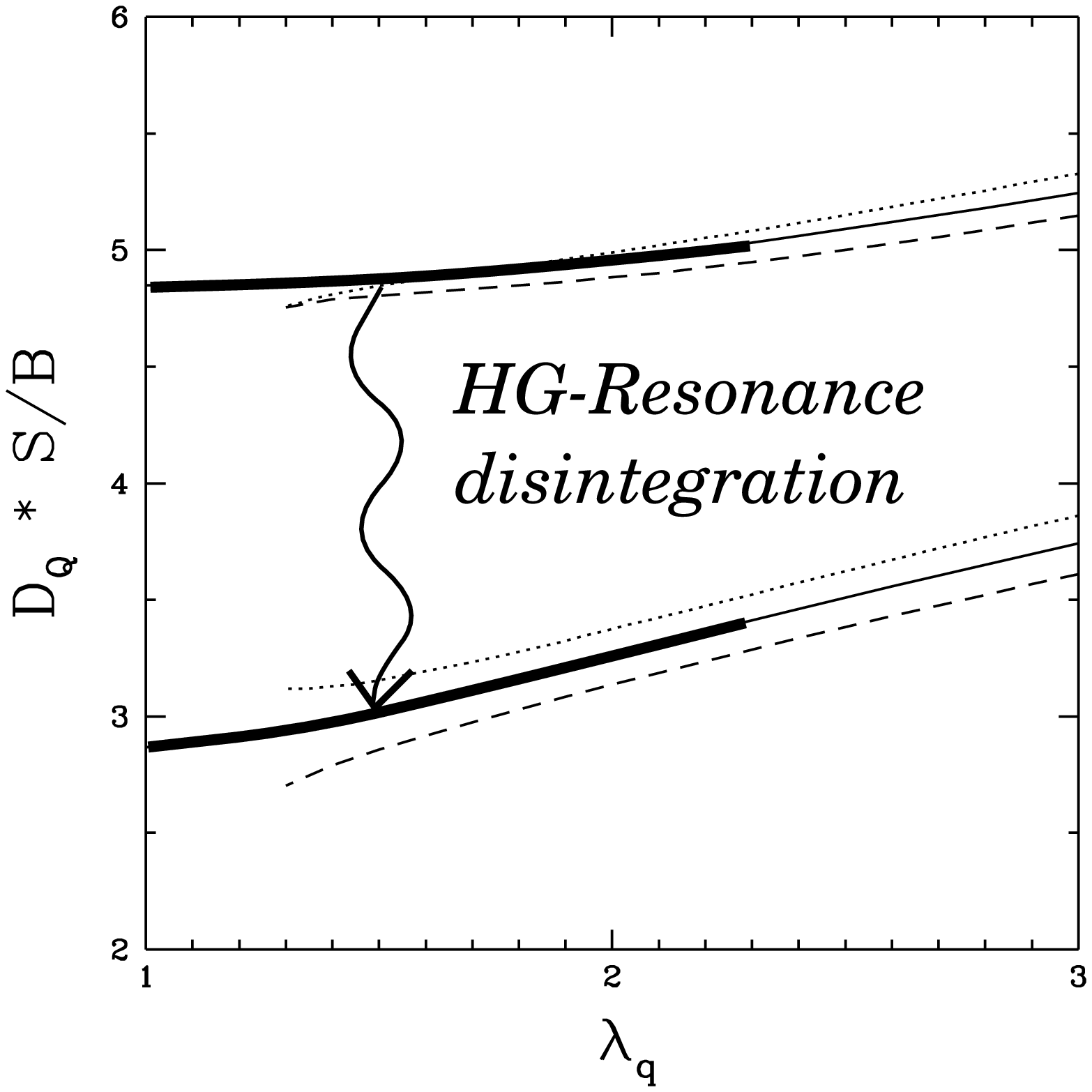}}
\vspace*{-4.cm}
\caption{ \small 
The product $D_{\rm Q}\cdot({\cal S/B})$ before (upper curves) and after 
(lower curves) resonance disintegration, as a function of $\lambda_{\rm
q}$, for fixed $\lambda_{\rm s}=1\pm0.05$  and conserved zero
strangeness in HG. Note the suppressed zero on the vertical axis.
\protect\label{fig2entro}}
\vfill
\end{figure}

The  EMU05 collaboration\cite{EMU05} has studied  S--Pb collision at 200
GeV A using a thin Pb-foil placed in front of a emulsion stack. They have
concentrated the analysis on central collision events requiring a total
final state charged multiplicity to be greater than 300,
corresponding to a total central particle multiplicity between 450--1000.
They present the multiplicities per interval in rapidity, separately for
positive and negative particles. From these data, one can determine the
number of protons in the central  fireball 
\begin{equation}    
 N_{\rm p}={\int}_{\!\!\hbox{\rm\scriptsize CR}}\frac{d(N^+-N^-)}
{dy}dy \simeq 28\,.\end{equation}
This gives a baryonic number of the fireball $B$ nearly equal to 60 and 
corresponds to a stopping $\eta_B\simeq$  50\%   for the participant
nucleons, the reference value being obtained from the geometric 
tube-like interaction region  model, in this case: 
\begin{equation}    
 N_F=\frac{3}{2}A^{2/3}_PA^{1/3}_T+A_P
=\frac{3}{2}32^{2/3}207^{1/3}+32\simeq 120\,.\end{equation}
If we take for the entropy per particle the thermal value for an isolated 
system of massless particles  $S/N\sim 4$, (in fact at high $T$ for
massive  hadrons $S/N$ is $ \ge 4$), we obtain taking the full
multiplicity $700\pm250$: $$\ds S/B\sim 50\pm20\,.$$
A more precise analysis of this experiment can be perform by starting 
from the ratio:
\begin{equation}    
   D_{\rm Q} \equiv \frac{\displaystyle{dN^+\over dy} - {dN^- \over dy}}
               {\displaystyle{dN^+\over dy} + {dN^- \over dy}}\,,
\end{equation}
which is found in  this experiment to be $0.085\pm 0.010$ in the central
region (see Fig.\,\ref{fig1entro}).
 
We note that in the numerator of $D_{\rm Q} $ the charge of particle
pairs produced cancels and hence this value is effectively a measure of
the baryon number, but there is a significant correction arising from
the presence of strange particles. The denominator is a measure of the
total multiplicity --- its value is different before or after disintegration
of the produced unstable hadronic resonances. Using as input the
distribution of final state particles as generated within the hadron gas
final state it is found \cite{entropy} that $D_Q\cdot S/B$  is nearly
independent of the thermal  parameters and varies between 4.8, before
disintegration of the  resonances,  to 3 after disintegration (see 
Fig.\,\ref{fig2entro}). It is less than clear that 
the conservative assumption of
the hadronic gas final state is justified, and if we were to assume that,
e.g., the deconfined state is hadronizing into the final hadronic
particles suddenly,
production of resonances would be largely suppressed, and thus the
greater value 4.8 would apply. Using the above value of $D_Q$ 
we find for the entropy per baryon of the final state we obtain for the
entropy content of the emulsion events analyzed by EMU05:
$$35<S/B<60\,,$$
with the upper limit applying in the case that few heavy meson resonances
are produced.

We recall that in table~\ref{sollresult1} we have listed in the bottom entry
the entropy content of the final state in different hadronization scenarios. 
These are the chemical equilibrium results (except for strangeness flavor)
for the hadronic gas formed at the given statistical conditions. We note 
that low temperature hadronization scenario appears to be consistent with 
the constraints of the EMU05 experiments. Hadronization at these conditions
cannot, however, be considered seriously to lead to chemically equilibrated 
meson and baryon abundances, since the observed final state meson abundance
in on average 50\% greater than the equilibrium HG value \cite{analyze}. 
Several features of final state particle abundances thus suggest that the 
dissociation of the fireball of dense hadronic matter is not leading to 
chemical equilibrium abundances of final state particles 
--- this is consistent
with  the sudden `explosion'  picture of this process we are employing.

It should be clearly observed at this point that the presence of the high
entropy phase precludes a reaction picture of relativistic heavy ion 
collisions based on conventional thermal hadron gas and adiabatic fireball 
evolution: the observed high inverse transverse mass slope $T\simeq 230$ MeV
in S--W interactions at 200A GeV implies a initial specific entropy content 
(see also table~\ref{sollresult1}) which is 1/3 of the final 
state value. Such a 
state would have to undergo very entropy generating expansion, and to best 
of our knowledge there is no experimental evidence for this, nor is there 
a suggestion of a mechanism that could accomplish such a task. 
On the other hand, 
we find that the entropy content of a dense hot QGP fireball formed at 
the required chemical conditions and later evolving adiabatically (see 
table \ref{bigtable}) just corresponds to the expected value of specific 
entropy. This allows to conclude that the only currently known reaction 
picture of relativistic heavy ions involves formation of the thermal 
QGP fireball.

\begin{figure}[tb]
\vspace*{-0.8cm}
\centerline{\hspace*{7.6cm}\psfig{height=9.5cm,figure=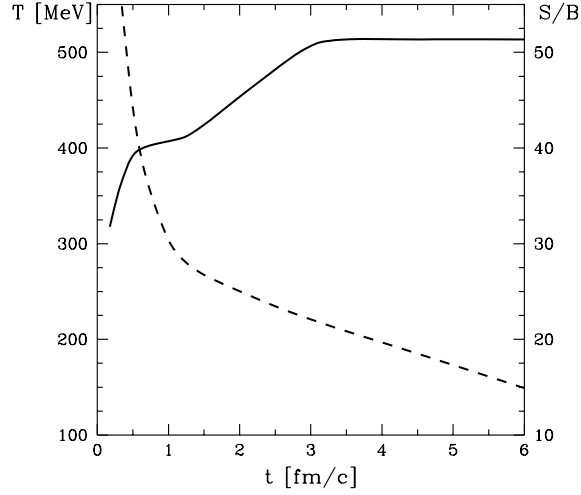}}
\vspace*{-2.5cm}
\caption{ \small 
The qualitative evolution of $T$ (dashed line) and $S/B$ (solid line)
versus  time, for the S--W case at $E/B=8.8$  GeV and stopping parameter
$\eta =0.5$. \protect\label{fig3entro}}
\end{figure}

\baselineskip=12.9pt
In our model of thermal fireball evolution the excess entropy is present in 
the early formation stage of the fireball, and as discussed, the
mechanisms of (rapid) entropy formation are not understood. 
We show in Fig.\,\ref{fig3entro}, for the  S--W case at $E/B=8.8$ GeV and
stopping parameter $\eta=0.5$, the qualitative evolution as function
of time of {$T$} and {$S/B$}. This result was obtained wit the hypothesis
$$\ds\gamma_{\rm g}=\gamma_{\rm q}\,,
\qquad \gamma_{\rm s}=\frac{1}{5}\gamma_{\rm q}
\quad{\rm when}\quad \gamma_q\le 1.$$  
We observe the considerable drop in temperature from the initial stage to the
freeze-out point. {\it A contrario}, the entropy content which determines
the final particle multiplicities evolves very little and {70\%}  of the
entropy is already present when quarks and gluons are still far from the
equilibrium abundance. Practically all the rise in entropy 
is due to the formation
of the strange flavor ({20\%}), the remaining 10\% arise as consequence
of the slight change in the value of $\lambda_{\rm q}$ given that we
enforce during the collision period the condition of dynamical pressure
equilibrium.
 
\vspace{-3mm}
\subsection{Final state strange baryon yields} 
\label{results}
\vspace{-1mm}

\begin{figure}[tb]
\vspace*{-.7cm}
\centerline{\hspace*{-0.6cm}
\psfig{width=10cm,figure=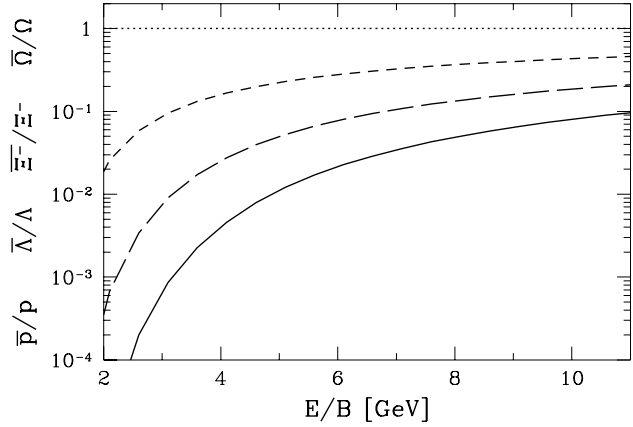}
}
\vspace*{-0.6cm}
\caption{ \small 
Antibaryon to baryon abundance ratios as function of energy per 
baryon $E/B$ in a QGP-fireball: $R_{\rm N}=\bar p/p$ (solid line),
$R_\Lambda=\overline{\Lambda}/\Lambda$ (long-dashed line), 
$R_\Xi=\overline{\Xi}/\Xi$ (short-dashed line)
 and $R_\Omega=\overline{\Omega}/\Omega$  (dotted line)
\protect\label{eqratios}}
\end{figure}

The ratios of strange baryon  to strange
antibaryon abundance, considering the same type of particles,
depends only on the chemical properties of the source. 
We show in Fig.\,\ref{eqratios} the three ratios and also $\bar
p/p$\,. Since we assume $\lambda_{\rm s}=1$\,, we obtain here in
particular $R_\Omega=\lambda_{\rm s}^{-6}=1$\,.
However, since  some re-equilibration  is
to be expected towards the HG behavior $\lambda_{\rm s}>1$, we expect
$\lambda_{\rm s}=1+\epsilon$, with $\epsilon$ small, 
and thus for this ratio $R_\Omega=1-6\epsilon<1$.\\ 
\begin{figure}[tb]
\vspace*{-0.5cm}
\centerline{\hspace*{-0.7cm}
\psfig{width=10cm,figure=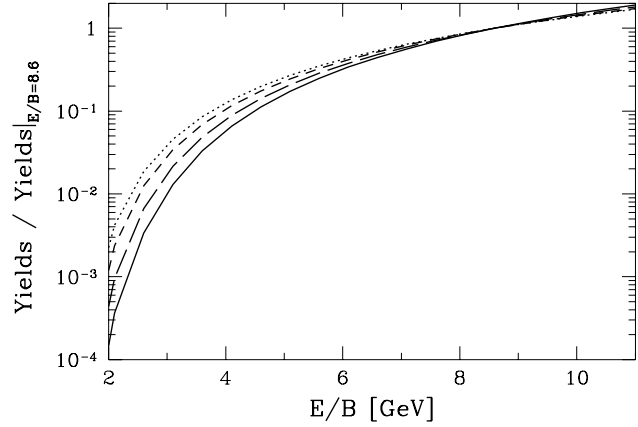}}
\vspace*{-0.6cm}
\caption{ \small 
Relative antibaryon yields as function of $E/B$ in a QGP-fireball
with $\gamma_{\rm s}=1$\,.
$\overline{p}$ (solid line), $\overline{\Lambda}$  (long-dashed line)
$\overline{\Xi^-}$ (short-dashed line) and $\overline{\Omega}$ (dotted
line), all normalized to their respective yields at $E/B=8.6$ GeV\,.
\protect\label{PLYIELDSNORM}}
\end{figure}
\baselineskip=12.5pt
 \indent An interesting question which arises quite often is how the individual
particle and in particular total antibaryon yields vary with energy. 
Eq.\,(\ref{4a}) allows to determine the absolute particle 
yields as function of fireball energy. Considerable uncertainty
is arising from the off-equilibrium nature of the hadronization 
process, which in particular makes it hard to estimate how the 
different heavy particle resonances are populated, and also, 
how the abundance factors $C_{\rm B}^{\rm s}$ vary as function 
of energy. Some of this uncertainties are eliminated when we 
normalize the yields at an energy, which we take here to be 
the value $E/B=8.6$ GeV which is applicable to the 
SPS experiments. In  Fig.\,\ref{PLYIELDSNORM} 
the so normalized yields of antibaryons  taking the freeze-out 
temperature $T=150$ MeV are shown (we also assume $\gamma_{\rm s}=1,\, 
\eta_{\rm p}=1$ and absence of any re-equilibration after particle
emission/production). 

These yields are decreasing in qualitatively 
similar systematic fashion with energy, as would be expected from the 
microscopic considerations, but the decrease of more strange antibaryons  is
less pronounced. The quantitative point to note is that at AGS 
($E/B=2.5$ GeV) the yield from a disintegrating QGP-fireball is a  factor
100--400 smaller compared to yields at $E/B=$8.6 GeV. Since the  particle
rapidity density $dN/dy$ is not that much smaller at the lower
energies (recall that the specific entropy, see~table \ref{bigtable}\,,
drops only by factor 3.5, implying a reduction in specific multiplicity 
by a factor 5), it is considerably more difficult at the lower energies 
to search for antibaryons than it is at higher energies. We should 
remember that the results presented in Fig.\,\ref{PLYIELDSNORM} 
are obtained assuming formation of the QGP-fireball and same 
freeze-out and hadronization conditions for all energies shown.
We have obtained the result presented in Fig.\,\ref{PLYIELDSNORM} assuming
that the strange phase space saturation is given by the dynamical evolution
of a three dimensional radial expansion and freeze-out at $T=140$ MeV. Our 
result was that all particle yields behave in the same qualitative fashion,
with the curves falling nearly directly on top of each other. This implies, as
we shall see in more detail below, that in a dynamical calculation
there is very little, if any variability in particle ratios with energy. 
 
\begin{figure}[t]
\vspace*{1.5cm}
\centerline{\hspace*{1.2cm}
\psfig{width=10cm,figure=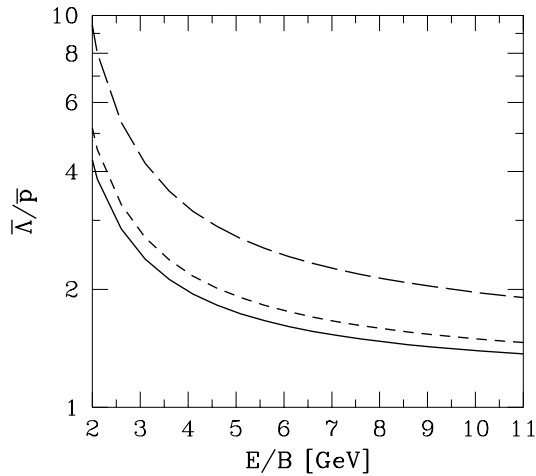}}
\vspace*{-1.8cm}
\caption{ \small 
Strange antibaryon ratio $\overline{\Lambda}/\overline{p}$,
as function of $E/B $ in a QGP-fireball for $\gamma_{\rm s}=1$;  
solid lines are for full phase space coverage, 
short dashed line  for particles with $p_\bot\ge 1$ GeV and
long dashed line for particles with $m_\bot \ge 1.7$ GeV.
 \protect\label{BARLP}}
\end{figure}

\begin{figure}[t]
\vspace*{1.9cm}
\centerline{\hspace*{1.2cm}
\psfig{width=10cm,figure=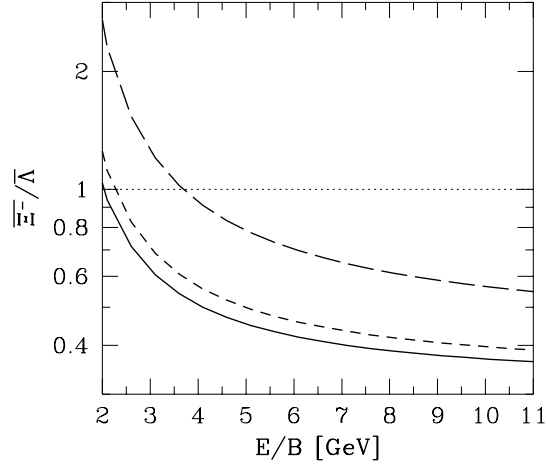}}
\vspace*{-1.8cm}
\caption{ \small 
Strange antibaryon ratio
$\overline{\Xi^-}/\overline{\Lambda}$ for $\gamma_{\rm s}=1$, 
with the same conventions as in Fig.\,\protect\ref{BARLP}.
 \protect\label{BARXL}}
\end{figure}

\begin{figure}[t]
\vspace*{1.5cm}
\centerline{\hspace*{1.2cm}
\psfig{width=10cm,figure=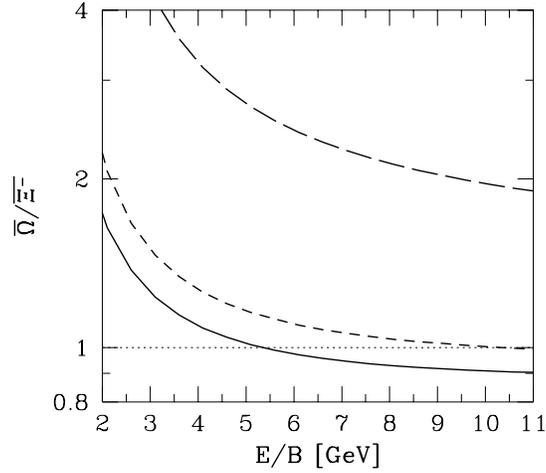}}
\vspace*{-1.9cm}
\caption{ \small 
Strange antibaryon ratio
$\overline{\Omega}/\overline{\Xi^-}$  for $\gamma_{\rm s}=1$, 
with the same conventions as in Fig.\,\protect\ref{BARLP}.
 \protect\label{BAROX}}
\end{figure}

We next present the particle ratio results 
assuming first $\gamma_{\rm s}=1$\,, and we turn to consider the variation of
$\gamma_{\rm s}$ with energy below. The choice $\gamma_{\rm s}=1$ is 
appropriate if we had a relatively large, long-lived QGP
fireball created in central collisions of 
largest available nuclei, or if our computation of the variation 
of $\gamma_{\rm s}$ with energy was underestimating strongly the 
actual production rates. In the Figs.\,\ref{BARLP}--\ref{BAROX}  
we show three ratios 
and for each ratio three results:  solid lines depicts the result for 
the full phase space coverage, short dashed line for particles with 
$p_\bot\ge 1$ GeV  and long dashed line for particles with $m_\bot  \ge
1.7$ GeV. In  Fig.\,\ref{BARLP} we show the ratio 
$\overline{\Lambda}/\bar p$, in  Fig.\,\ref{BARXL} the ratio 
$\overline{\Xi^-}/\overline{\Lambda}$ and in  Fig.\,\ref{BAROX}  the
ratio $\overline{\Omega}/\overline{\Xi^-}$. Because $\lambda_{\rm q}$
rises with decreasing $E/B$ and we have kept $\gamma_{\rm
s}=1$\,, we find that these three ratios increase quite strongly {\it as the
collision energy is reduced}. 
The behavior of particle ratios shown in Figs.\,\ref{BARLP}--\ref{BAROX} 
may be of considerable importance, since in reaction models in
which QGP is not assumed and the particles 
are made in a sequence of microscopic collisions these ratios
{\it do increase} from production thresholds with the collision
energy, reflecting in this  behavior the phase space factors 
inherent in the reaction cross section. 

\begin{figure}[tb]
\vspace*{2.cm}
\centerline{\hspace*{1.5cm}
\psfig{width=11cm,figure=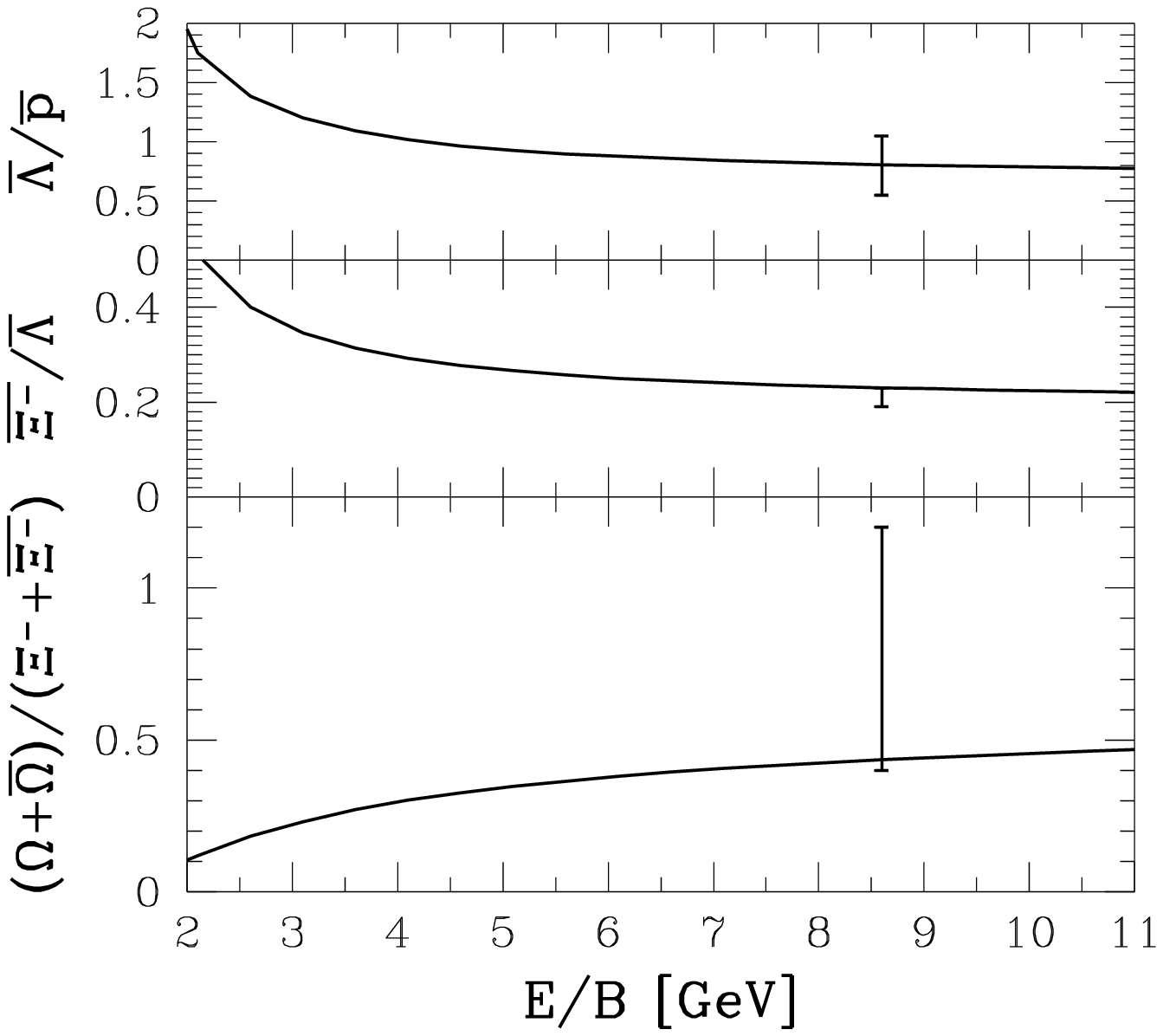}}
\vspace*{-2.1cm}
\caption{ \small 
Strange antibaryon ratios for S--W/Pb collisions as function 
of $E/B$ in a QGP-fireball:
$\overline{\Lambda}/\overline{p}$ (full phase space),
$\overline{\Xi^-}/\overline{\Lambda}$ for $p_\bot>1.2$ GeV and
$(\overline{\Omega}+\Omega)/(\overline{\Xi^-}+\Xi^-)$ for
$p_\bot>1.6$ GeV; experimental results 
shown are from experiments NA35, WA85.
 \protect\label{sratios}}
\end{figure}

\baselineskip=12.9pt
We now directly compare our theoretical results with the
experimental data we have reported in section \ref{expres}:
the WA85 $\Omega/\Xi^-$ production ratio obtained  for the S-W at 200A GeV 
\cite{Omega}; the $\overline{\Lambda}/\bar p$ ratio
of the NA35 collaboration obtained  for the S--Au system at 200A GeV 
\cite{NA35pbar}.  Fig.\,\ref{sratios} shows a comparison of our ab initio
calculation and the pertinent experimental results. We use the same cuts
on the range of $p_\bot$ as in the experiment: the experimental points
show the results $\overline{\Lambda}/\bar p\simeq 0.8\pm 0.25$  (NA35)
for full phase space,
$\overline{\Xi^-}/\overline{\Lambda}=0.21\pm0.02$ (WA85) for $p_\bot>1.2$
GeV; and $(\Omega+\overline{\Omega})/(\Xi^-+\overline{\Xi^-})=0.8\pm0.4$
(WA85) for $p_\bot>1.6$ GeV. The values $\gamma_{\rm s}=0.70 $
and $\eta_{\rm p}=0.5$  also bring about good agreement of our model 
with the precise value of $\overline{\Xi^-}/\overline{\Lambda}$. 
  Fig.\,\ref{sratios} shows also the impact of the change of the
collision energy on these results, using 50\% stopping, rather than 
$\eta=1$ used in  Figs.\,\ref{BARLP}--\ref{BAROX}. We can
conclude that the fact that the  two ratio
$\overline{\Lambda}/\bar p$ (NA35) and 
$(\Omega+\overline{\Omega})/(\Xi^-+\overline{\Xi^-})$ (WA85)
are satisfactorily  explained, provides a very  nice confirmation of  the
consistency of the thermal QGP fireball model. Moreover, considering that we
have now the power to compute  without ad hoc assumptions 
within the framework of dynamical model, it is quite remarkable that such a 
good agreement with the two very recent results could be attained. 

\begin{figure}[b]
\vspace*{1.6cm}
\centerline{\hspace*{1.2cm}
\psfig{width=10cm,figure=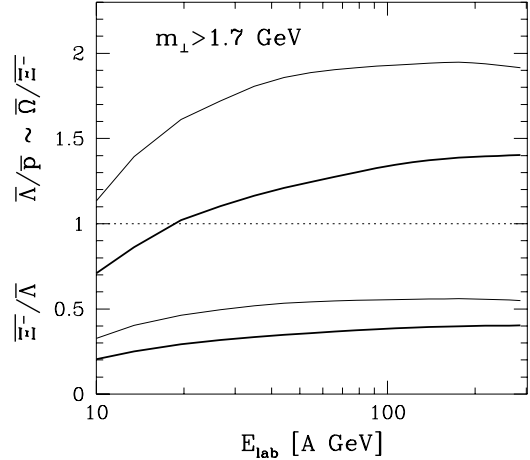}}
\vspace*{-1.9cm}
\caption{ \small 
Fixed $m_\bot$ $\overline{\Lambda}/\overline{p}$ 
and $\overline{\Xi^-}/\overline{\Lambda}$ as function of
beam energy $E_{\rm lab}$ in the collision (logarithmic scale)
arising from a Pb--Pb central interaction fireball, 
taking into account
variation of $\gamma_{\rm s}$ computed with running 
$\alpha_{\rm s},\,m_{\rm s}(M_Z)=90$ MeV, for thick 
lines $ \alpha_{\rm s}(M_Z)=0.102$, for thin lines 
$\alpha_{\rm s}(M_Z)=0.115$.
\protect\label{stratiogam}} 
\end{figure}

It is worthwhile to note that even when we incorporate in 
these strange antibaryon ratios in Figs.\,\ref{BARLP}--\ref{BAROX} 
the variation of $\gamma_{\rm s}$ shown in Fig.\,\ref{gammasr},
with considerable decrease of $\gamma_{\rm s}$ with decreasing 
energy (here shown as function of lab energy in the collision,
(logarithmic scale), we still retain the remarkable
behavior that the ratios do not decrease significantly 
with decreasing energy down to
the energy thresholds for the production of the
(multi)strange (anti)baryons. There can be little
doubt that if this behavior should be observed down to some low
energy in heavy ion collisions, and subsequently a sudden drop 
should occur, we could safely conclude about a change in the reaction
scenario, and probably even pinpoint the mechanisms presented here
as being at the origin of this result.

\begin{figure}[tb]
\vspace*{-3.7cm}
\centerline{\hspace*{-2.cm}
\psfig{width=12.5cm,figure=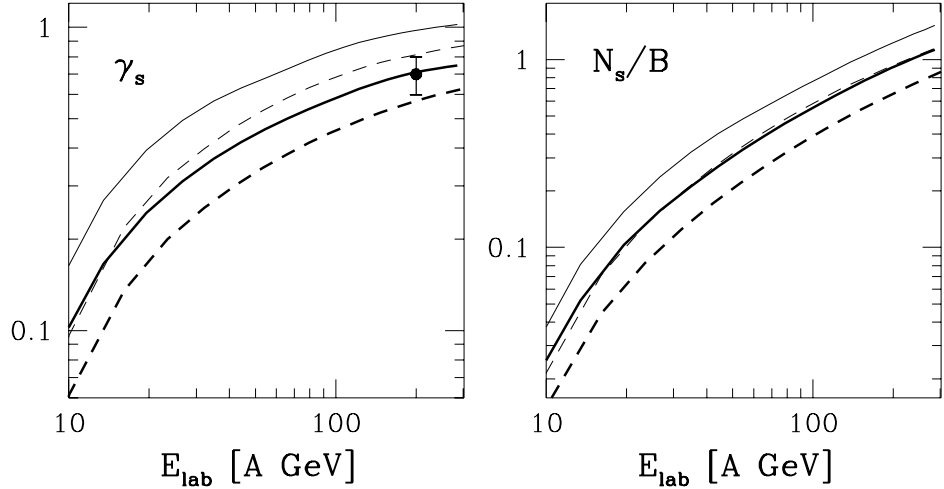}
}
\vspace*{-0.3cm}
\caption{ \small
$\gamma_{\rm s}(t_{\rm f})$ and $N_{\rm s}/B$ 
as function of beam energy for 
central  S--W/Pb collisions (dashed lines) and  Pb--Pb collisions 
(solid lines) assuming $m_{\rm s}(M_Z)=90$ MeV, for thick 
lines $ \alpha_{\rm s}(M_Z)=0.102$, for thin lines 
$\alpha_{\rm s}(M_Z)=0.115$, 
three dimensional expansion of the fireball with $v=c/\protect\sqrt(3)$, and 
stopping 50\% (S--W/Pb), 100\% (Pb--Pb). For $\gamma_{\rm s}$ 
we  take freeze-out at $T_{\rm f}=140$ MeV --- the vertical bar corresponds 
to the value of  $\gamma_{\rm s}$ found in S--W data 
analysis{\protect\cite{Raf91}}\label{figvarelab}.}
\end{figure}

Finally, let us redraw some of the above results in Fig.\,\ref{figvarelab} 
as function of the beam energy, for the central collisions of symmetric 
(Pb--Pb, Au--Au)  and asymmetric (S--W/Pb) systems. 
To the left in Fig.\,\ref{figvarelab}  we present 
$\gamma_{\rm s}(t_{\rm f}) $ used in Fig.\,\ref{stratiogam}, 
which depends as discussed on both the 
initial production and the enrichment of the strange phase space 
occupancy by the dilution effect, and thus on freeze-out conditions, 
here assumed to occur at $T_{\rm f}=140$~MeV. High values of
$\gamma_{\rm s}$ should accompany low freeze-out temperature,
provided that there has been extreme initial conditions allowing
to produce strangeness. To the right we present the specific 
strangeness yield. We recall that for S--Ag collisions at 200A GeV
a recent evaluation of the strangeness yield leads to 
$N_{\rm s}/B=0.86\pm0.14$ (see table~4 of Ref.\cite{GR96}) which 
fits also nicely into the range of values shown at 200 GeV in 
Fig.\,\ref{figvarelab}. 

\section{   
Summary and conclusions}\label{endrem}

While it is today impossible, based solely on current experimental
results, to claim that the `macroscopic', deconfined QCD-phase has been
discovered in the 200A GeV collisions, we
have developed the experimental and theoretical 'strangeness'
tools which allow to
resolve in the foreseeable future this question. Today, we can affirm
that the natural hypothesis of a QGP-fireball provides us within the
thermal fireball model with a very
comprehensive and satisfactory explanation of all available experimental
data both at 200A GeV and 10-15A GeV. While the lower energy is
also well described by the HG type models, the 200A GeV results are
incompatible with an assumed HG structure of the interacting hadronic
matter in the thermal fireball, or cascades of conventional hadrons 
studied in transport approaches \cite{Csernai}.

We have described in detail how production and 
final state manifestation of  strangeness in antibaryon yields,
 can help today to identify and study the properties of the deconfined
phase. Our exploration of thermal charm production has shown that open
charm could become an interesting probe of initial conditions 
reached at LHC energies. 

We have shown that the experimental results 
suggest that the thermal (kinetic) 
equilibrium is established, while the chemical (particle abundance)
equilibrium in the processes governing final state particle freeze-out is
 not achieved in 200A GeV collisions. Motivated by the absence of chemical
particle abundance equilibrium, we developed and used here a picture of
final state hadron production which involves rapid disintegration of the 
QGP-fireball. Central to the particle abundances are then
the chemical properties of the QGP-fireball and we have discussed these 
comprehensively as function of collision energy and stopping. 

We have shown how a kinetic reaction picture model allows to determine 
 the thermal conditions reached in high density deconfined matter
created in heavy ion collisions. It is based on the observation that
during the collision the compression of the quark-gluon matter can
proceed until the internal pressure exerts sufficiently
strong counter force. Using so established initial conditions and adiabatic 
fireball expansion, we have shown that the thermal conditions we
find at the end of the evolution,
see bottom of table~\ref{bigtable}\,,
are in good agreement  with our expectations derived from
particle yields seen in {S--Pb/W} 200 GeV A collisions. To wit 
we needed to make a reasonable choice of the physical parameters: at
$T=250$--$300$ MeV we took $\alpha_{\rm s}=0.6$, justified for the purpose
of strangeness production 
by our study of running QCD properties in section \ref{runalfasec},
for  stopping we adopted $\eta=50$\% based on experimental data \cite{stop},
 about equal for baryon number,
energy and momentum. Given these assumptions, we were able to study 
 the current strange particle data at 200A GeV and
have reached good agreement with experiment.

We studied in detail  the production and evolution of total strangeness
abundance in a dynamical QGP fireball evolution model. As expected we 
found that  the large strangeness abundance produced in 
the early stages will not be reannihilated. We presented the yields
 as function of both, the collision energy and number of participating 
baryons (impact parameter). We saw that the yield is rising linearly
with the CM-energy per baryon, and that as far as data is available, 
it is in excellent agreement with the 200A GeV experiment S--Ag \cite{GR96}.
We also explored in detail the evolution of the strangeness phase 
space occupancy. Strangeness can overpopulate the 
available phase space at plasma disintegration at low freeze-out temperatures, 
and thus strange antibaryon abundances could 
show $\gamma_{\rm s}>1$, even though 
our reaction picture leads to values $\gamma_{\rm s}\simeq 1$\, --- as this
discussion shows, $\gamma_{\rm s}$ is  a sensitive probe of both the 
initial, and final freeze-out conditions.

The last figures shown above in many ways are fruit of the many 
individual developments we presented here. 
The computation of relative strange antibaryon yields 
shown in Fig.\,\ref{stratiogam} was done without ad hoc
parameters but there remains some uncertainty about 
$m_{\rm s}(M_Z)$ and the reaction mechanism, here in particular 
stopping fraction $\eta$\,. This result is special as it
 shows that the strange antibaryon ratio 
as the energy of the interaction is reduced, but deconfined phase still
reached, is at worse slightly declining, but still anomalously large at
AGS (10A GeV) or at minimal energies reachable at SPS (40A GeV). This
justifies an exploration of the energy behavior of this
 central rapidity observable. 

Similar comments apply to the less directly measured observables, the  
freeze-out strangeness occupancy $\gamma_{\rm s}(t_{\rm f})$ and the 
specific strangeness yield, shown above in Fig.\,\ref{figvarelab}.
In particular the large strangeness yield deserves attention, which as we 
noted earlier is essentially linearly proportional to the available CM-energy
in the fireball.

Our results overall imply that in key features 
the strange particle production results obtained at 200A GeV, are
consistent with the QGP hypothesis of the central, thermal fireball. 
However, in order to ascertain the possibility that indeed
the QGP phase is already formed at 200A GeV a more systematic exploration
as function of collision energy of these observables would be needed --- 
conclusions drawn from a small set of experimental results
suffer from the possibility that some coincidental and unknown 
features in the reaction mechanisms could simulate just the observed QGP-like
properties. It is highly unlikely that this would remain the case, should
a key feature such as collision energy be varied. 
We stress that our description and hence the anomalous behavior
of particle production discussed here is based on
collective mechanisms (QGP-fireball), which is intrinsically 
different from  microscopic approaches, in
particular when these are based on a hadronic cascade picture. 
 
We hope to have conveyed to the reader the reason why we 
firmly believe, considering the results we
have reported, that experimental data on strange (anti)baryon 
production provides the best hadronic 
signatures, and diagnostic tools, of the deconfined matter. It seems
that the discovery of the deconfined QGP-phase of hadronic
matter is just around the corner.

\vspace{0.5cm}
\noindent J.R. acknowledges partial support by  DOE, grant
               DE-FG03-95ER40937\,. \\
Laboratoire de Physique Th\'eorique et Hautes Energies (LPTHE)
is: Unit\'e  associ\'ee au CNRS UA 280\,.

\end{document}